\documentclass[fleqn,usenatbib]{mnras}

\usepackage[T1]{fontenc}
\usepackage{ae,aecompl}

\usepackage{longtable,lscape}
\usepackage{multicol}
\usepackage{afterpage}
\usepackage{natbib}
\usepackage{txfonts}
\usepackage{hyperref}
\usepackage[labelformat=simple,labelfont=bf,labelsep=period]{caption}
\usepackage{booktabs}

\def\m2s2{\,m$^{2}$\,s$^{-2}$} 

\usepackage{graphicx}	
\usepackage{amsmath}	
\usepackage{amssymb}	
\usepackage{multirow}
\usepackage{afterpage}
\usepackage{pdflscape}

\title[The binarity of massive close-in giant planets]{A high binary fraction for the most massive close-in giant planets and brown dwarf desert members}

\author[C. Fontanive et al.]{
C. Fontanive$^{1,2}$\thanks{E-mail: \href{mailto:fontan@roe.ac.uk}{fontan@roe.ac.uk}},
K. Rice$^{1,2}$,
M. Bonavita$^{1,2}$,
E. Lopez$^{3,4}$,
K. Mu{\v z}i{\'c}$^{5}$
and B. Biller$^{1,2}$\\
\\
$^{1}$SUPA, Institute for Astronomy, University of Edinburgh, Blackford Hill, Edinburgh EH9 3HJ, UK\\
$^{2}$Centre for Exoplanet Science, University of Edinburgh, Edinburgh EH9 3HJ, UK\\
$^{3}$NASA Goddard Space Flight Center, 8800 Greenbelt Rd, Greenbelt, MD 20771, USA\\
$^{4}$GSFC Sellers Exoplanet Environments Collaboration, NASA GSFC, Greenbelt, MD 20771, USA\\
$^{5}$CENTRA, Faculdade de Ci\^{e}ncias, Universidade de Lisboa, Ed. C8, Campo Grande, P-1749-016 Lisboa, Portugal
}

\date{Accepted 2019 March 5. Received 2019 February 27; in original form 2019 January 17}

\pubyear{2019}

\begin{document}

\label{firstpage}
\pagerange{\pageref{firstpage}--\pageref{lastpage}}
\maketitle

\begin{abstract}
Stellar multiplicity is believed to influence planetary formation and evolution, although the precise nature and extent of this role remain ambiguous. We present a study aimed at testing the role of stellar multiplicity in the formation and/or evolution of the most massive, close-in planetary and substellar companions. Using past and new direct imaging observations, as well as the \textit{Gaia} DR2 catalogue, we searched for wide binary companions to 38 stars hosting massive giant planets or brown dwarfs ($M > 7$ M$_\mathrm{Jup}$) on orbits shorter than $\sim$1 AU. We report the discovery of a new component in the WASP-14 system, and present an independent confirmation of a comoving companion to WASP-18. From a robust Bayesian statistical analysis, we derived a binary fraction of $79.0^{+13.2}_{-14.7}$\% between 20$-$10,000 AU for our sample, twice as high as for field stars with a 3-$\sigma$ significance. This binary frequency was found to be larger than for lower-mass planets on similar orbits, and we observed a marginally higher binary rate for inner companions with periods shorter than 10 days. These results demonstrate that stellar companions greatly influence the formation and/or evolution of these systems, suggesting that the role played by binary companions becomes more important for higher-mass planets, and that this trend may be enhanced for systems with tighter orbits. Our analysis also revealed a peak in binary separation at 250 AU, highlighting a shortfall of close binaries among our sample. This indicates that the mechanisms affecting planet and brown dwarf formation or evolution in binaries must operate from wide separations, although we found that the Kozai-Lidov mechanism is unlikely to be the dominant underlying process. We conclude that binarity plays a crucial role in the existence of very massive short-period giant planets and brown dwarf desert inhabitants, which are almost exclusively observed in multiple systems.
\end{abstract}



\section{Introduction}
\label{intro}

In the search for analogues to the planets in our own Solar System, exoplanet studies originally firmly excluded known binary systems, despite the fact that about half of Solar-type stars are found in multiple systems \citep{Raghavan2010}. Serendipitous discoveries and subsequent dedicated surveys later revealed that a significant fraction of exoplanets are actually found in binary star systems (e.g. \citealp{Patience2002,Desidera2004,Mugrauer2006,Mugrauer2009}), mostly with binary separations of at least a few hundred AU. These findings led to numerous studies investigating how stellar binarity affects planet formation and the characteristics and demographics of planetary populations (e.g. \citealp{Desidera2007,Eggenberger2007,Eggenberger2011,Daemgen2009,Adams2012,Adams2013,Ginski2012}). The dominant results that emerged from these surveys were a strong deficit of binary companions within $\sim$50$-$100 AU for planet hosts \citep{Roell2012,Bergfors2013,Wang2014a,Wang2014b,Kraus2016}, and the idea that massive short-period planets appear to be preferentially found in multiple-star systems \citep{ZuckerMahez2002,Eggenberger2004}.

These studies, however, focused primarily on systems in which the planet had a mass less than $\sim$4 M$_{\rm Jup}$. Theoretical calculations \citep{Kratter2010,Forgan2011} and numerical simulations \citep{Stamatellos2008,Stamatellos2013,Hall2017} both suggest that planets that form via disc fragmentation in gravitationally unstable discs \citep{Boss1998} typically have masses above $\sim$4 M$_{\rm Jup}$. Therefore the planets in these existing studies probably formed via the standard core accretion scenario (CA; \citealp{Pollack1996}), rather than via gravitational instability (GI). 

When it comes to planets that formed via core accretion, binarity on close separations is generally considered to have a negative influence (see \citealp{Thebault2015} for a review of planet formation in binaries and the issues introduced by the presence of a close binary companion). Theoretical studies have concluded that close stellar companions can hinder planet formation by stirring up protoplanetary discs (e.g. \citealp{Mayer2005}), tidally truncating the discs (e.g. \citealp{Pichardo2005,Kraus2012}), or leading to the ejection of planets \citep{Kaib2013,Zuckerman2014}. More compact, truncated discs generally have just enough mass left to form a low-mass Jovian planet \citep{Jang-Condell2008}, and planet formation is then further complicated by the very short lifetime ($\lesssim$1 Myr) of these truncated discs \citep{Kraus2012}.

On the other hand, \citet{Batygin2011} and \citet{Rafikov2013} predict that stellar companions should have little influence on planetesimal growth. It has also been proposed that the presence of an outer companion could raise spiral arms in protoplanetary discs, creating regions of high gas and particle densities, favourable for planetesimal formation \citep{Youdin2005} and pebble accretion \citep{Johansen2007,Lambrechts2014}. For example, the spiral arm structures observed in the disc around HD 100453 \citep{Wagner2015} may be due to perturbations from the M-dwarf companion \citep{dong2016}, located at 120 AU from the primary and originally reported by \citet{Chen2006}. Similarly, the asymmetric disc of HD 141569 is attributed to the stellar companions in this triple system \citep{Augereau2004}. In the ``Friends of hot Jupiters'' series of papers, \citet{Knutson2014}, \citet{Piskorz2015} and \citet{Ngo2015,Ngo2016} found a binary fraction 3 times higher for hosts to hot Jupiters (mostly up to $M_p \sim 4$ M$_{\rm Jup}$) than for field stars on separations of 50$-$2000 AU, and concluded that wide binary systems may either facilitate the formation of Jovian planets, or help the inward migration of planets formed at wider separations.

It has also been suggested that binary companions could induce the inward migration of planets through secular interactions such as the Kozai-Lidov mechanism \citep{Kozai1962,Lidov1962}. In this scenario, an outer companion with a large mutual inclination between the planetary and binary orbits can excite large periodic oscillations of the eccentricity and inclination of the planet. Tidal interactions between the planet and its host star can then drive the planet onto a final orbit with a very small orbital separation when compared to its initial location (e.g. \citealp{Fabrycky2007,Naoz2012,Petrovich2015}). In particular, the Kozai-Lidov migration process has been proposed to explain the high obliquities often observed in hot Jupiters, although recent studies indicate that this mechanism can only account for about 20$-$30\% of the observed hot Jupiter population \citep{Naoz2012,Ngo2016}. Similarly, it has been suggested \citep{Rice2015} that Kozai-Lidov oscillations could drive planetary-mass bodies that form on wide orbits via GI onto short-period orbits. Since disc fragmentation preferentially forms massive planets or brown dwarfs \citep{Kratter2010,Stamatellos2008,Forgan2011}, such a process would tend to be associated with more massive planets ($M_p \gtrsim 4$ M$_{\rm Jup}$), or brown dwarfs. 

Although the true influence of binarity on planet formation and evolution is still unclear, systems hosting planets with masses up to a few Jupiter masses have been extensively surveyed.
In contrast, systems with more massive planets ($M_p > 4$ M$_{\rm Jup}$), and objects within the brown dwarf mass regime, have yet to be studied in detail in the context of stellar multiplicity.
\citet{ZuckerMahez2002} were the first to point out that massive ($M_p > 4$ M$_\mathrm{Jup}$) short-period planets tend to be predominantly found orbiting one component of a multiple star system and possess distinctive characteristics compared to planets orbiting single stars \citep{Eggenberger2004}.

Such massive planetary and substellar companions are very challenging to form at small separations. Giant planet formation, whether by CA or GI, is thought to occur preferentially in the relatively cool outer regions of protoplanetary discs, from a few AU for CA \citep{Pollack1996}, to several tens of AU for GI \citep{Rafikov2005}. Massive hot Jupiters are thus expected to have formed at wide orbital separations from their host stars, where the lower temperatures in the protoplanetary disc allow for planet formation to proceed \citep{Bell1997}, or be born under very different conditions than currently encompassed by most planet formation models. Recently, \citet{Schlaufman2018} found evidence for two distinct populations of close-in giant planets, with a suggestion of a transition between CA and GI companions occurring at around $\sim$4$-$10 M$_\mathrm{Jup}$. This is consistent with both semi-analytic \citep{Kratter2010,Forgan2011} and numerical simulations \citep{Stamatellos2008, Stamatellos2013,Hall2017} which suggest that objects that form via GI have masses above $\sim$3$-$5 M$_\mathrm{Jup}$, and might suggest that these more massive close-in planets formed by GI rather than via CA.

In this work, we aim to constrain the multiplicity statistics of hosts to the most massive giant planets ($M_p > 7$ M$_\mathrm{Jup}$) and brown dwarfs found within $\sim$1 AU, in order to investigate the role of binarity in the formation or evolution of these systems. This will allow us to assess if a wide binary companion could be responsible for the observed orbital configurations of these objects, which are both scarce and challenging to explain with current formation theories. Our investigation will provide an indication of whether the Kozai-Lidov mechanism could play a role in the origin of the most massive short-period gas giant planets and brown dwarfs. This study will also help us determine if these massive companions are an extension of the population of lower-mass, CA giant planets, or if they belong to a separate population formed through a distinct mechanism (i.e. GI on wide orbits, followed by inward migration; \citealp{Nayakshin2010,Rice2015}). In particular, we will explore the binary properties of hosts to members of the ``brown dwarf desert'' \citep{Marcy2000}, depicting the significant deficit of brown dwarf companions found within a few AU around Sun-like stars (e.g. \citealp{Grether2006,Ma2014}).

In Section \ref{sample_selection} we present our selected sample of targets. Section \ref{observations} describes the direct imaging observations acquired for this project and the data reduction. In Section \ref{companion_search} we detail our search for wide companions, using past imaging surveys found in the literature to complement our new direct imaging data, as well as the \textit{Gaia} Data Release 2 (GR2; \citealp{GaiaCollaboration2016,GaiaCollaboration2018}) catalogue. Section \ref{statistical_analysis} describes our approach to the statistical analysis of our survey, and we present our results in Section \ref{results}. Finally, we discuss our interpretation of the obtained results in Section \ref{discussion} and summarise the main results of our project in Section \ref{conclusion}.

\section{Sample selection}
\label{sample_selection}

\afterpage{
\begin{landscape}
\begin{table}
\begin{small}
\caption{Orbital properties of the planets considered.}
\begin{tabular}{ l c c c c c c c l}
\hline\hline
Planet ID & $N_\mathrm{p}$ & $P$ & $a$ & $M_2$ & $M_2 \sin i$ & $e$ & $\tau_\mathrm{circ}$ & Ref.\\
 & & (days) & (AU) & (M$_\mathrm{Jup}$) &  (M$_\mathrm{Jup}$) &  & ($\log_{10}$[yr]) & \\
 \hline
4 UMa b         & 1 & $269.30\pm1.96$ & $0.87\pm0.04$ &  ... & $7.1\pm1.6$ & $0.432\pm0.024$ & 16.4 & \citet{Dollinger2007} \\
11 Com b        & 1 & $326.03\pm0.32$ & $1.29\pm0.05$ & ... & $19.4\pm1.5$ & $0.231\pm0.005$ & 15.9 & \citet{Liu2008} \\
30 Ari B b      & 1 & $335.1\pm2.5$ & $0.995\pm0.012$ & ... & $9.88\pm0.94$ & $0.289\pm0.092$ & 16.7 & \citet{Guenther2009} \\
59 Dra b     & 1 & $28.44\pm0.01$ & 0.2 & ... & 25 & $0.20\pm0.01$ & 12.5 & \citet{Galland2006} \\
70 Vir b        & 1 & $116.6884\pm0.0044$ & $0.484\pm0.028$ & ... & $7.49\pm0.61$ & $0.4007\pm0.0035$ & 14.5 & \citet{Butler2006} \\
$\tau$ Gem b    & 1 & $305.5\pm0.1$ & $1.17$ & ... & $20.6$ & $0.031\pm0.009$ & 15.6 & \citet{Mitchell2013} \\
$\upsilon$ And c & 4 & $240.9402\pm0.047$ & $0.8259\pm0.0043$ & $13.98^{+2.3}_{-5.3}$ & $1.96\pm0.05$ & $0.245\pm0.006$ & 15.9 & \citet{McArthur2010} \\
AS 205A b       & 1 & $24.84\pm0.03$ & $0.162\pm0.04$ & ... & $19.25\pm1.96$ & $0.34\pm0.06$ & 11.9 & \citet{Almeida2017} \\
BD+24 4697 b    & 1  & $145.081\pm0.016$ & $0.50\pm0.08$ & ... & $53\pm3$ & $0.50048\pm0.00043$ & 14.9 & \citet{Wilson2016} \\
CI Tau b        & 1 & $8.9891\pm0.0202$ & 0.079 & $12.29\pm2.13$ & $8.08\pm1.53$ & $0.28\pm0.16$ & 10.2 & \citet{Johns-Krull2016} \\
EPIC 219388192 b & 1 & $5.292569\pm0.000026$ & $0.0593\pm0.0029$ & $36.84\pm0.97$ & ... & $0.1929\pm0.0019$ & 9.4 & \citet{Nowak2017} \\
HAT-P-2 b       & 1 & $5.6334729\pm0.0000061$ & $0.06878\pm0.00068$ & $9.09\pm0.24$ & ... & $0.5171\pm0.0033$ & 8.5 & \citet{Pal2010} \\
HAT-P-20 b      & 1 & $2.875317\pm0.000004$ & $0.0361\pm0.0005$ & $7.246\pm0.187$ & ... & $0.015\pm0.005$ & 8.3 & \citet{Bakos2011} \\
HD 5891 b       & 1 & $177.11\pm0.32$ & $0.76\pm0.02$ & ... & $7.6\pm0.4$ & $0.066\pm0.022$ & 16.4 & \citet{Johnson2011} \\
HD 33564 b      & 1 & $388\pm3$ & $1.1$ & ... & $9.1$ & $0.34\pm0.02$ & 16.8 & \citet{Galland2005} \\
HD 39392 b      & 1 & $394.3^{+1.4}_{-1.2}$ & $1.08\pm0.03$ & ... & $13.2\pm0.8$ & $0.394\pm0.008$ & 16.9 & \citet{Wilson2016}\\
HD 41004 B b    & 1 & $1.328300\pm0.000012$ & $0.0177$ & ... & $18.37\pm0.22$ & $0.081\pm0.012$ & 8.0 & \citet{Zucker2004} \\
HD 77065 b      & 1 & $119.1135\pm0.0026$ & $0.438\pm0.013$ & ... & $41\pm2$ & $0.69397\pm0.00036$ & 13.3 & \cite{Wilson2016} \\
HD 87646 A b    & 2 & $13.481\pm0.001$ & $0.117\pm0.003$ & ... & $12.4\pm0.7$ & $0.05\pm0.02$ & 11.4 & \citet{Ma2016} \\
HD 89744 b      & 1 & $256.78\pm0.05$ & $0.918\pm0.010$ & ... & $8.44\pm0.23$ & $0.673\pm0.007$ & 14.7 & \citet{Wittenmyer2009} \\
HD 104985 b     & 1 & $199.505\pm0.085$ & $0.95$ & ... & $8.3$ & $0.090\pm0.009$ & 16.8 & \citet{Sato2008} \\
HD 112410 b     & 1 & $124.6$ & $0.565$ & ... & $9.18$ & $0.23$ & 16.1 & \citet{Jones2013} \\
HD 114762 b     & 1 & $83.9151\pm0.0030$ & $0.353\pm0.001$ & ... & $10.98\pm0.09$ & $0.3354\pm0.0048$ & 14.2 & \citet{Kane2011b} \\
HD 134113 b     & 1 & $201.680\pm0.004$ & $0.638\pm0.010$ & ... & $47^{+2}_{-3}$ & $0.891\pm0.002$ & 11.2 & \citet{Wilson2016} \\
HD 156279 b     & 1 & $131.05\pm0.54$ & $0.495\pm0.017$ & ... & $9.71\pm0.66$ & $0.708\pm0.018$ & 12.9 & \citet{Diaz2012} \\
HD 156846 b     & 1 & $359.5546\pm0.0071$ & $1.096\pm0.021$ & ... & $10.57\pm0.29$ & $0.84785\pm0.00050$ & 13.1 & \citet{Kane2011a} \\
HD 160508 b     & 1 & $178.9049\pm0.0074$ & $0.68\pm0.02$ & ... & $48\pm3$ & $0.5967\pm0.0009$ & 14.5 & \citet{Wilson2016} \\
HD 162020 b     & 1 & $8.428198\pm0.000056$ & $0.0751\pm0.0043$ & ... & $14.4\pm2.1$ & $0.277\pm0.002$ & 10.2 & \citet{Udry2002} \\
HD 168443 b     & 2 & $58.1112\pm0.0009$ & $0.290\pm0.005$ & ... & $7.48\pm0.27$ & $0.530\pm0.001$ & 12.6 & \citet{Udry2002, Wittenmyer2007} \\ 
HD 178911 B b   & 1 & $71.484\pm0.002$ & $0.339\pm0.006$ & ... & $7.03\pm0.28$ & $0.114\pm0.003$ & 14.4 & \citet{Wittenmyer2009} \\
HD 180314 b     & 1 & $396.03\pm0.62$ & $1.34^{+0.02}_{-0.08}$ & ... & $20.3^{+0.6}_{-2.4}$ & $0.257\pm0.010$ & 15.8 & \citet{Sato2010} \\
HD 203949 b     & 1 & $184.2\pm0.5$ & $0.81\pm0.03$ & ... & $8.2\pm0.2$ & $0.02\pm0.03$ & 15.5 & \citet{Jones2014} \\
KELT-1 b        & 1 & $1.217514\pm0.000015$ & $0.02466\pm0.00016$ & $27.23^{+0.50}_{-0.48}$ & ... & $0.0099^{+0.0100}_{-0.0069}$ & 8.0 & \citet{Siverd2012} \\
Kepler-13 A b   & 1 & $1.763588\pm0.000001$ & $0.03641\pm0.00087$ & $9.28\pm0.16$ & ... & $0.00064\pm0.00015$ & 8.0 & \citet{Esteves2015} \\
NLTT 41135 b    & 1 & $2.889475\pm0.000025$ & $0.024\pm0.001$ & $33.7^{+2.8}_{-2.6}$ & ... & $< 0.02$ & 8.8 & \citet{Irwin2010} \\
WASP-14 b       & 1 & $2.24376524\pm0.00000044$ & $0.0371\pm0.0011$ & $7.76\pm0.47$ & ... & $0.0830\pm0.0030$ & 8.1 & \citet{Wong2015} \\
WASP-18 b       & 1 & $0.94145299\pm0.00000087$ & $0.02026\pm0.00068$ & $10.30\pm0.69$ & ... & $0.0092\pm0.0028$ & 8.0 & \citet{Hellier2009} \\
XO-3 b          & 1 & $3.1915239\pm0.0000068$ & $0.04540\pm0.00082$ & $11.79\pm0.59$ & ... & $0.260\pm0.017$ & 8.4 & \citet{Winn2008} \\

\hline \\ [-2.5ex]
\multicolumn{9}{l}{
  \begin{minipage}{1.3\textwidth}
    \textbf{Notes.}
    $N_\mathrm{p}$ is the number of known planets in the system. Tidal circularisation timescales were calculated in this paper (see text). All other parameters come from the given references and references therein.\\
  \end{minipage}}

\label{t:planet_properties}
\end{tabular}
\end{small}
\end{table}
\end{landscape}

\begin{landscape}
\begin{table}
\begin{small}
\caption{Stellar properties for the selected systems.}
\begin{tabular}{ l c c c c c c c c c l}
\hline\hline
Object ID & RA   & Dec.  & SpT & $V$ & Distance & [Fe/H] & $M_*$ & Age & Ref. & Other name \\
   & (J2000) & (J2000) &   & (mag) & (pc) & (dex) & (M$_\odot$) & (Gyr)  &  & \\
\hline
4 UMa & 08:40:12.82 & $+$64:19:40.6 & K1 III & $4.787\pm0.005$ & $73.5\pm1.2$ & $-0.25\pm0.04$ & $1.234\pm0.15$ & $4.6\pm2.0$ & (1) & HD 73108, HIP 42527 \\
11 Com b & 12:20:43.03 & $+$17:47:34.3 & G8 III & $4.74\pm0.02$ & $93.2\pm1.9$ & $-0.31\pm0.02$ & $2.02\pm0.11$ & $1.17\pm0.28$ & (2) & HD 107383; HIP 60202 \\
30 Ari B & 02:36:57.74 & $+$24:38:53.0 & F6 V & $7.020\pm0.011$ & $44.7\pm0.1$ & $+0.245\pm0.195$ & $1.16\pm0.04$ & $0.91\pm0.83$ & (3) & HD 16232, HIP 12184 \\
59 Dra & 19:09:09.88 & $+$76:33:37.8 & A9 V & $5.107\pm0.009$ & $27.4\pm0.1$ & $+0.016$ & $1.447\pm0.015$ & $0.436\pm0.200$ & (4) & HD 180777, HIP 94083 \\
70 Vir & 13:28:25.81 & $+$13:46:43.6 & G5 V & $5.045\pm0.009$ & $17.9\pm0.1$ & $-0.012$ & $1.07\pm0.01$ & $8.1\pm0.4$ & (5,6) & HD 117176, HIP 65721 \\
$\tau$ Gem & 07:11:08.37 & $+$30:14:42.6 & K2 III & 4.42 & $112.5\pm4.1$ &  $+0.14\pm0.10$ & $2.3\pm0.3$ & $1.22\pm0.76$ & (7) & HD 54719, HIP 34693 \\
$\upsilon$ And & 01:36:47.84 & $+$41:24:19.6 & F8 V & $4.10\pm0.05$ & $13.4\pm0.1$ & $+0.131\pm0.067$ & $1.31\pm0.02$ & $3.12\pm0.22$ & (8) & HD 9826, HIP 7513 \\
AS 205A & 16:11:31.34 & $-$18:38:26.0 & K5 V & $12.63\pm0.21$ & $127.5\pm1.6$ & $-0.043\pm0.060$ & $1.086\pm0.100$ & 0.001 & (9) & V866 Sco, EPIC 205249328 \\
BD+24 4697 & 23:01:39.32 & $+$25:47:16.5 & K2 V & $9.847\pm0.022$ & $44.7\pm0.3$ & $-0.16\pm0.03$ & $0.754\pm0.016$ & $5.207\pm4.150$ & (2) & HIP 113698 \\
CI Tau & 04:33:52.01 & $+$22:50:30.1 & K7 V & 13.80 & $158.0\pm1.2$ & $-0.727\pm0.050$ & $0.80\pm0.02$ & 0.002 & (10) & EPIC 247584113 \\
EPIC 219388192 & 19:17:34.03 & $-$16:52:17.8 & G V & $12.535\pm0.020$ & $305.0\pm4.6$ & $+0.03$ & $1.01\pm0.04$ & $3.6^{+1.8}_{-1.5}$ & (11) & \\
HAT-P-2 & 16:20:36.36 & $+$41:02:53.1 & F8 V & $8.743\pm0.013$ & $127.8\pm0.5$ & $+0.14\pm0.08$ & $1.36\pm0.04$ & $2.6\pm0.5$ & (12) & HD 147506, HIP 80076 \\
HAT-P-20 & 07:27:39.95 & $+$24:20:11.5 & K3 V & $11.346\pm0.030$ & $71.0\pm0.2$ & $+0.35\pm0.08$ & $0.756\pm0.028$ & $6.7^{+5.7}_{-3.8}$ & (13) & \\
HD 5891 & 01:00:33.19 & $+$20:17:33.0 & G5 III & $8.11\pm0.01$ & $283.5\pm4.9$ & $+0.13\pm0.08$ & $1.242\pm0.041$ & $1.0^{+0.8}_{-0.5}$ & (14) & HIP 4715 \\
HD 33564 & 05:22:33.53 & $+$79:13:52.1 & F6 V & $5.140\pm0.009$ & $21.0\pm0.1$ & $-0.12$ & $1.25\pm0.04$ & $3.0^{+0.6}_{-0.3}$ & (15) & HIP 25110 \\
HD 39392 & 05:53:19.00 & $+$22:04:19.7 & F8 V & $8.449\pm0.013$ & $102.5\pm0.7$ & $-0.54\pm0.01$ & $0.94\pm0.04$ & $9.06\pm1.40$ & (2) & HIP 27828 \\
HD 41004 B & 05:59:49.65 & $-$48:14:22.9 & M2 V & 12.63 & $41.6\pm0.5$ & $+0.10$ & 0.4 & 1.6 & (16) & HIP 28393 \\
HD 77065  & 09:00:47.45 & $+$21:27:13.4 & G5 V & $8.786\pm0.021$ & $32.7\pm0.1$ & $–0.42\pm0.02$ & $0.71\pm0.01$ & $7.59\pm3.69$ & (2) & HIP 44259 \\
HD 87646 A & 10:06:40.77 & $+$17:53:42.4 & G1 IV & $8.143\pm0.013$ & $137.1\pm15.6$ & $-0.17\pm0.08$ & $1.12\pm0.09$ & $4.75\pm1.1$ & (17,18) & HIP 49522 \\
HD 89744 & 10:22:10.56 & $+$41:13:46.3 & F7 V & $5.782\pm0.009$ & $38.6\pm0.1$ & $+0.26\pm0.03$ & $1.558\pm0.048$ & $2.50\pm0.30$ & (6,19) & HIP 50786 \\
HD 104985 & 12:05:15.12 & $+$76:54:20.6 & G8.5 III & $5.785\pm0.009$ & $100.6\pm0.7$ & $-0.15$ & $1.2\pm0.1$ & $4.9\pm1.2$ & (6,20) & HIP 58952 \\
HD 112410 & 12:57:31.96 & $-$65:38:47.3 & G8 III & $6.86\pm0.01$ & $156.5\pm0.7$ & $-0.28\pm0.05$ & $1.21\pm0.25$ & $4.17\pm2.34$ & (21) & HIP 63242 \\
HD 114762 & 13:12:19.74 & $+$17:31:01.6 & F9 V & $7.361\pm0.013$ & $40.2\pm0.4$ & $-0.774\pm0.030$ & $0.83\pm0.01$ & $12.4\pm0.6$ & (6,22) & HIP 64426 \\
HD 134113 & 15:07:46.50 & $+$08:52:47.2 & F9 V & $8.339\pm0.018$ & $72.2\pm0.4$ & $-0.92\pm0.02$ & $0.85\pm0.02$ & $10.98\pm0.66$ & (2) & HIP 74033 \\
HD 156279 & 17:12:23.20 & $+$63:21:07.5 & K0 V & $8.167\pm0.013$ & $36.2\pm0.1$ & $+0.14\pm0.01$ & $0.93\pm0.02$ & $7.40\pm2.20$ & (6,23) & HIP 84171 \\
HD 156846 & 17:20:34.31 & $-$19:20:01.5 & G1 V & $6.564\pm0.010$ & $47.7\pm0.1$ & $+0.18\pm0.02$ & $1.38\pm0.05$ & $2.78\pm0.37$ & (2,21) & HIP 84856 \\
HD 160508 & 17:39:12.70 & $+$26:45:27.1 & F8 V & $8.177\pm0.012$ & $111.6\pm0.8$ & $-0.16\pm0.02$ & $1.14\pm0.04$ & $5.55\pm0.57$ & (2) & HIP 86394 \\
HD 162020 & 17:50:38.36 & $-$40:19:06.1 & K3 V & $9.227\pm0.022$ & $30.8\pm0.1$ & $+0.01\pm0.11$ & $0.75\pm0.01$ & $3.10\pm2.70$ & (6,24) & HIP 87330 \\
HD 168443 & 18:20:03.93 & $-$09:35:44.6 & G6 V & $7.000\pm0.011$ & $39.6\pm0.1$ & $+0.06\pm0.05$ & $1.02\pm0.01$ & $10.00\pm0.30$ & (6,25) & HIP 89844 \\ 
HD 178911 B & 19:09:04.39 & $+$34:36:01.6 & G5 V & $7.494\pm0.010$ & $41.0\pm0.1$ & $+0.34\pm0.03$ & $1.03\pm0.02$ & $4.80\pm1.30$ & (6,19) & HIP 94075 \\
HD 180314 & 19:14:50.21 & $+$31:51:37.3 & K0 III & $6.721\pm0.010$ & $122.4\pm0.5$ & $+0.24\pm0.07$ & $2.13\pm0.13$ & $1.14\pm0.24$ & (2) & HIP 94576 \\
HD 203949 & 21:26:22.87 & $-$37:49:46.0 & K2 III & $5.620\pm0.009$ & $78.6\pm0.8$ & $+0.28\pm0.06$ & $1.99\pm0.10$ & $1.23\pm0.19$ & (21) & HIP 105854 \\
KELT-1 & 00:01:26.92 & $+$39:23:01.8 & F5 V & $10.701\pm0.057$ & $268.4\pm3.0$ & $-0.85$ & $1.335\pm0.063$ & $1.75\pm0.25$ & (26) & \\
Kepler-13 A & 19:07:53.15 & $+$46:52:05.9 & A5 V & $10.349\pm0.037$ & $473.3\pm18.3$ & $-0.50\pm0.10$ & $1.72\pm0.10$ & $1.12\pm0.10$ & (27,28) & KOI-13 \\
NLTT 41135 & 15:46:04.26 & $+$04:41:30.0 & M5.1 V & 18 & $34.1\pm0.1$ & $-0.5$ & $0.164\pm0.020$ & 5 & (29) & \\
WASP-14 & 14:33:06.36 & $+$21:53:41.0 & F5 V & $9.798\pm0.026$ & $162.0\pm0.8$ & $-0.13\pm0.08$ & $1.35\pm0.12$ & $2.4^{+1.5}_{-1.0}$ & (30,31) & \\
WASP-18 & 01:37:25.03 & $-$45:40:40.4 & F6 IV/V & $9.357\pm0.018$ & $123.5\pm0.4$ & $+0.00\pm0.09$ & $1.20\pm0.01$ & $0.90\pm0.20$ & (6,32) & HD 10069, HIP 7562 \\
XO-3 & 04:21:52.71 & $+$57:49:01.9 & F5 V & $9.904\pm0.027$ & $213.1\pm2.7$ & $-0.177\pm0.080$ & $1.213\pm0.066$ & $2.82^{+0.58}_{-0.82}$ & (33) & \\

\hline \\ [-2.5ex]
\multicolumn{11}{l}{
  \begin{minipage}{1.3\textwidth}
    \textbf{Notes.}
    Distances are based on the estimates from \citet{Bailer-Jones2018} derived from Gaia DR2 parallax measurements. All other parameters come from the given references and references therein.\\
    \textbf{References:}
    (1) \citet{Dollinger2007};
    (2) \citet{Manaldo2017};
    (3) \citet{Guenther2009};
    (4) \citet{Jones2015};
    (5) \citet{Butler2006};
    (6) \citet{Bonfanti2016};
    (7) \citet{Mitchell2013};
    (8) \citet{McArthur2010};
    (9) \citet{Almeida2017};
    (10) \citet{Guilloteau2014};
    (11) \citet{Nowak2017};
    (12) \citet{Pal2010};
    (13) \citet{Bakos2011};
    (14) \citet{Johnson2011};
    (15) \citet{Galland2005};
    (16) \citet{Santos2002};
    (17) \citet{Aguilera-Gomez2018};
    (18) \citet{Ma2016};
    (19) \citet{Wittenmyer2009};
    (20) \citet{Sato2008};
    (21) \citet{Jofre2015};
    (22) \citet{Kane2011b};
    (23) \citet{Diaz2012};
    (24) \citet{Udry2002};
    (25) \citet{Wittenmyer2007};
    (26) \citet{Siverd2012};
    (27) \citet{Morton2016};
    (28) \citet{Shporer2014}
    (29) \citet{Irwin2010};
    (30) \citet{Southworth2012};
    (31) \citet{Knutson2014};
    (32) \citet{Hellier2009};
    (33) \citet{Winn2008}.
  \end{minipage}}

\label{t:stellar_properties}
\end{tabular}
\end{small}
\end{table}
\end{landscape}

}

The aim of this project is to search for wide, substellar or stellar companions to stars hosting a massive planet or brown dwarf on a very short orbit.
Recent findings suggest that GI forms planets with masses larger than $\sim$4 M$_{\rm Jup}$ \citep{Stamatellos2013,Stamatellos2015,Hall2017}, and the transition between CA and GI companions is thought to occur around $\sim$4$-$10 M$_{\rm Jup}$ \citep{Schlaufman2018}. Studies of core accretion populations found that CA rarely forms planets with masses larger than $\sim$5 M$_\mathrm{Jup}$ \citep{Matsuo2007, Mordasini2018}, and shows a steep drop and a strong metallicity dependence in the formation of higher-mass planets \citep{Mordasini2012, Jenkins2017}.
In order to investigate the higher-mass planetary population, which likely formed by disc GI rather than through CA, we choose for this survey a lower limit on inner companion mass $M_2$ of 7 M$_\mathrm{Jup}$, based on the studies mentioned above. This allows us to avoid the main region of overlap between CA and GI, while keeping a sufficiently large sample size for our study.
We place an upper limit of 70 M$_\mathrm{Jup}$ (the hydrogen-burning limit) on the mass (or projected mass) of the inner companions, so as to limit our sample to likely substellar objects.

We place an upper limit of $P < 400$ days (about 1 AU around a Sun-like star) on the orbital period of the close-in companions. It is now well-accepted that if planet formation via GI does occur, it typically takes place in the outer regions ($>$30 AU) of protostellar discs \citep{Rafikov2005,Clarke2009,Rice2009}.
This thus ensures that all selected companions have undergone significant migration between their expected GI formation location and their current configurations, or that they had to be formed under considerably different natal environments than for standard planet formation in order to be born \textit{in-situ}.
We set an upper limit of 500 pc on the distances of our targets in order to be sensitive to wide companions from 50$-$100 AU around most stars in our sample. We use the distance estimates from \citet{Bailer-Jones2018} to infer distances for our targets. These distances are derived from the highly-precise parallax measurements provided by the \textit{Gaia} DR2 catalogue, correcting for the nonlinearity of the transformation between parallax and distance. Finally, we only consider stellar primaries and place a limit on the host's mass of $M_* > 0.1$ M$_\odot$.

Based on the arguments presented above, we selected all systems from the NASA Exoplanet Archive\footnote{\url{https://exoplanetarchive.ipac.caltech.edu}}, the Exoplanet Data Explorer\footnote{\url{http://exoplanets.org}} and the Extrasolar Planets Encyclopaedia\footnote{\url{www.exoplanet.eu}} with confirmed transiting or radial velocity companions with well-constrained orbits that satisfy the following criteria:
\begin{description}
    \item $-$ inner companion mass $M_2$ (or $M_2 \sin i$) between 7$-$70 M$_\mathrm{Jup}$.
    \item $-$ inner companion orbital period $P < 400$ days.
    \item $-$ distance within 500 pc based on \textit{Gaia} DR2 parallax.
    \item $-$ primary mass $M_* > 0.1$ M$_\odot$.
\end{description}
Our final sample consists of 38 objects, and includes some very short period ($P < 10$ days) transiting systems, together with radial velocity objects extending to larger separations. Properties of the inner companions are presented in Table \ref{t:planet_properties} and the host stars are listed in Table \ref{t:stellar_properties}. We selected our sample without regard to the targets' multiplicity, known or unknown. However, radial velocity and transit surveys are typically biased against binaries, excluding known multiple systems in target selection processes. As these biases are difficult to quantify and account for, our obtained results may somewhat underestimate the multiplicity rate of the population probed here, but we consider that our study is in no way biased towards the presence of wide companions.

About three quarters of the selected systems were discovered and characterised via radial velocity measurements.
Mass estimates for companions discovered through this method only allow for the determination of a lower limit on the companion mass due to the unknown inclination $i$ of the system. Radial velocity systems are therefore expected to be more massive than the estimated $M_2 \sin i$ as a result of the projection factor. Selected systems discovered via this method are thus likely to be more massive than the minimum masses reported in Table \ref{t:planet_properties}. Given the projected masses of these companions and assuming a uniform distribution of inclinations between 0 and 90 degrees, we can easily show with a Monte-Carlo approach that an average of 72\% of the radial velocity systems considered here are statistically likely to be above the deuterium burning limit at 13 M$_\mathrm{Jup}$. Combining this with our transiting systems, this means that more than $\sim$60\% of our targets are likely in the brown dwarf mass regime, and close to 80\% of our sample is expected to have a true mass $>$10 M$_\mathrm{Jup}$. We therefore consider that our gathered sample of objects is representative of the population of the most massive planets found on tight orbits and provides a robust insight into short-period brown dwarf desert members.

We define in Table \ref{t:planet_properties} a tidal circularisation timescale $\tau_{\rm circ}$ for each planet and brown dwarf companion in our sample, given in $\log_{10}$[yr]. We estimate this parameter using the formalism presented in \citet{Rice2012}, which is based on that developed by \citet{Eggleton1998} (see also \citealp{Mardling2002} and \citealp{Dobbs-Dixon2004}). We assume the star has a tidal quality factor of $Q’_* = 5 \times 10^6$ and the planet has a tidal quality factor of $Q’_p = 10^5$. We take the star mass, planet mass, orbital semi-major axis, and orbital eccentricity from Tables \ref{t:planet_properties} and \ref{t:stellar_properties}. We assume that the star has a rotation period of 20 days and that the planet is rotating synchronously. We estimate the circularisation timescale by simply evolving each system for a short period of time and determining the resulting change in eccentricity (i.e., $\tau_{\rm circ} = e/\dot{e}$).

\citet{Petrovich2015} found that planets migrating via the Kozai-Lidov mechanism \citep{Kozai1962,Lidov1962}, under the influence of a distant companion, spend most of their lifetimes undergoing eccentric oscillations at separations $>$2 AU, or as Hot Jupiters at $<$0.1 AU. All the inner companions considered here have orbital separations smaller than 1 AU. If they migrated from wider separations to their current locations through the Kozai-Lidov scenario, they should be able to circularise onto Hot Jupiter orbits fairly rapidly.
Inner companions with circularisation timescales longer than the age of the Universe are thus unlikely to be driven by secular perturbations such as the Kozai-Lidov mechanism.
On the other hand, objects with timescales smaller than the age of the Universe (i.e., less than $\sim$10.2 in $\log_{10}$[yr]) could have migrated inwards via the Kozai-Lidov scenario. A total of 12 targets have tidal circularisation timescales shorter than that and may thus be consistent with a Kozai-Lidov migration process. The subset of Kozai-consistent objects corresponds to all the inner companions in our sample with an orbital period shorter than 10 days. This is in good agreement with the idea that planets migrating via the Kozai-Lidov mechanism spend most of their lifetime around their initial, wide separations, or on hot Jupiter orbits, as discussed in \citet{Petrovich2015}.

\section{New Observations}
\label{observations}

\subsection{Observations and data reduction}

We used direct imaging facilities at the Very Large Telescope (VLT), Gemini North and the WIYN Observatory to acquire data for six objects in the sample presented in Section \ref{sample_selection}, four of which did not have any previously reported direct imaging observations. The observations are summarised in Table \ref{t:observations}.

\begin{table*}
    \centering
    \caption{Summary of our new observations.}
    \begin{tabular}{p{1.8cm} c c c c c c}
        \hline \hline
        Target & Observation Date & Telescope / Instrument & Filter & Field of View & Pixel Scale & Previous Observations \\
        \hline
        WASP-18 & September 4, 2017 & VLT / NACO & \textit{L'} & $28\arcsec\times28\arcsec$ & 0\farcs027  & \citet{Ngo2015} \\
        HD 162020 & September 6, 2017 & VLT / NACO & \textit{L'} & $28\arcsec\times28\arcsec$ & 0\farcs027  & \citet{Eggenberger2007} \\
        BD+24 4697 & September 6, 2017 & Gemini North / NIRI & \textit{Ks} & $22\arcsec\times22\arcsec$ & 0\farcs022  & $-$ \\
        HD 77065 & December 12, 2017 & Gemini North / NIRI & \textit{Ks} & $22\arcsec\times22\arcsec$ & 0\farcs022  & $-$ \\
        HD 134113 & June 22, 2018 & WIYN / NESSI & 562nm, 832nm & $4.6\arcsec\times4.6\arcsec$ & 0\farcs040  & $-$ \\
        HD 160508 & June 24, 2018 & WIYN / NESSI & 562nm, 832nm & $4.6\arcsec\times4.6\arcsec$ & 0\farcs040  & $-$ \\
        \hline
    \end{tabular}
    \label{t:observations}
\end{table*}

\subsubsection{VLT / NACO observations}

We obtained images in the $L'$ filter ($3.8\,\mu$m) using the AO-assisted imager NACO at VLT \citep{Lenzen2003,Rousset2003} for HD 162020 and WASP-18 (programme 099.C-0728, PI Fontanive). These new data were acquired with the aim to confirm or refute a candidate reported in \citet{Eggenberger2007} around the former target, and to achieve deeper detection limits than in currently available imaging data of the latter object (\citealt{Ngo2015}; see Appendix \ref{A:binary_search}).
The observing setup included the L27 camera, and the data were taken in the pupil tracking mode, where the telescope pupil is held fixed, and the field rotates. 
Each target was observed using a three-point dither pattern, designed to avoid a bad quadrant of the NACO detector. We used short integration time (0.2\,s) in order not to saturate the primaries, allowing photometric and astrometric calibrations.

Standard near-infrared data reduction techniques were applied using our custom IDL routines, including sky subtraction, flat-fielding and bad-pixel correction. Some of the frames were affected by the horizontal additive noise pattern, that sporadically appeared in the NACO data, and was variable in intensity and time. The pattern was removed following the procedure described in \citet{Hussmann12}. Individual frames were de-rotated according to the parallactic angle, and finally stacked together.
We retrieved in our final images the unconfirmed candidate companion around HD 162020 reported by \citet{Eggenberger2007} and were able to refute the bound nature of this source based on our new data. The detailed analysis of the rejected candidate is presented in Section \ref{HD_162020}. No companion was detected around WASP-18 within the field of view of our images.

\subsubsection{Gemini North / NIRI observations}

We acquired images in $K_s$ band (1.95$-$2.30 $\mu$m) using the Gemini Near-Infrared Imager (NIRI; \citealp{Hodapp2003}) instrument at the Gemini North telescope for BD+24 4697 and HD 77065 (programme GN-2017B-Q-40, PI Fontanive).
Targets were observed in the standard imaging mode, using the Gemini North adaptive optics (AO) system ALTAIR \citep{Herriot2000} to obtain diffraction-limited images with the f/32 camera. Both our target were bright enough to be used as natural guide stars. 
The observing strategy adopted was similar to the one described in \citet{Lafreniere2008} and \citet{Daemgen2015}. Each target was observed at five dither positions to allow for sky subtraction and bad pixel correction. At each dither position we acquired one non-saturated short exposure (divided into many coadds) in high read noise mode, followed by a longer exposure in low read noise mode. This prevents our observations from being limited by the high read out noise, resulting in a high observing efficiency and a large dynamic range, providing sensitivity at both small and large separations. Our targets were not saturated, even in the deeper exposures.

We followed standard procedures for near-infrared data reduction, using the Gemini NIRI IRAF package and our dedicated IDL routines. A sky frame was constructed by taking the median of the dithered images, masking the regions dominated by the target's signal. The individual images were then sky-subtracted and divided by a normalised flat field, and bad pixels were replaced by a median calculated over their good neighbours. For all images, field distortion was corrected as described in \citet{Lafreniere2014}.
No candidate companion was identified around either target.

\subsubsection{WIYN / NESSI observations}

We acquired observations of HD 160508 and HD 134113 with the WIYN 3.5-m telescope at Kitt Peak National Observatory (KPNO). We used the NASA-NSF Exoplanet Observational Research (NN-Explore) Exoplanet and Stellar Speckle Imager (NESSI) in diffraction-limited speckle imaging mode. NESSI is based on an upgraded design of the Differential Speckle Survey Instrument (DSSI; \citealp{Horch2009,Horch2012}). Each target was observed simultaneously in two cameras, with a filter centered on 562 nm ($r$-narrow) on the blue channel and a bandpass at 832 nm ($z$-narrow) on the red channel. The standard NESSI observing strategy was followed, with typical integration times of 40 ms (see \citealp{Scott2018}). Data were reduced by the KPNO speckle reduction pipeline that generates reconstructed images and contrast limit curves for each observation \citep{Scott2018}. We did not identify any candidate companion in the obtained data around these two targets.

\subsection{Achieved sensitivities}
\label{sensitivities}

\begin{figure}
    \centering
    \includegraphics[width=0.45\textwidth]{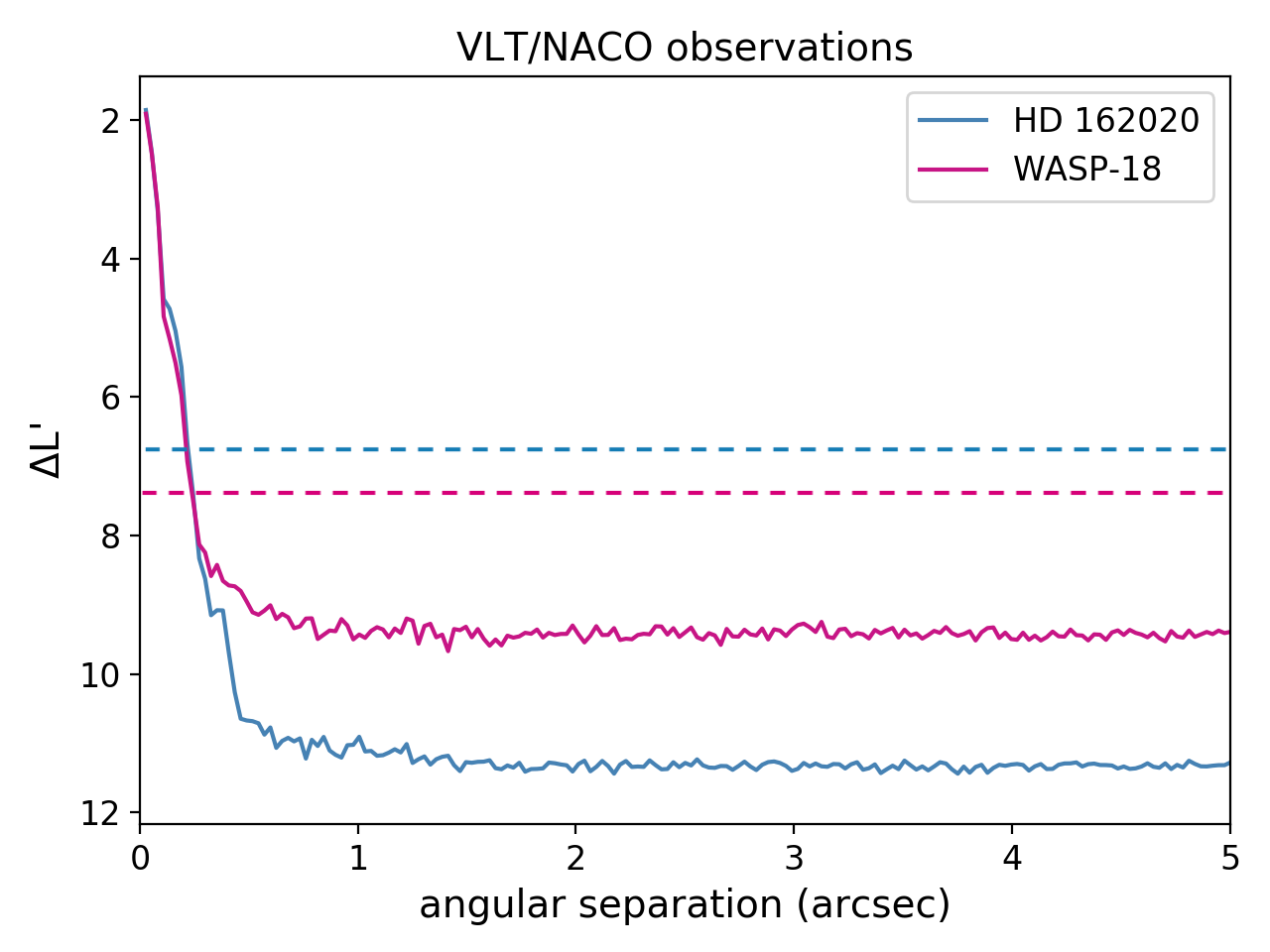}
    \caption{Achieved sensitivities showing the 5-$\sigma$ magnitude differences in $L'$ as a function of angular separation for our two targets observed with VLT/NACO. The dashed lines indicate the magnitude differences corresponding to the hydrogen-burning limit for each target.}
    \label{f:NACO_limits}
\end{figure}

\begin{figure}
    \centering
    \includegraphics[width=0.45\textwidth]{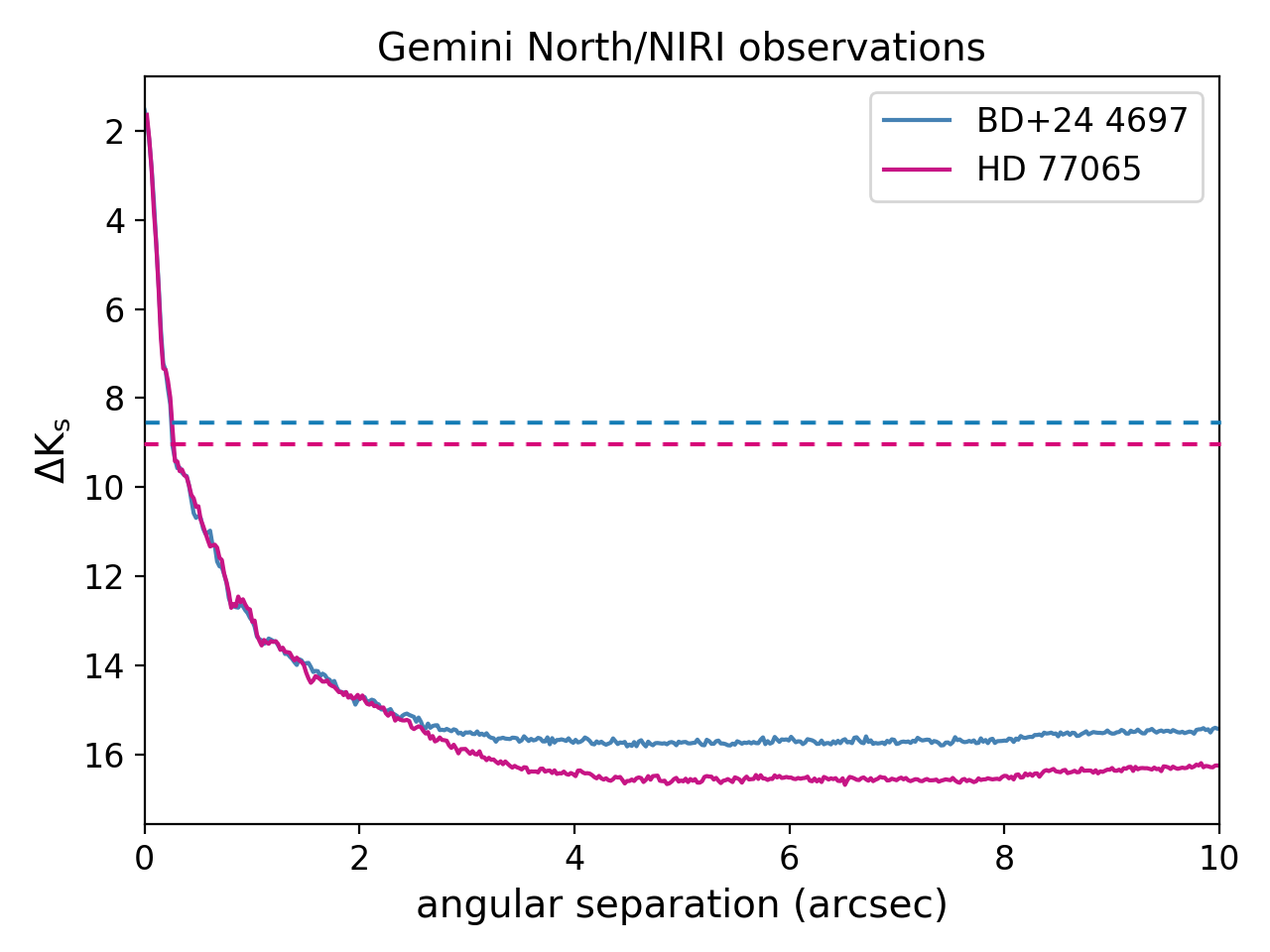}
    \caption{5-$\sigma$ magnitude differences achieved in $K_s$ for our two targets observed with Gemini North/NIRI, showing the corresponding hydrogen-burning limits (dashed lines).}
    \label{f:NIRI_limits}
\end{figure}

\begin{figure}
    \centering
    \includegraphics[width=0.45\textwidth]{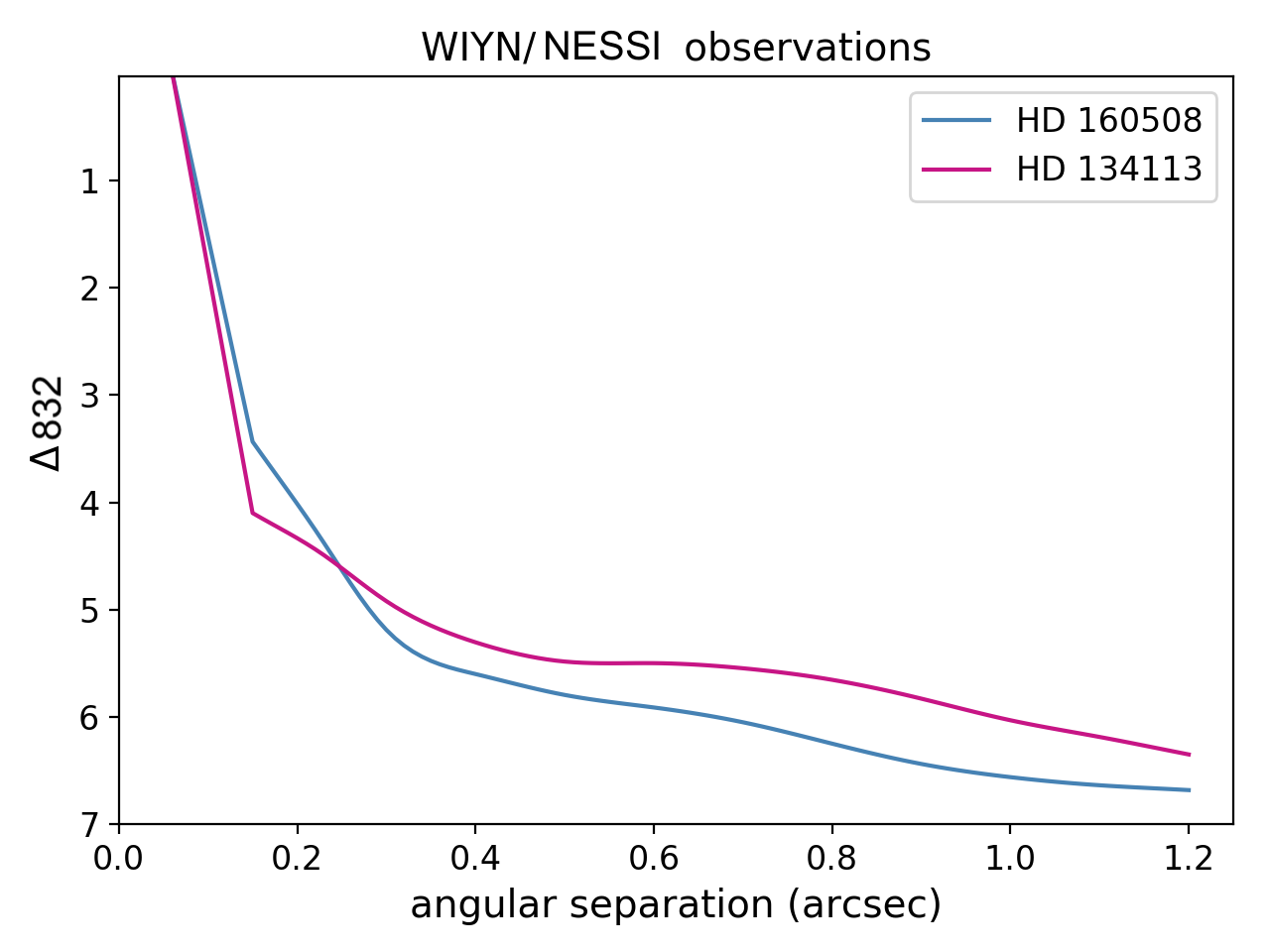}
    \caption{5-$\sigma$ magnitude differences achieved in the 832 nm filter for our two targets observed with WIYN/NESSI. These observations do not reach the hydrogen-burning limit.}
    \label{f:WIYN_limits}
\end{figure}

We estimated the limiting sensitivities reached around our observed targets in order to establish the full range of detectable companions covered by the obtained data. For the VLT / NACO and Gemini North / NIRI data, detection limits were determined from the final images described above. The 5-$\sigma$ noise curves were calculated as a function of radius by computing the standard deviation in circular annuli with 1-pixel widths, centred on the primary targets. Noise levels were then converted into magnitude contrasts by dividing by the peak pixel values of the targets (which were not saturated), and converting the obtained flux ratios into magnitude differences in the considered filters. The contrast curves generated from the custom KPNO pipeline in the 832 nm filter were considered for the WIYN data, as the redder filter is better suited for the detection of warm, low-mass companions.
The achieved magnitude contrasts are presented in Figures \ref{f:NACO_limits}, \ref{f:NIRI_limits} and \ref{f:WIYN_limits} for our NACO, NIRI and NESSI observations, respectively. The hydrogen-burning limits are shown for the first two data sets, showing that we are sensitive to substellar companions around these stars. We did not reach the stellar/substellar boundary in the WIYN observations and are only able to detect low-mass stellar companions in these data.

\subsection{Refuted candidate around HD 162020}
\label{HD_162020}

HD 162020 had previously been observed with NACO as part of the survey conducted in \citet{Eggenberger2007}. \citet{Eggenberger2007} reported two point sources within 5\arcsec from the star, one of which was found by the authors to clearly be a background source. They found that the second candidate, at 4\farcs98 $\pm$ 0\farcs03, was more likely unbound than bound, although the low significance level of this result led them to report the companionship of this candidate as inconclusive based on their data alone. Both sources are retrieved in our new NACO images (Section \ref{observations}). The positions of the detected sources were extracted using the {\sc StarFinder} PSF-fitting algorithm \citep{Diolaiti2000}, employing an empirical PSF extracted from the primary.  

To calibrate the pixel scale and the True North (TN) of the detector, we used the astrometric calibrator system $\theta_1$ Orionis C, observed on October 6 2017. Using the same procedure as described in \citet{Chauvin2012}, we obtain the pixel scale of $27.10\pm0.05\,$mas, and the TN position of $-0.45\pm 0.10\,$deg. However, as previously pointed out by \citet{Eggenberger2007} and \citet{Chauvin2012}, additional systematic errors might be present in the determination of the TN of the NACO detector in a case where different sets of calibrator stars were used between the epochs. Since we do not know which calibrators were used to derive astrometry in the previous epochs by \citet{Eggenberger2007}, we add $0.5\,$deg to the TN error budget (G. Chauvin, priv. comm.).

\begin{figure}
    \centering
    \includegraphics[width=0.38\textwidth]{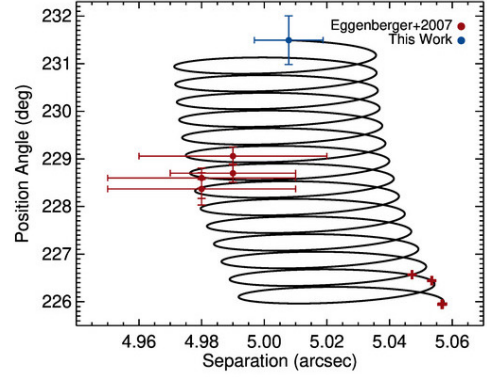}
    \caption{Common proper motion analysis of the faint companion around HD 162020, originally identified by \citet{Eggenberger2007}. The black solid line represents the motion of a background object relative to the primary, computed using the proper motion and parallax measurements of HD 162020 from \textit{Gaia} DR2. The blue and red circles mark the relative positions of the components in our new NACO observations and in the data from \citet{Eggenberger2007}, respectively. The red crosses indicate the expected positions of a background source at the dates of the observations used by \citet{Eggenberger2007}. The relative motion of the candidate over the available epochs is not consistent with a comoving pair.}
    \label{f:CPM_HD_162020}
\end{figure}

With a baseline larger than a decade between the observations used in \citet{Eggenberger2007} and ours, and given the proper motion of the primary from \textit{Gaia} DR2, we were able to refute the bound nature of this companion. Figure \ref{f:CPM_HD_162020} shows the relative positions of the primary and candidate at the various epochs available, together with the expected motion of a background object. The plot clearly demonstrates that the candidate does not share common proper motion with the primary. The fact that the relative positions at the old epochs of observation are not consistent with the primary, nor with the expected background positions, suggests that the source has a non-negligible proper motion of its own.

\section{Search for wide companions}
\label{companion_search}

\begin{figure}
    \centering
    \includegraphics[width=0.45\textwidth]{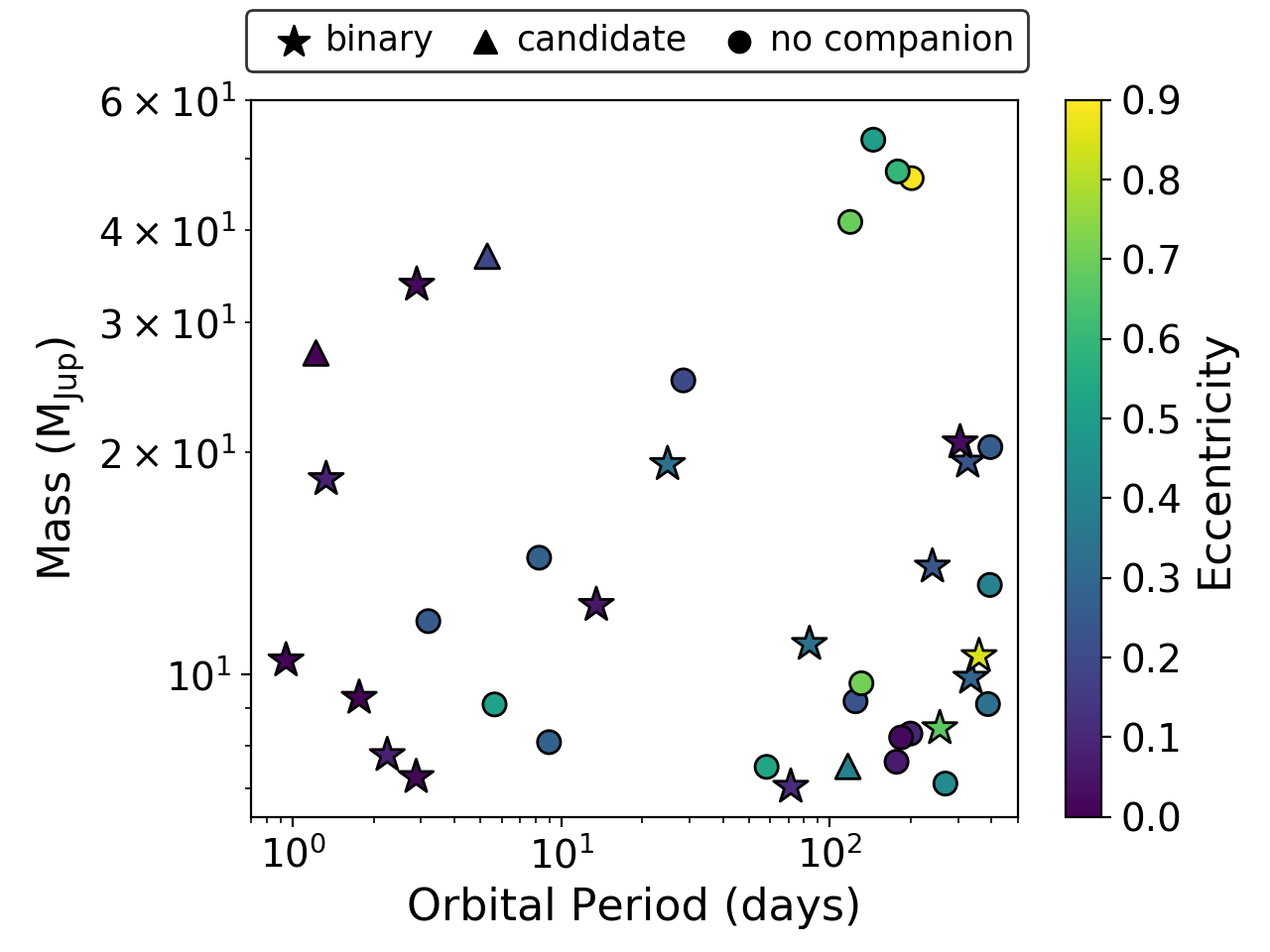}
    \caption{Orbital properties of the inner companions in our sample showing the systems that are known to be binaries or higher-order multiples (stars), those with a candidate companion (triangles) and stars that are apparently single (circles).}
    \label{f:planet_binarity}
\end{figure}

\begin{figure}
    \centering
    \includegraphics[width=0.42\textwidth]{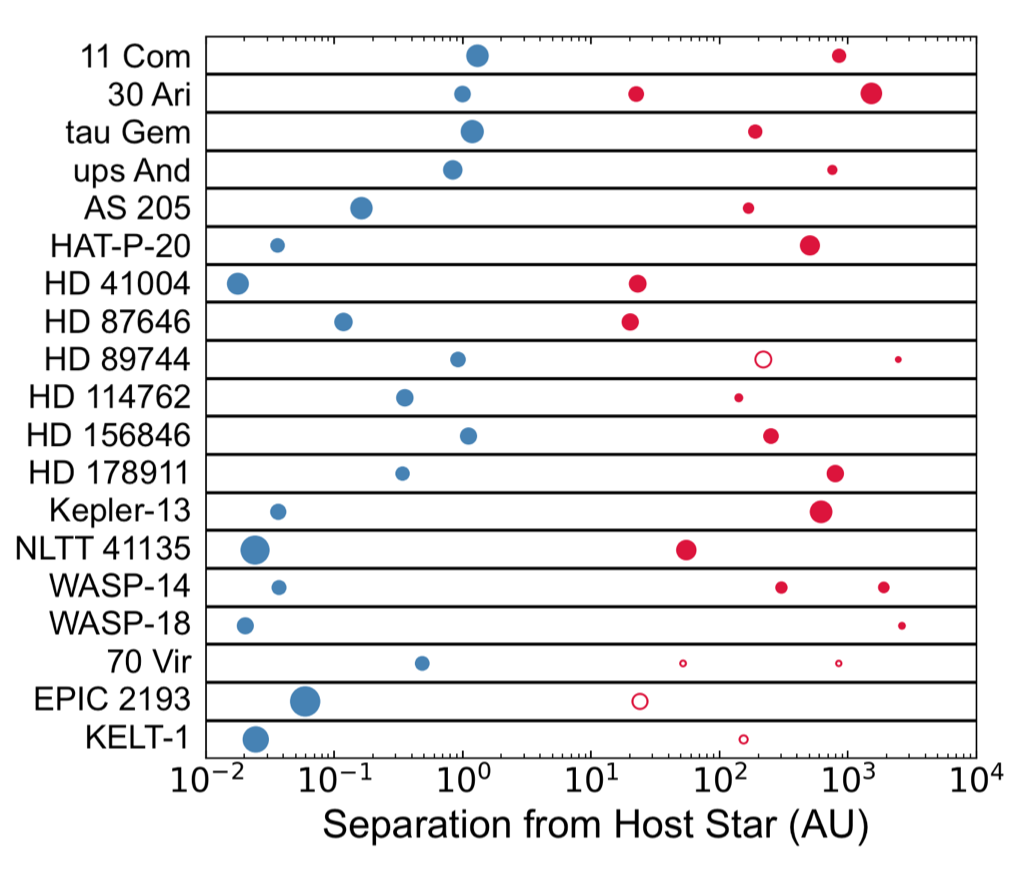}
    \caption{Architecture of all binary or higher-order multiple systems found in our sample, following the order of the targets shown in Tables \ref{t:confirmed_binaries} and \ref{t:candidate_binaries}. The blue circles represent the inner brown dwarfs and planets, with symbol sizes proportional to their masses. Red circles show the positions of all known confirmed (filled symbols) and candidate (open symbols) wide companions, with radii proportional to the mass ratios of these outer companions to the planet hosts. Separations of inner companions correspond to semi-major axes, while observed projected separations are displayed for the wide binary companions.}
    \label{f:architecture}
\end{figure}

We searched for wide companions to the 38 systems from our core sample using our new data and published direct imaging observations, as well as the \textit{Gaia} DR2 catalogue.
A total of 16 objects were found to have at least one wide stellar or substellar companion confirmed to be comoving, listed in Table \ref{t:confirmed_binaries}. Another 7 candidate companions are reported in the literature around 4 of our targets and are presented in Table \ref{t:candidate_binaries}. One of the targets with a reported candidate is already a confirmed wide binary (HD 89744). Figure \ref{f:planet_binarity} displays the properties of the inner planets and brown dwarfs, showing the positions in the planet period-mass space of confirmed binaries (star symbols), targets with a candidate companion (triangles) and apparently single objects (circles). In Figure \ref{f:architecture} we present the architecture of each identified hierarchical system, plotting the semi-major axes of the inner companions in blue and the projected separations of detected wide binary components in red, with symbol sizes proportional to the planetary masses and binary mass ratios, respectively.

\subsection{Literature search and imaging surveys}
\label{lit_search}

We conducted an extensive literature search to compile available observations of all objects in our sample and gather existing knowledge about the multiplicity of our targets. We present our findings for each individual target in Appendix \ref{A:binary_search}, providing detailed information about every companion, candidate or confirmed, reported around our targets in imaging surveys or catalogues, as well as null-detections. A total of 30 targets are mentioned in the literature in the context of a search for wide companions (with or without detections), to which we add 4 of our 6 observed targets that had no previously reported observations (see Section \ref{observations}).

Of these 34 objects, we found 21 targets with reported detections in the literature. Among those, 15 are confirmed bound systems (11 Com AB, 30 Ari ABC, $\tau$ Gem AB, $\upsilon$ And AB, AS 205 ABC, HAT-P-20 AB, HD 41004 AB, HD 87646 AB, HD 89744 AB, HD 114762 AB, HD 156846 AB, HD 178911 ABC, Kepler-13 AB, NLTT 41135 AB and WASP-14 AB), which we detail in Appendix \ref{A:binaries}. 14 of these binaries or higher-order multiple were demonstrated to form physical pairs in the literature and we confirm the true companionship of the $\tau$ Gem AB system in this work (see Figure \ref{f:tau_Gem_cpm}). We list all confirmed multiples in Table \ref{t:confirmed_binaries}. Three of these systems, HD 87646, HD 41004 and HD 178911, were identified as binaries in the Tycho-Hipparcos catalogues and for the latter two systems, our planet-host stars correspond to the fainter component of the binary system.

\afterpage{
\onecolumn
\begin{landscape}

\begin{small}
\begin{longtable}[l]{ p{1.2cm} c c c c c c c c c c l}

\caption{Confirmed common proper motion systems.}\\
\toprule \toprule
System & \multicolumn{2}{c}{Separation} & Comp. & SpT & Photometry & Mass & $G$ & $\pi$ & $\mu_\alpha$ & $\mu_\delta$ & References \\
\cmidrule{2-3}
 & (\arcsec) & (AU) & & & (mag) & (M$_\odot$) & (mag) & (mas) & (mas yr$^{-1}$) & (mas yr$^{-1}$) & \\
\midrule
\endfirsthead

\caption{(Continued.)}\\
\toprule \toprule
System & \multicolumn{2}{c}{Separation} & Comp. & SpT & Photometry & Mass & $G$ & $\pi$ & $\mu_\alpha$ & $\mu_\delta$ & References \\
\cmidrule{2-3}
 & (\arcsec) & (AU) & & & (mag) & (M$_\odot$) & (mag) & (mas) & (mas yr$^{-1}$) & (mas yr$^{-1}$) & \\
\midrule
\endhead

\multirow{2}{*}{11 Com}  & \multirow{2}{*}{9.1} & \multirow{2}{*}{850} & \textbf{A} & G8 III & $V$=4.8 & 2.02 & 4.37 & $10.71\pm0.22$ & $-109.24\pm0.32$ & $88.17\pm0.28$ &\multirow{2}{*}{CCDM; GDR2} \\
 &   &  & B & ... & $V$=12.9 & \textit{0.7} & 12.2 & $9.56\pm0.05$ & $-108.09\pm0.08$ & $89.64\pm0.05$ & \\     [0.4cm]

\multirow{3}{*}{30 Ari} & \multirow{2}{*}{0.536} & \multirow{2}{*}{22} & \textbf{B} & F8 V & $V$=7.1,$i$=6.9 & 1.16 & 6.96& $22.36\pm0.05$ & $141.41\pm0.08$ & $-10.68\pm0.09$ & \multirow{2}{*}{B-C pair: \citet{Riddle2015, Roberts2015}} \\
 & & & C & M1 V & $i$=11.2 & 0.5 & ... & ... & ... & ... & \\
& 38.2 & 1520 & A & F5 V & $V$=6.5 & 1.32  & 6.38 & $22.13\pm0.07$ & $136.86\pm0.14$ & $-15.19\pm0.14$ & A-BC pair: \citet{Guenther2009}; GDR2 \\     [0.4cm]

\multirow{2}{*}{$\tau$ Gem} & \multirow{2}{*}{1.9} & \multirow{2}{*}{187} & \textbf{A} & K2 III & $V$=4.5 & 2.3 & 3.95 & $8.88\pm0.32$ & $-31.49\pm0.49$ & $-48.28\pm0.43$ & \multirow{2}{*}{CCDM; WDS; \textit{this work}} \\
& & & B & K0 V & $V$=11 & 0.8 & 9.42 & ... & ... & ... & \\     [0.4cm]

\multirow{2}{*}{$\upsilon$ And} & \multirow{2}{*}{55} & \multirow{2}{*}{750} & \textbf{A} & F8 V & $J$=3.2 & 1.31 & 3.90 & $74.57\pm0.35$ & $-172.25\pm0.52$ & $-382.90\pm0.54$ & \multirow{2}{*}{\citet{Lowrance2002}; GDR2} \\
& & & B & M4.5 V & $J$=9.4 & 0.2 & 12.51 & $74.21\pm0.09$ & $-172.08\pm0.13$ & $-383.90\pm0.14$ & \\     [0.4cm]

\multirow{2}{*}{AS 205} & \multirow{2}{*}{1.3} & \multirow{2}{*}{166} & \textbf{A} & K5 V & $J$=8.1 & 1.086 & 12.37 & $7.82\pm0.10$ & $-7.45\pm0.20$ & $-26.89\pm0.14$ & A-BC pair: \citet{Ghez1993, Prato2003}; GDR2 \\
& & & BC & K7+M0 V & $J$=9.2 & 0.22 & 13.40 & $6.38\pm0.19$ & $-9.48\pm0.47$ & $-23.17\pm0.43$ & BC pair: \citet{Eisner2005} \\     [0.4cm]

\multirow{2}{*}{HAT-P-20} & \multirow{2}{*}{6.86} & \multirow{2}{*}{500} & \textbf{A} & K3 V & $J$=9.3 & 0.756 & 10.99 & $14.05\pm0.04$ & $-4.99\pm0.07$ & $-96.23\pm0.06$ &  \multirow{2}{*}{WDS; \citet{Bakos2011}} \\
 & & & B & M V & $J$=10.2 & \textit{0.57} & 12.80 & ... & ... & ... & \\     [0.4cm]

\multirow{2}{*}{HD 41004} & \multirow{2}{*}{0.54} & \multirow{2}{*}{23} & \textbf{B} & M2 V & $Hp$=12.5 & 0.4 & ... & ... & ... & ... & \multirow{2}{*}{\citet{Zucker2003}; WDS; CCDM; Hipparcos-Tycho} \\
& & & A & K1 V & $Hp$=8.8 & 0.7 & 8.38 & $24.04\pm0.26$ & $-41.52\pm0.48$ & $59.65\pm0.56$ & \\     [0.4cm]

\multirow{2}{*}{HD 87646} & \multirow{2}{*}{0.26} & \multirow{2}{*}{20} & \textbf{A} & G1 IV & $Hp$=8.3 & 1.12 & 7.93 & $7.40\pm0.78$ & $-82.88\pm1.55$ & $-27.08\pm1.35$ & \multirow{2}{*}{\citet{Ma2016}; Hipparcos-Tycho, WDS; CCDM} \\
& & & B & K V & $Hp$=11.0 & 0.6 & ... & ... & ... & ... & \\     [0.4cm]

\multirow{2}{*}{HD 89744} & \multirow{2}{*}{63} & \multirow{2}{*}{2460} & \textbf{A} & F7 V & $J$=4.9 & 1.558 & 5.59 & $25.85\pm0.07$ & $-120.57\pm0.11$ & $-138.14\pm0.13$ & \multirow{2}{*}{\citet{Mugrauer2004}; GDR2} \\
& & & B & L0 V & $J$=14.9 & 0.08 & 19.66 & $25.96\pm0.95$ & $-119.93\pm0.98$ & $-140.08\pm1.07$ & \\     [0.4cm]

\multirow{2}{*}{HD 114762} & \multirow{2}{*}{3.2} & \multirow{2}{*}{140} & \textbf{A} & F9 V & $J$=6.2 & 0.83 & 7.14 & $24.86\pm0.27$ & $-586.29\pm0.71$ & $2.26\pm0.24$ & \multirow{2}{*}{\citet{Patience2002}} \\
& & & B & M9 V & $J$=13.8 & 0.09 & ... & ... & ... & ... & \\     [0.4cm]

\multirow{2}{*}{HD 156846} & \multirow{2}{*}{5.1} & \multirow{2}{*}{250} & \textbf{A} & G1 V & $J$=5.5 & 1.38 & 6.36 & $20.92\pm0.05$ & $-137.10\pm0.10$ & $-143.20\pm0.07$ & \multirow{2}{*}{\citet{Tamuz2008}; WDS; GDR2} \\
& & & B & M4 V & $J$=9.4 & 0.59 & 12.19 & $21.43\pm0.42$ & $-129.62\pm0.70$ & $-132.84\pm0.58$ & \\     [0.4cm]

\multirow{2}{*}{HD 178911} & \multirow{2}{*}{16.1} & \multirow{2}{*}{790} & \textbf{B} & G5 V & $Hp$=8.3 & 1.03 & 7.87 & $24.38\pm0.03$ & $57.18\pm0.04$ & $195.90\pm0.05$ & AC-B pair: \citet{Tokovinin2000}; Hipparcos-Tycho; GDR2 \\
& & & AC & G1+K1 V & $Hp$=6.8 & 1.9 & 6.57 & $20.23\pm0.38$ & $76.62\pm0.69$ & $207.13\pm0.66$ & A-C pair: \citet{McAlister1987} \\     [0.4cm]

\multirow{2}{*}{Kepler-13} & \multirow{2}{*}{1.15} & \multirow{2}{*}{610} & \textbf{A} & A5 V & $V$=10.35 & 1.72 & 10.55 & $2.09\pm0.08$ & $-3.95\pm0.18$ & $-15.05\pm0.26$ & A-BC pair: \citet{Szabo2011, Shporer2014}; GDR2 \\
& & & BC & A+G V & $V$=10.48 & 1.68 & 10.37 & $1.91\pm0.11$ & $-4.40\pm0.19$ & $-15.78\pm0.24$ & B-C pair: \citet{Santerne2012, Johnson2014} \\     [0.4cm]

\multirow{2}{*}{NLTT 41135} & \multirow{2}{*}{2.4} & \multirow{2}{*}{55} & \textbf{B} & M5.1 V & $z$=13.1 & 0.164 & 14.94 & $29.27\pm0.12$ & $162.51\pm0.18$ & $-282.72\pm0.18$ & \multirow{2}{*}{\citet{Irwin2010}; GDR2} \\
& & & A & M4.2 V & $z$=12.4 & 0.21 & 13.97 & $29.11\pm0.16$ & $153.67\pm0.24$ & $-281.98\pm0.24$ &  \\     [0.4cm]

\multirow{3}{*}{WASP-14} & \multirow{2}{*}{1.45} & \multirow{2}{*}{300} & \textbf{A} & F5 V & $J$=8.9 & 1.35 & 9.65 & $6.14\pm0.03$ & $29.24\pm0.06$ & $-6.95\pm0.06$ & \multirow{2}{*}{A-B pair: \citet{Ngo2015}} \\
& & & B & ... & $J$=14.1 & 0.33 & ...& ... & ... & ... & \\
& 11.5 & 1900 & C & \textit{K5 V} & ... & \textit{0.280} & 17.32 & $6.08\pm0.10$ & $27.97\pm0.20$ & $-6.15\pm0.18$ & AB-C pair: \textit{this work}; GDR2. \\     [0.4cm]

\multirow{2}{*}{WASP-18} & \multirow{2}{*}{26.7} & \multirow{2}{*}{3300} & \textbf{A} & F6 IV/V & $V$=9.3 & 1.20 & 9.17 & $8.07\pm0.02$ & $25.24\pm0.03$ & $20.60\pm0.03$ & \multirow{2}{*}{This work; GDR2} \\
& & & B & \textit{M7.5 V} & ... & \textit{0.092} & 20.92 & $9.43\pm1.52$ & $23.65\pm1.98$ & $18.38\pm2.40$ & \\ [0.1cm]

\bottomrule \\ [-1ex]
\multicolumn{12}{l}{
  \begin{minipage}{1.3\textwidth}
    \textbf{Notes.} The top component of each system (marked in bold) is the planet host considered in the sample studied here. Spectral types and masses in italic were derived in this work. $G$-magnitudes, parallaxes and proper motions come from the Gaia DR2 catalogue.\\
    CCDM: Catalog of Components of Double and Multiple stars, \citet{Dommanget2000}.\\
    GDR2: Gaia Data Release 2, \citet{GaiaCollaboration2018}.\\
    WDS: the Washington Double Star Catalog, \citet{Mason2001}.
  \end{minipage}}

\label{t:confirmed_binaries}
\end{longtable}
\end{small}

\begin{small}
\begin{longtable}[l]{ l c c c c c c c c l}

\caption{Candidate binary companions.}\\
\toprule \toprule
System & Comp. & SpT & Photometry & Mass & \multicolumn{2}{c}{Separation} & Prob. & Companionship & References \\
\cmidrule{6-7}
 & & & (mag) & (M$_\odot$) & (\arcsec) & (AU) & (\%) & & \\
\midrule
\endfirsthead

70 Vir & \textbf{A} & G5 V & $I=3.98, J=3.80$ & 1.07 & -- & -- & -- & -- & \\
& B & $>$ M5 V & $\Delta I=11.4\pm1.2$ & \textit{0.08} & 2.86 & 52 & 99.67 & likely bound & \citet{Roberts2011}  \\
& C & L V & $J=15.84\pm0.16$ & \textit{0.07} & 42.7 & 848 & 76.91 & inconclusive & \citet{Pinfield2006} \\ [0.4cm]

EPIC 219388192 & \textbf{A} & G V & $H=10.734, K=10.666$ & 1.01 & -- & -- & -- & -- & \\
& B & M V & $\Delta K=2.24$ & 0.52 & 0.082 & 24 & 99.99 & likely bound &  Curtis et al. (private communication) \\
& C & $>$ M8 V & $\Delta H=7.087\pm0.032$ & $<$ 0.1 & 5.988 & 1769 & 14.33 & likely background & \citet{Nowak2017} \\
& D & $>$ M8 V & $\Delta H=7.663\pm0.057$ & $<$ 0.1 & 7.538 & 2224 & 4.65 & likely background & \citet{Nowak2017} \\ [0.4cm]

HD 89744 & \textbf{A} & F7 V & $I=5.2$ & 1.558 & -- & -- & -- & -- & \\ 
& C & ... & $\Delta I=13\pm2$ & \textit{0.08} & 5.62 & 219 & 99.17 & likely bound  & \citet{Roberts2011} \\[0.4cm]

KELT-1 & \textbf{A} & F5 V & $H=9.534\pm0.030, K=9.437\pm0.019$ & 1.335 & -- & -- & -- & -- & \\
& B & M4-5 V & $\Delta H=5.90\pm0.10, \Delta K=5.59\pm0.12$ & \textit{0.2} & 0.588 & 154 & 99.95 & likely bound & \citet{Siverd2012} \\ [0.4cm]

\bottomrule \\ [-1ex]
\multicolumn{10}{l}{
  \begin{minipage}{1.3\textwidth}
    \textbf{Notes.} The top component of each system (marked in bold) is the planet host considered here.
  \end{minipage}}

\label{t:candidate_binaries}
\end{longtable}
\end{small}

\end{landscape}
\twocolumn

}

The remaining 6 targets are mentioned to have unconfirmed candidate companions and are discussed in Appendices \ref{A:good_candidates} and \ref{A:rejected_candidates}. We discarded the 3 point sources reported by \citet{Moutou2017} around HD 168443, which are highly likely to be background contaminants given the crowded galactic latitude of the target (\citealp{Moutou2017}; see discussion of HD 168443 in Appendix \ref{A:rejected_candidates}). In Section \ref{HD_162020}, we showed that the faint candidate reported around HD 162020 by \citet{Eggenberger2007} does not share common proper motion with the primary and thus rejected this candidate. We were also able to refute the candidate companion reported around XO-3 by \citet{Bergfors2013} and \citet{WollertBrandner2015} based on the inconsistent parallax and proper motion of this source and XO-3 in \textit{Gaia} DR2. This leaves 3 targets with unconfirmed candidates, namely, 70 Vir (two candidates), EPIC 219388192 (three candidates) and KELT-1 (one candidate). A candidate companion is also reported by \citet{Roberts2011} around HD 89744, already known to be a confirmed wide binary \citep{Mugrauer2004}. These final 7 candidate companions retained for this study are presented in Table \ref{t:candidate_binaries}.

We are not able to make any clear statement on the physical association of these candidates based on the available data. We can however make a statistical argument on the chance of finding an unrelated background source at close angular separation from the primaries. For each source, we used the Trilegal galaxy models \citep{Vanhollebeke2009} to calculate a probability of the observed candidates being true companions. This was done by estimating the surface density $\rho$ of background sources expected to be found within 30\arcmin from the primary targets, given the galactic latitude and longitude of the objects and the depth and wavelength of the obtained observations. From \citet{Brandner2000}, the probability $P(\Theta, m)$ of detecting one or more background stars within an angular separation $\Theta$ (in arcsec) and down to a limiting magnitude $m$ is then given by:
\begin{equation}
    P(\Theta, m) = 1 - e^{-\pi \Theta^2 \: \rho(m)}.
\end{equation}
The probability of an observed candidate being physically associated to the primary is then given by the complement of the chance of alignment, that is, $1 - P(\Theta, m)$. The resulting probabilities are listed in Table \ref{t:candidate_binaries} for each candidate. The two faint candidates identified beyond 7\arcsec from EPIC 219388192 by \citet{Nowak2017} were found to likely be background sources, with probabilities $<$15\% of being physically associated. With the exception of the wider candidate around 70 Vir, most other candidates were found to have very high probabilities of being bonafide companions: the close candidates around 70 Vir, EPIC 219388192, HD 89744 and KELT-1 have $>$99\% probabilities of being bound. While additional observations will be required to confirm their true companionship through common proper motion analyses, these objects are therefore highly likely to be true companions.

Finally, a total of 9 targets from our core sample are mentioned as single objects in the literature and have reported null detections from direct imaging surveys (CI Tau, HAT-P-2, HD 5891, HD 33564, HD 104985, HD 156279, HD 180314, HD 203949 and WASP-18; see Appendix \ref{A:null_detecitons}). We add to these objects 4 of our observed targets that had no previous observations (BD+24 4697, HD 77065, HD 134113 and HD 160508) and around which we did not find any companion. No published data were found in the literature for the targets 4 UMa, 59 Dra, HD 39392 and HD 112410, which we were not able to observe either.

\subsection{Companions in \textit{Gaia} DR2}
\label{gaia_dr2}

\citet{Raghavan2010} found that the period distribution of binary companions to nearby FGK stars is approximately a Gaussian in the logarithm of the period, with a broad peak around 300 yrs ($\sim$50 AU), and a 1-$\sigma$ Gaussian interval spanning from 2 to 1500 AU, in reasonable agreement with previous studies by \citet{Duquennoy1991}. Most of the imaging data considered here only allow for the detection of companions out to several hundred AU. Hence a significant number of wider companions could remain outside the field of view of these observations and be missed by direct imaging surveys. At these wide separations, outer companions are expected to be massive (i.e. a stellar binary) in order to be able to affect the formation or evolution of close-in planets or brown dwarfs. Such wide stellar companions are expected to be found in the \textit{Gaia} Data Release 2 (DR2; \citealp{GaiaCollaboration2016, GaiaCollaboration2018}), which may thus be used to search for widely-separated comoving components to the objects in our sample. We therefore searched for \textit{Gaia} DR2 sources with parallaxes and proper motions consistent with those of our targets, to complement our direct imaging search for wide companions in Section \ref{lit_search}.

The recent release of the \textit{Gaia} DR2 catalogue provides unprecedentedly-precise astrometric measurements on the parallaxes and proper motions of stars. However, these highly-precise measurements must be considered and handled with caution in the context of a search for common proper motion systems. \citet{Shaya2011} accurately pointed out that astrometric missions spanning a few years only (e.g. Hipparcos, 3.5 year baseline) capture the reflex motions of multiple systems in their kinematics measurements. Indeed, the components of a binary wobble around the centre of mass of the system and a short-term proper motion measurement is highly likely to reflect this orbital motion. Longer time spans are required to ensure that the observed proper motions correspond to the true barycentric motion (e.g. the Tycho-2 catalogue which uses data from over a century timescale). The apparent changes in proper motions between short and long-term measurements can be as large as several tens of mas yr$^{-1}$, based on the components masses, binary separation, orbital phase and parallax of the system \citep{Shaya2011}. These changes in proper motion may even be exploited as a way to search for hidden companions, as was done by \citet{Makarov2005} with the Hipparcos and Tycho-2 catalogues.
Despite the excellent precision of \textit{Gaia} DR2, the catalogue is based entirely on data collected between July 2014 and May 2016, spanning a period of only 22 months, and the same problem as for Hipparcos is encountered.

While these effects are reduced at very wide separations, further complications can also arise from the presence of an unresolved binary. \citet{Shaya2011} estimated that a tight system separated by a few AU could induce proper motion fluctuations of several mas yr$^{-1}$ on the primary, orders of magnitude larger than the errors on \textit{Gaia} DR2 measurements. Close binaries not resolved in \textit{Gaia} are treated as single objects in the second Data Release, which can lead to specious astrometric solutions \citep{Arenou2018}. A third component at a wide separation around an unresolved binary is therefore likely to show somewhat different astrometric parameters (proper motion and parallax) compared to its comoving, unresolved primary.

As a result, we adopted rather loose selection criteria to search for comoving companions to the objects in our sample using the \textit{Gaia} DR2 catalogue. We considered the relative differences in parallax $\pi$, i.e. $\Delta \pi / \pi_0 \equiv | (\pi_0 - \pi_i)/\pi_0 |$, where the subscript 0 corresponds to our science target and $i$ to other \textit{Gaia} sources. We then defined similar relative differences for the proper motion components, $\mu_{\alpha*}$ and $\mu_\delta$. To account for the uncertainties in the \textit{Gaia} measurements, we generated, for each pair of objects, $10^5$ parallaxes and proper motions drawn from Gaussian distributions centred on the measured values, with a standard deviation set to the \textit{Gaia} uncertainties. We then calculated $10^5$ corresponding fractional differences in $\pi$, $\mu_{\alpha*}$ and $\mu_\delta$ and set the final relative differences and associated uncertainties to the mean and standard deviation of the output distributions.

We selected sources that were consistent with relative differences of less than 20\% in parallax and in at least one of the two proper motion components (including the correct direction), with a maximum relative discrepancy of 50\% in the other proper motion component. We searched for such companions in the \textit{Gaia} DR2 catalogue for all targets in our sample, out to angular separations corresponding to projected separations of $10^4$ AU. We found a total of 11 systems fulfilling the above selection criteria, 9 of which were previously known systems. These systems are listed in Table \ref{t:confirmed_binaries}, in which we give the \textit{Gaia} DR2 parallaxes and proper motions for each binary component. The characterisation of the two newly-identified \textit{Gaia} systems, WASP-14 AB-C and WASP-18 A-B, is detailed in Sections \ref{WASP-14} and \ref{WASP-18} below.

\begin{table}
    \centering
    \caption{Relative differences in parallax and proper motion, with their associated errors, between the components of all \textit{Gaia} binaries. Fractional differences are calculated relative to the first component listed for each system (single or binary), and our science targets always correspond to the first component given. The two new binaries identified in this work are marked in bold.}
    \begin{tabular}{p{2.3cm} c c c}
        \hline \hline
        System & $\Delta \pi / \pi_0$ & $\Delta \mu_{\alpha*} / \mu_{\alpha*,0}$ & $\Delta \mu_{\delta} / \mu_{\delta,0}$ \\
         & (\%) & (\%) & (\%) \\
        \hline
        11 Com\dotfill A-B      & 10.74 $\pm$ 1.92 & 1.05 $\pm$ 0.30 & 1.67 $\pm$ 0.33 \\
        30 Ari\dotfill BC-A     & 1.03 $\pm$ 0.39 & 3.21 $\pm$ 0.11 & 42.23 $\pm$ 1.81 \\
        $\upsilon$ And\dotfill A-B & 0.48 $\pm$ 0.38 & 0.10 $\pm$ 0.19 & 0.26 $\pm$ 0.14 \\
        AS 205\dotfill A-BC     & 18.41 $\pm$ 1.55 & 27.25 $\pm$ 7.14 & 13.83 $\pm$ 1.66 \\
        HD 89744\dotfill A-B    & 0.43 $\pm$ 2.24 & 0.53 $\pm$ 0.58 & 1.40 $\pm$ 0.75 \\
        HD 156846\dotfill A-B   & 2.44 $\pm$ 1.70 & 5.46 $\pm$ 0.51 & 7.23 $\pm$ 0.41 \\
        HD 178911\dotfill B-AC  & 17.02 $\pm$ 1.58 & 33.40 $\pm$ 1.20 & 5.73 $\pm$ 0.34 \\
        Kepler-13\dotfill A-BC  & 8.61 $\pm$ 5.46 & 11.39 $\pm$ 6.55 & 4.85 $\pm$ 2.38 \\
        NLTT 41135\dotfill B-A  & 0.55 $\pm$ 0.51 & 5.44 $\pm$ 0.18 & 0.26 $\pm$ 0.11 \\
        \textbf{WASP-14\dotfill AB-C}    & 0.98 $\pm$ 1.17 & 4.34 $\pm$ 0.71 & 11.51 $\pm$ 2.69 \\
        \textbf{WASP-18\dotfill A-B}     & 16.85 $\pm$ 14.60 & 6.30 $\pm$ 5.94 & 10.78 $\pm$ 9.03 \\
        \hline
    \end{tabular}
    \label{t:Gaia_frac_diff}
\end{table}

\begin{figure}
    \centering
    \includegraphics[width=0.45\textwidth]{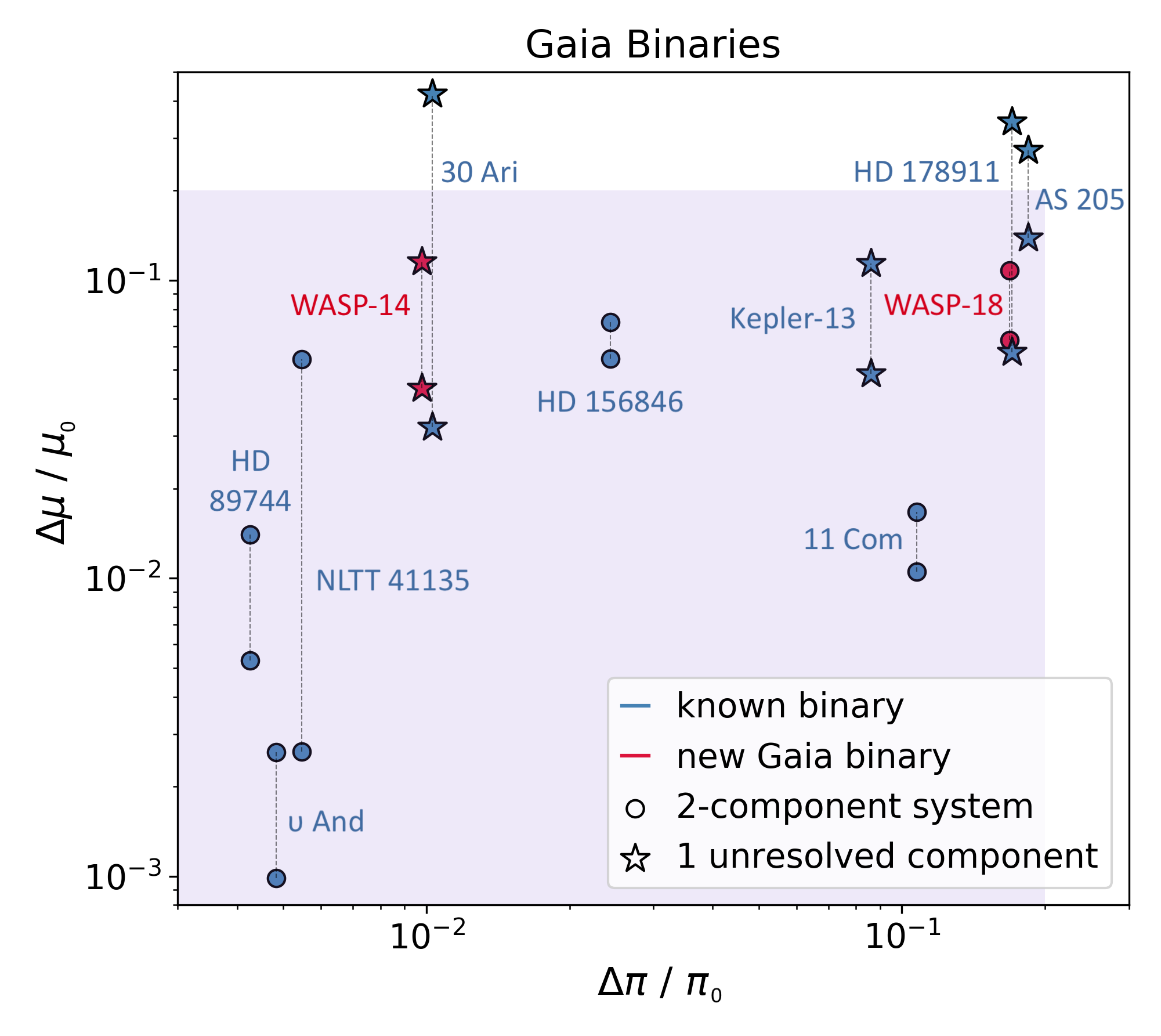}
    \caption{Binaries identified in \textit{Gaia} DR2, showing the relative difference in parallax (x-axis) against the relative differences in proper motion (y-axis, both RA and Dec components) between the science target and the selected companion. Our selection criteria correspond to the shaded area (see text). Systems marked with stars rather than circles indicate binaries that have a component known to be unresolved in \textit{Gaia}. The two new binaries identified in this work are marked in red. We do not plot error-bars for clarity of the figure but show them in Table \ref{t:Gaia_frac_diff} instead.}
    \label{f:Gaia_binaries}
\end{figure}

In Figure \ref{f:Gaia_binaries} we plot the relative differences in parallax and proper motion (RA and Dec coordinates) between the components of all identified \textit{Gaia} binaries. The shaded area represents our arbitrary cut at 20\% in the relative differences in parallax and proper motion. The 9 previously known systems are marked in blue and our the two new systems discovered here are shown in red. The obtained values and their associated uncertainties are given in Table \ref{t:Gaia_frac_diff} and are all consistent with our chosen constraints at the 1-$\sigma$ level.

In Appendix \ref{A:Gaia_analysis} we examined those systems more carefully, as well as other known binaries in our sample, in order to assess our selection criteria. We found that our selection method successfully identified all known binaries that were recoverable given the sensitivity and completeness of \textit{Gaia} DR2, and consider that we unlikely missed additional binaries present in the \textit{Gaia} DR2 catalogue. Based on the location of systems with a known unresolved component in Figure \ref{f:Gaia_binaries} (marked with stars), we conclude that most binaries should have relative discrepancies of $<$10\% in all astrometric parameters ($\pi$, $\mu_{\alpha*}$ and $\mu_\delta$), while systems agreeing to within 20\% in parallax and in one of the proper motion coordinates are likely to be hierarchical systems with an unresolved component (see Appendix \ref{A:Gaia_analysis} for details).

\subsubsection{WASP-14}
\label{WASP-14}

WASP-14 A is already known to have a 0.33 M$_\odot$ bound companion at 300 AU (\citealp{WollertBrandner2015, Ngo2015}) as discussed in Appendix \ref{A:binaries} (see Table \ref{t:confirmed_binaries}). This companion is not detected in \textit{Gaia} due to the small angular separation (1\farcs45) and large magnitude difference ($\Delta$\textit{J}$=$5.2 mag) of the WASP-14 A-B system (see discussion in Appendix \ref{A:Gaia_analysis}). We report a new companion to this system, WASP-14 C (\textit{Gaia} DR2 1242084166679297920), at a separation of 11\farcs5397$\pm$0\farcs0001 and a position angle of 4.5827$\pm$0.0003 deg. The measured angular separation corresponds to a wide projected separation of 1900 AU at the distance of WASP-14 (see Table \ref{t:confirmed_binaries}). WASP-14 AB and C have measured \textit{Gaia} DR2 parallaxes in excellent agreement, with a relative difference $<$1\%. The relative discrepancies in proper motion are slightly larger but still in very good agreement: 4.34\% in $\mu_{\alpha*}$ and 11.51\% in $\mu_{\delta}$. Given the consistent parallax and small offsets in proper motion, we conclude that the two objects are comoving and form a physical pair. Comparing the placement of WASP-14 in Figure \ref{f:Gaia_binaries} to the other \textit{Gaia} binaries in our sample also reinforces the idea that WASP-14 is a true binary and confirms our intuition that systems with an unresolved component tend to show larger disparities in their observed short-term proper motions.

WASP-14 C has a \textit{Gaia} $G$-band magnitude of 17.32 mag, for a magnitude difference of $\Delta$\textit{G}$=$7.67 mag with the unresolved WASP-14 AB primary. Photometry in the blue ($G_\mathrm{BP}$) and red ($G_\mathrm{RP}$) filters of \textit{Gaia} indicate fairly red colours for this object, with $G_\mathrm{BP} - G_\mathrm{RP} = 2.67$ mag. This suggests a mid-K to early-M main sequence star according to the \textit{Gaia} DR2 HR diagram analysis in \citet{GaiaCollaboration2018}. The new companion to WASP-14 is also found in the 2MASS catalogue, with magnitudes of $J=14.297\pm0.054$, $H=13.801\pm0.049$ and $K_s=13.592\pm0.058$. According to \citet{Schmidt-Kaler1982}, the 2MASS colours correspond to a K5 V spectral type. This implies a bolometric correction of BC$_K$ of $\sim$2.3$\pm$0.1 mag \citep{Masana2006}. From these values, we calculated a bolometric luminosity and used the BT-Settl models \citep{Allard2012} to infer a mass for WASP-14 C. Adopting a distance based on the \textit{Gaia} DR2 parallax of the target and the age of the system given in Table \ref{t:stellar_properties}, we derived a mass $0.280\pm0.016$ M$_\odot$ for the newly discovered stellar component of the triple system WASP-14, making this companion the lowest-mass component of the system.

\begin{figure}
    \centering
    \includegraphics[width=0.47\textwidth]{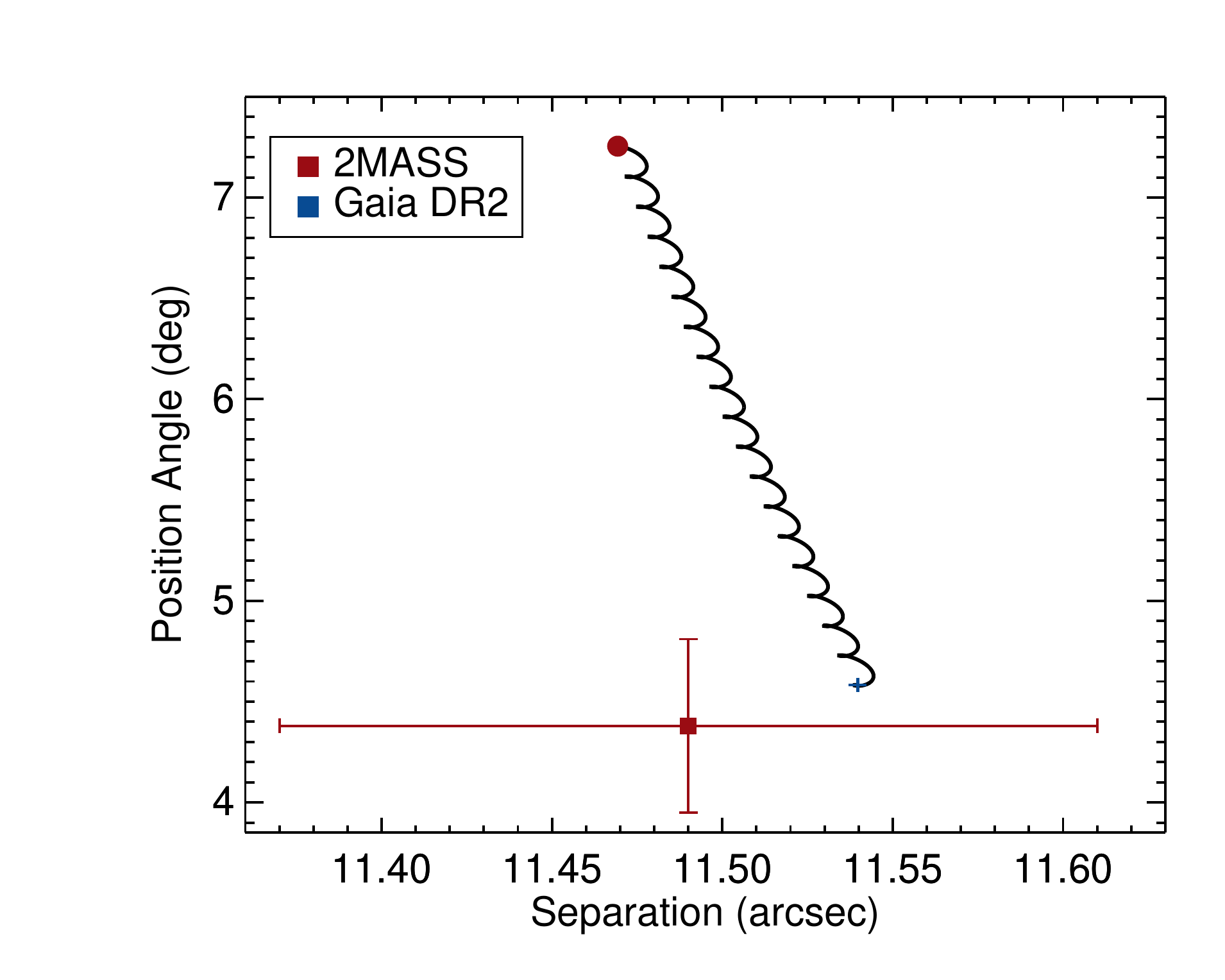}
    \caption{Common proper motion analysis of the WASP-14 AB-C system. The black solid line represents the motion of a background object relative to the primary, WASP-14 AB, computed using the proper motion and parallax measurements of our science target the \textit{Gaia} DR2 catalogue. The blue cross and red square mark the measured position of the components in \textit{Gaia} DR2 and in 2MASS, respectively. The red circle indicates the expected position of a background source at the date of the 2MASS observations (June 1997). The \textit{Gaia} DR2 epoch is 2015.5, providing an 18-year baseline between the two epochs available. As expected from the \textit{Gaia} astrometry of the system, the relative motion of the companion between the two epochs is consistent with a comoving pair.}
    \label{f:CPM_WASP-14}
\end{figure}

With its low mass and extremely wide separation, WASP-14 C is unlikely to have played a role in the formation or evolution of the 7.8 M$_\mathrm{Jup}$ planet on a 2.2 day orbit around WASP-14 A, as the closer and more massive WASP-14 B component would have had a much stronger influence (if any) on the inner substellar companion.
In Figure \ref{f:CPM_WASP-14} we show the relative positions of WASP-14 AB and C from the 2MASS and \textit{Gaia} DR2 catalogues, confirming over a $\sim$20-year baseline that the companion is indeed comoving. The WASP-14 A-B pair is unresolved in both 2MASS and \textit{Gaia}.

\subsubsection{WASP-18}
\label{WASP-18}

We report here the detection of a faint comoving companion at 26\farcs728$\pm$0\farcs001 ($\sim$3300 AU projected separation) and a position angle of 200.520$\pm$0.001 deg, found in the \textit{Gaia} DR2 catalogue (\textit{Gaia} DR2 4931352153572401152), outside the field of view of our VLT/NACO data (see Section \ref{observations}). The same companion was independently identified by \citet{Csizmadia2019} at the same time. The source has a \textit{Gaia} magnitude $G = 20.92$ mag, and a $G - G_\mathrm{RP}$ colour of 1.79 mag. We do not consider the $G_\mathrm{BP}$ photometry of this source because of the known poor quality of fluxes in this bandpass for red sources with \textit{G} $\sim$ 20$-$21 mag \citep{GaiaCollaboration2018}. At the distance of WASP-18, this places the companion on the M/L-dwarf sequence in the HR diagram for \textit{Gaia} DR2 sources presented in \citet{GaiaCollaboration2018}.

\begin{figure}
    \centering
    \includegraphics[width=0.47\textwidth]{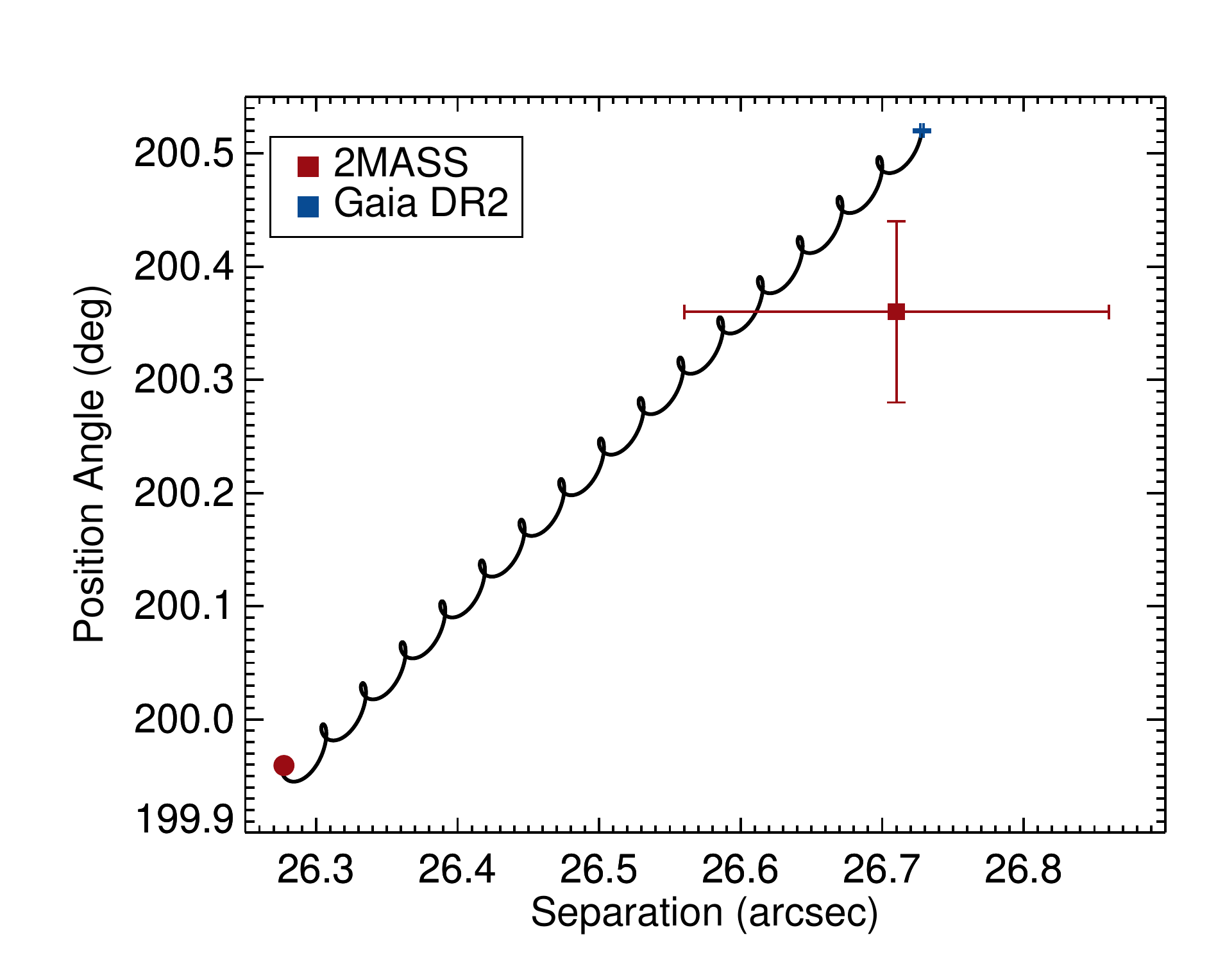}
    \caption{Same as Figure \ref{f:CPM_WASP-14} for the WASP-18 A-B system. The \textit{Gaia} DR2 epoch is again 2015.5 and the 2MASS observations date from August 1999, providing a 16-year time span, clearly demonstrating that the two objects share common proper motion.}
    \label{f:CPM_WASP-18}
\end{figure}

We searched for the same object in the 2MASS catalogue and retrieved a source at the same relative position (26\farcs71$\pm$0\farcs15 and 200.36$\pm$0.08 deg), in excellent agreement with the consistent proper motions of the two sources in \textit{Gaia} DR2. Figure \ref{f:CPM_WASP-18} shows the relative positions of WASP-18 A and B at the time of the 2MASS and \textit{Gaia} DR2 observations, together with the expected motion of a background source between the two epochs. The two objects are clearly found to be comoving based on this analysis, confirming our findings from the \textit{Gaia} DR2 catalogue. The 2MASS photometry of WASP-18 B is $J=16.289\pm0.096$, $H=15.513\pm0.083$ and $K_s=15.146\pm0.121$. Using the relation between spectral type and M$_J$ from \citet{Filippazzo2015}, we infer a spectral type of M7.5 for the companion, in agreement with our rough estimate from the \textit{Gaia} HR diagram. We used this spectral type to estimate a bolometric correction BC$_J$ from the relations in \citet{Filippazzo2015} for field objects, and derived a corresponding bolometric luminosity. We finally interpolated the obtained luminosity into the BT-Settl evolutionary models \citep{Allard2012} at an age of $0.90\pm0.20$ Gyr \citep{Bonfanti2016}, to obtain a mass of $0.092\pm0.003$ M$_\odot$ for WASP-18 B.

\begin{table*}
\centering
\begin{small}
\caption{Detection limits used.}
\begin{tabular}{ l l c c c c l}
\hline\hline
Object ID & Facilities / Instrument & Filter & Phot. (mag) & Limits units & Curve flag & Data Set \\
\hline
30 Ari B    & Keck / NIRC2          & $J$   & $6.080\pm0.02$    & $\Delta$mag   & 0     & \citet{Kane2015} \\
70 Vir      & AEOS                  & $I$   & 3.98              & mag           & 2     & \citet{Roberts2011}$^\mathrm{a}$ \\
            & 2MASS                 & $J$   & 3.80              & mag           & 2     & \citet{Pinfield2006} \\
$\upsilon$ And  & AEOS              & $I$   & 3.35              & mag           & 2     & \citet{Roberts2011}$^\mathrm{a}$ \\
                & Lick / LIRC II    & $K'$  & 2.86              & $\Delta$mag   & 0     & \citet{Patience2002} \\
BD+24 4697  & Gemini North / NIRI   & $Ks$  & $7.474\pm0.023$   & $\Delta$mag   & 0     & This paper \\
CI Tau      & Subaru / HiCIAO       & $H$   & $8.43\pm0.04$     & $\Delta$mag   & 0     & \citet{Uyama2017} \\
EPIC 219388192  & Keck / NIRC2      & $K$   & $10.666\pm0.0216$ & $\Delta$mag   & 0     & Curtis et al. (private communication) \\
                & Subaru / IRCS     & $H$   & $10.734\pm0.021$  & flux ratio    & 0     & \citet{Nowak2017} \\
HAT-P-2     & Calar Alto / AstraLux & $z'$  & $9.506\pm0.001$   & $\Delta$mag   & 1     & \citet{Bergfors2013} \\
HAT-P-20    & Keck / NIRC2          & $K$   & $8.601\pm0.019$   & $\Delta$mag   & 1     & \citet{Ngo2015} \\
HD 5891     & Calar Alto / AstraLux & $i'$  & $7.47\pm0.01$     & $\Delta$mag   & 0     & \citet{Ginski2016} \\
HD 33564    & Calar Alto / AstraLux & $i'$  & 4.6               & $\Delta$mag   & 1     & \citet{Ginski2012} \\
HD 41004 B  & Hipparcos             & $Hp$  & $8.785\pm0.014 ^*$   & $\Delta$mag   & 2     & Hipparcos and Tycho Catalogues \\ 
HD 77065    & Gemini North / NIRI   & $Ks$  & $6.638\pm0.020$   & $\Delta$mag   & 0     & This paper \\
HD 87646 A  & Hipparcos             & $Hp$  & $8.203\pm0.002$   & $\Delta$mag   & 2     & Hipparcos and Tycho Catalogues \\
HD 89744    & AEOS                  & $I$   & 5.2               & mag           & 2     & \citet{Roberts2011}$^\mathrm{a}$ \\
            & UKIRT / UFTI          & $H$   & 4.53              & mag           & 0     & \citet{Mugrauer2004} \\
HD 104985   & Calar Alto / AstraLux & $i'$  & $8.302\pm0.145$   & $\Delta$mag   & 1     & \citet{Ginski2012} \\
HD 114762   & Keck / NIRC2          & $K$   & $5.888\pm0.017$   & $\Delta$mag   & 0     & \citet{Patience2002} \\
HD 134113   & WIYN / NESSI          & $z'$  & 7.6               & $\Delta$mag   & 0     & This paper \\
HD 156279   & Calar Alto /AstraLux  & $i'$  & $7.65\pm0.03$     & $\Delta$mag   & 0     & \citet{Ginski2016} \\
HD 160508   & WIYN / NESSI           & $z'$  & 7.4               & $\Delta$mag   & 0     & This paper \\
HD 162020   & VLT / NACO            & $L'$  & $6.539\pm0.024$   & $\Delta$mag   & 0     & This paper \\
HD 168443   & VLT / SPHERE          & $H$   & $5.325\pm0.016$   & $\Delta$mag   & 0     & \citet{Moutou2017} \\
            & VLT / SPHERE          & $Ks$  & $5.211\pm0.015$   & $\Delta$mag   & 0     & \citet{Moutou2017} \\
HD 178911 B & Hipparcos             & $Hp$  & $6.835\pm0.013 ^*$   & $\Delta$mag   & 2     & Hipparcos and Tycho Catalogues \\ 
HD 180314   & Calar Alto /AstraLux  & $i'$  & $6.14\pm0.05$     & $\Delta$mag   & 0     & \citet{Ginski2016} \\
HD 203949   & VLT / SPHERE          & $H$   & $3.107\pm0.200$   & $\Delta$mag   & 0     & \citet{Moutou2017} \\
            & VLT / SPHERE          & $Ks$  & $2.994\pm0.232$   & $\Delta$mag   & 0     & \citet{Moutou2017} \\
KELT-1      & Keck / NIRC2          & $K'$  & $9.437\pm0.019$   & $\Delta$mag   & 2     & \citet{Siverd2012}$^\mathrm{b}$ \\
Kepler-13 A & Palomar HALE / PHARO  & $K'$  & 9.958             & $\Delta$mag   & 0     & \citet{Adams2012} \\
WASP-14     & Keck / NIRC2          & $K$   & $8.621\pm0.019$   & $\Delta$mag   & 1     & \citet{Ngo2015} \\
WASP-18     & VLT / NACO            & $L'$  & $8.131\pm0.027$   & $\Delta$mag   & 0     & This paper \\
XO-3        & Calar Alto / AstraLux & $z'$  & 9.798             & $\Delta$mag   & 0     & \citet{Bergfors2013} \\

\hline \\ [-2.5ex]
\multicolumn{7}{l}{
  \begin{minipage}{0.95\textwidth}
    \textbf{Notes.} Curve flags: 0 is contrast curve specific to the observations of the target; 1 is average limits of observed sample (or subset); 2 is typical sensitivity of instrument for the observational set up used (see text for more detail).\\
    $^\mathrm{a}$ we used the typical performance curves given in \citet{Turner2006} for the same observational set up.\\
    $^\mathrm{b}$ as no limits are provided in \citet{Siverd2012}, we assumed a similar performance as in \citet{Ngo2015} which used a comparable observing strategy on the same instrument, and used the average detection level from that paper.\\
    $^*$ magnitude of the primary (see text).
  \end{minipage}}

\label{t:limits}
\end{tabular}
\end{small}
\end{table*}

WASP-18 B is our \textit{Gaia} source with the largest uncertainties in its parallax and proper motion measurements, with significant errors of 1.52 mas in $\pi$, 1.98 mas in $\mu_{\alpha*}$ and 2.40 mas $\mu_\delta$. This is due to the fact that this source only has 169 observations in the Along-Scan direction and none in the Across-Scan direction. In comparison, WASP-14 A has 412 observations in both the Along-scan and Across-scan directions, allowing a much higher precision on its astrometric measurements (0.02$-$0.03 mas). Propagating the measurement errors we found fractional differences in parallax and proper motion between WASP-18 A and B of $16.85\pm14.60$ \%, $6.30\pm5.94$ \% and $10.78\pm9.03$ \%, respectively. While these errors are all very large and comparable to the obtained value for $\Delta \pi$, $\Delta \mu_{\alpha*}$ and $\Delta \mu_\delta$, they are still consistent with our selection criteria at the 1-$\sigma$ level. In particular, the 1-$\sigma$ intervals in the fractional difference of both proper motion components remain within 20\%, which strongly indicates that both objects are travelling in the same direction. With the close angular proximity of the two sources (26\arcsec) and parallax measurements consistent with each other, we conclude that WASP-18 A and B form a physically associated pair.

This is further supported by the red colours of the companion, its placement on the \textit{Gaia} HR diagram, and the consistent relative positions of the two components over 16 years between 2MASS and \textit{Gaia} DR2 (Figure \ref{f:CPM_WASP-18}). The possible offset in the \textit{Gaia} DR2 parallaxes could indicate that one of the components is an unresolved binary, as we found in our analysis that systems with an unresolved component tend to show higher inconsistencies in the short-term astrometric measurements available in \textit{Gaia} DR2 (e.g. AS 205 A-BC and HD 178911 AC-B have relative differences in parallax close to 20\%). Alternatively, this discrepancy could also reflect a much wider binary separation, along the line of sight, with one component located at a slightly larger distance. Further observations will be required to reduce the uncertainties in the parallax of WASP-18 B, which will be available in future data releases of the \textit{Gaia} mission. Despite these large uncertainties, we are nevertheless able to confirm that WASP-18 A and B form of a common proper motion pair, in which the fainter component is a low-mass M dwarf close to the stellar/substellar boundary.

\section{Detection limits}
\label{limits}

In order to take into account observational biases and survey sensitivities in our analysis, we gathered and generated detection limits for each target in our sample. We searched for existing contrast curves from the literature for the targets with previous observations (Section \ref{imaging_limits}) and derived \textit{Gaia} detection limits for all targets in our sample (Section \ref{gaia_limits}). Combining those with our own sensitivity limits for our six observed targets (see Section \ref{sensitivities}), we are able to define a detection probability map which is presented and described in Section \ref{detection_prob_map}.

\subsection{Imaging contrast curves}
\label{imaging_limits}

The data used to derive detection limits for all targets with existing imaging data (new or from the literature) are summarised in Table \ref{t:limits}. Sensitivity limits were found to be available for a total of 29 objects, including our 6 observed targets. When multiple sets of observations were found, we chose the best limits available. For a few targets, the deepest contrast curves found only covered a limited range of separations. In those cases, we also consider the shallower detection limits and keep the best value available at any given projected separation.

We used the contrast curves presented in Section \ref{sensitivities} for the 6 targets observed as part of this survey. Most targets with archival observations have contrast curves provided in the literature that are specific to the best set of observations for each target. We flag those with a 0 in the ``curve flag'' column in Table \ref{t:limits}. A number of surveys only provide average detection limits for the observed sample (\citealp{Ginski2012, Bergfors2013, Ngo2015}), which we flag as 1. Finally, we considered the typical sensitivities achieved by specific facilities and instruments when no detection limits were available. These curves are flagged with a 2 in Table \ref{t:limits}, and are detailed below.

Three of our targets have past observations with the AEOS telescope presented in the survey by \citet{Roberts2011}. As the authors do not provide detection limits for their observations, we used the typical performance curves for the AEOS telescope given in \citet{Turner2006} for the same observation set up as that described in \citet{Roberts2011}, and consider that they are representative of those data. Similarly, \citet{Siverd2012} acquired NIRC2 images of KELT-1 with Keck but do not present their achieved sensitivities. We therefore assumed a similar performance to that achieved in \citet{Ngo2015} for comparable NIRC2 observations and used the average detection limits from that work for this target.

\citet{Pinfield2006} searched for companions to 70 Vir out to 30\arcsec using 2MASS. As no detection limits are available in that paper, we generated a 2MASS contrast curve based on the typical resolving and completeness limits of the 2MASS survey. According to the 2MASS documentation \citep{Skrutskie2006}, close doubles with separations $<$ 5\arcsec are not reliably resolved by 2MASS and stellar PSFs can contaminate neighbour sources up to 10\arcsec. The $J$-band completeness limit is given at 16.0 mag. We therefore start our 2MASS contrast curve at 5\arcsec with $\Delta J = 0$ (equal mass binary at the resolving limit). We then use the completeness limit $J = 16.0$ from 10\arcsec out to 300\arcsec, the radial search limit from \citet{Pinfield2006}, with a linear increase in $\Delta J$ between 5\arcsec$-$10\arcsec.

Finally, three targets in our sample were found to be Hipparcos-Tycho binaries, namely, HD 41004 B, HD 87646 A and HD 178911 B. For those three targets, we used the typical sensitivity to binaries in the Hipparcos catalogue based on the plot of separation against $\Delta H_p$ of all Hipparcos binaries found in the ESA documentation \citep{ESA1997}. We extend the separation range out to 30\arcsec, the given widest separation of identified Hipparcos-Tycho binaries. For two of these systems, our sample targets correspond to the fainter, lower-mass component of the binary system, which were detected as companions to the brighter primaries. We thus considered the magnitude of the primary of these systems to derive detection limits around the primary binary component.

\begin{figure}
    \centering
    \includegraphics[width=0.48\textwidth]{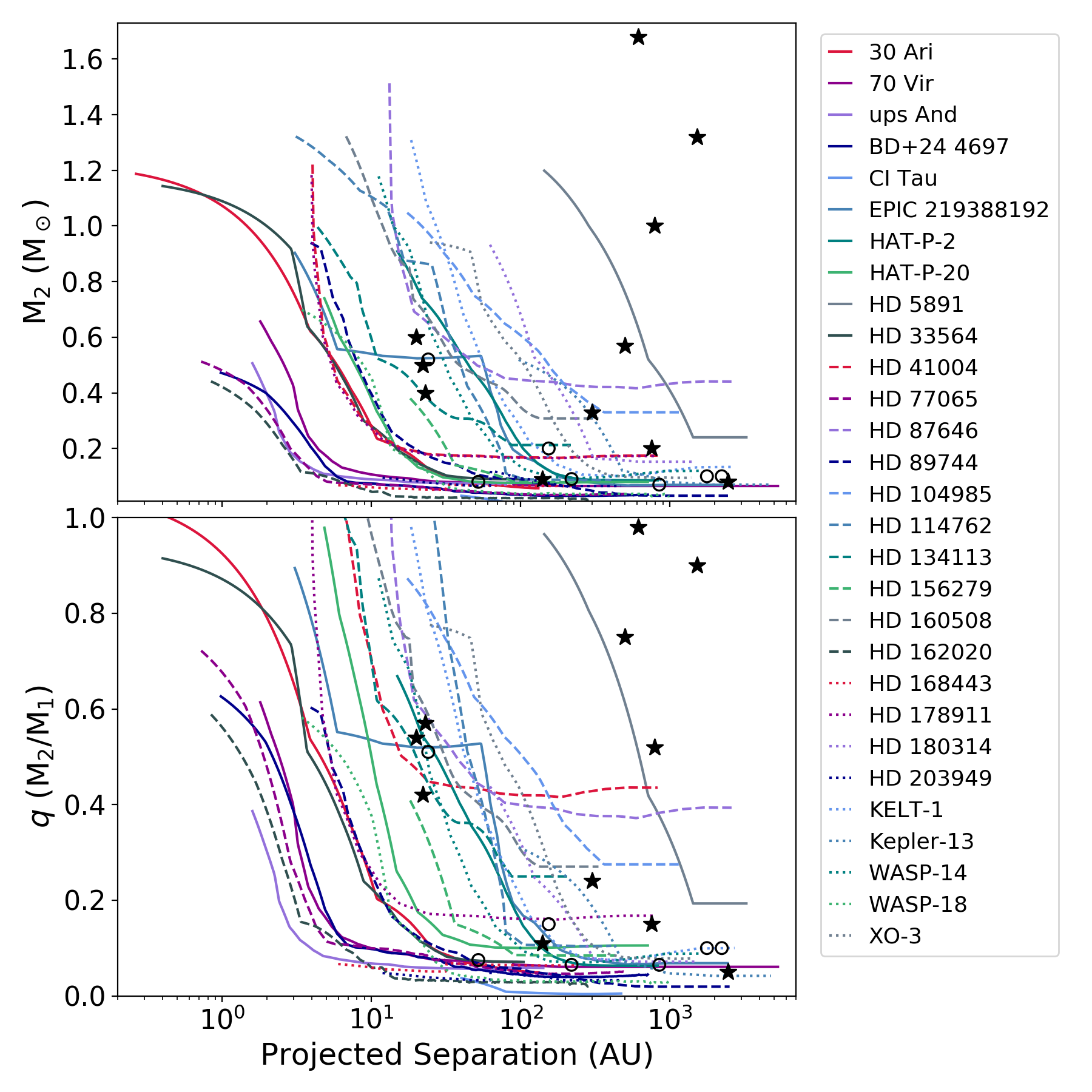}
    \caption{Detection limits for all targets in our sample with published or new direct imaging observations in terms of secondary mass (top) and system mass ratio (bottom). Limits were derived using the data listed in Table \ref{t:limits} and following the approach described in the text. The black stars indicate the positions of confirmed companions to the stars with imaging limits and the open circles correspond to direct imaging candidate companions.}
    \label{f:DI_limits}
\end{figure}

Most of the obtained detection limits are in units of magnitude difference, $\Delta$mag, while a few are provided in magnitudes and one in flux ratio. These are indicated in the ``Limits units'' column of Table \ref{t:limits}. All sensitivity limits are given as a function of angular separation. For all limits that were not in units of magnitudes we started by converting the contrast curves into apparent magnitudes using the photometry of our targets in the considered filters and given in Table \ref{t:limits}. Using the distances from Table \ref{t:stellar_properties}, we then converted all magnitude limits into absolute magnitudes and the angular separations into physical projected separations, in AU. 
Adopting the ages from Table \ref{t:stellar_properties} for our targets, the obtained absolute magnitude curves were then interpolated into the BT-Settl evolutionary models by \citet{Allard2012} to derive corresponding minimum detectable companion masses. The BT-Settl models provide isochrones for numerous photometric systems. We were therefore able to use models corresponding to the specific facilities and filters considered and infer mass limits for each target. Finally, we used the stellar masses listed in Table \ref{t:stellar_properties} for our sample to convert the obtained mass limits into mass ratios $q$ (using the masses of the binary primaries for the two Hipparcos systems mentioned above).
For the few targets with multiple entries in Table \ref{t:limits}, we considered the lowest mass ratio value in the overlapping regions of separations in order to define a unique sensitivity curve for each object. The final mass and mass ratio curves for each target with direct imaging data are shown in Figure \ref{f:DI_limits}, together with the positions of all confirmed (black stars) and candidate (open circles) companions around these objects.

\subsection{\textit{Gaia} detection limits}
\label{gaia_limits}

\begin{figure*}
    \centering
    \begin{minipage}[t]{0.48\textwidth}
    \includegraphics[width=\textwidth]{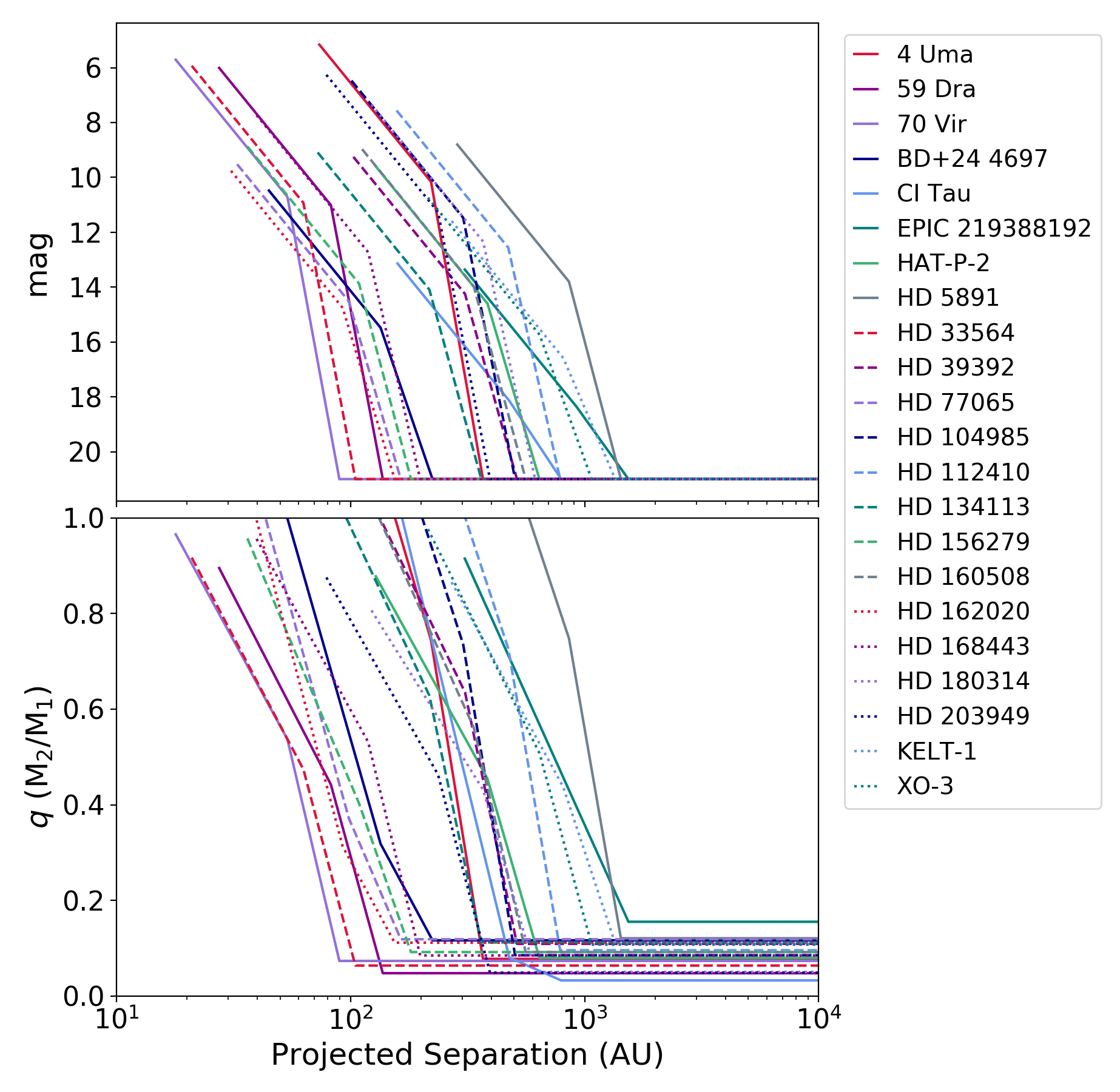}
    \end{minipage}\hfill
    \begin{minipage}[t]{0.48\textwidth}
    \includegraphics[width=\textwidth]{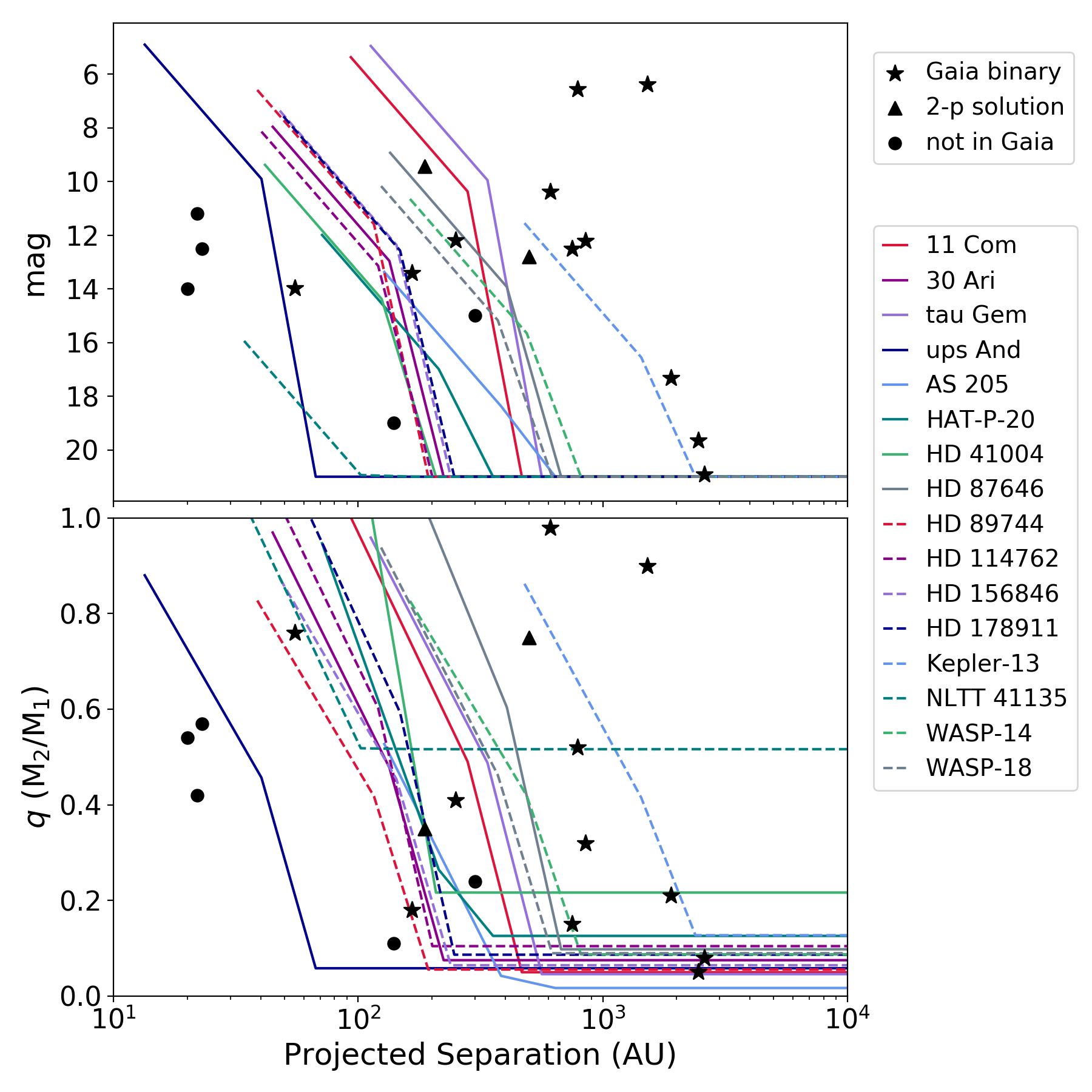}
    \end{minipage}
    \caption{\textit{Gaia} detection limits for all targets without a known companion (left panels) and those that are confirmed binaries or higher order multiples (right panels). The top panels show the \textit{Gaia} $G$-band detection limits as defined in the text. Bottoms panels correspond to the same magnitude sensitivities converted into mass ratios using the BT-Settl models \citep{Allard2012} and the stellar properties from Table \ref{t:stellar_properties}. On the right panels, we also show the positions of the confirmed companions to our targets. Companions detected in \textit{Gaia} are marked with stars. The two known companions present in \textit{Gaia} DR2 but with only a 2-parameter astrometric solution are shown by triangles. Binary companions not retrieved in \textit{Gaia} are indicated with filled circles.}
    \label{f:gaia_limits}
\end{figure*}

Since all objects in our sample are found in the \textit{Gaia} DR2 catalogue, we are able to derive \textit{Gaia} detection limits for each of our targets. \textit{Gaia} DR2 is found to be complete between $G=12$ and $G=17$, with a limiting magnitude of $G\sim21$ and a bright limit of $G\sim3$ \citep{GaiaCollaboration2018}. \citet{Ziegler2018} investigated the recoverability of close binaries in \textit{Gaia} DR2 looking for known binaries from the Robo-AO Kepler survey (\citealp{Law2014, Baranec2016, Ziegler2017}) in the \textit{Gaia} DR2 catalogue. They found that near equal-brightness binaries ($\Delta G < 1$) were consistently retrieved from separations of 1\arcsec and that binaries down to $\Delta G = 6$ were recovered at separations of $\sim$3\arcsec. Based on their results, we define our \textit{Gaia} DR2 sensitivity limits to start at 1\arcsec and $\Delta G = 1$, with a linear decrease to $\Delta G = 6$ from 1\arcsec$-$3\arcsec. We then adopt a linear decrease out to 5\arcsec from $\Delta G = 6$ to $G=21$, the \textit{Gaia} faint limit, and use that limiting magnitude at wider separations, out to projected separations corresponding to $10^4$ AU.

Figure \ref{f:gaia_limits} shows the obtained sensitivity limits in terms of apparent $G$ magnitude and mass ratio for all objects in our sample. We plot on the left panels the limits for the targets without a known companion and on the right panels the limits for all confirmed multiple systems, with the positions the known companions. Magnitude limits were converted into corresponding mass ratio curves adopting the properties of our targets listed in Table \ref{t:stellar_properties} and following the approach described in the previous section with BT-Settl isochrones specific to the \textit{Gaia} filter system.

While \textit{Gaia} is essentially complete in the range \textit{G}$\sim$12$-$17 mag, the catalogue has an ill-defined faint magnitude limit which depends on celestial position \citep{GaiaCollaboration2018}. In addition, the number of sources with a full 5-parameter astrometric solution (position, parallax and proper motion) decreases towards the faint end, where a larger fraction of sources only have positional measurements available, as discussed in the assessment of the \textit{Gaia} DR2 astrometric performance by \citet{Lindegren2018}. We must take into account the catalogue completeness in our detection limits to account for companions that are missed by \textit{Gaia}, but also for those like $\tau$ Gem B and HAT-P-20 B that only have a 2-parameter solution and which we were not able to identify as \textit{Gaia} companions in our analysis in Section \ref{gaia_dr2}. \citet{Arenou2018} report the completeness of the \textit{Gaia} DR2 catalogue as a function of $G$-band magnitude in their catalogue validation work. The provided completeness level for sources with full astrometric solutions decreases from $\sim$99\% at $G<17$ to $\sim$80\% at $G=20$, before sharply dropping to 0\% as $G$ approaches 21 mag (see figure A.1. in the appendix of \citealp{Arenou2018}). We thus use the completeness levels provided in that paper to account for these effects.

\begin{figure}
    \centering
    \includegraphics[width=0.45\textwidth]{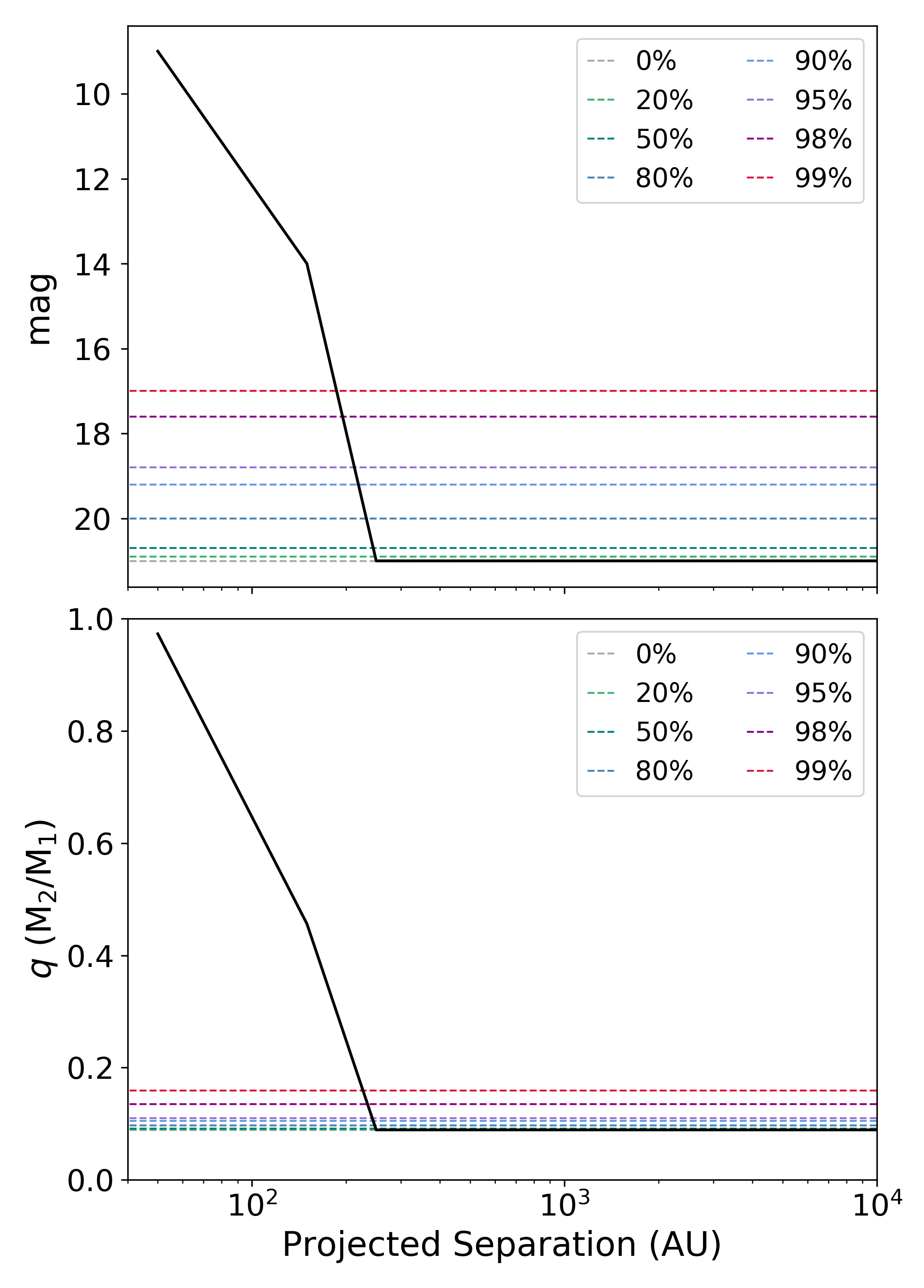}
    \caption{Completeness of \textit{Gaia} DR2 compared to the \textit{Gaia} detection limits for a representative target of 1 M$_\odot$ at 3 Gyr, with a parallax of 20 mas and a $G$-band magnitude of 8 mag. The top panel shows the contrast curve in terms of apparent magnitude of the secondary and the bottom panel displays the corresponding sensitivity in terms of mass ratio, computed in the same way as the \textit{Gaia} detection limits for our targets in Figure \ref{f:gaia_completeness}. The coloured dashed lines represent the completeness levels of \textit{Gaia} DR2 taken from \citet{Arenou2018} for sources with a 5-parameter solution.}
    \label{f:gaia_completeness}
\end{figure}

In Figure \ref{f:gaia_completeness} we show an example of \textit{Gaia} sensitivity curves for a primary of $G=8$ with a parallax of 20 mas, corresponding to a mass of 1 M$_\odot$ at 3 Gyr, representative of the targets in our sample. In the top panel, we show the detection limits for such an object in terms of apparent magnitude, defined as described above. The horizontal dashed lines indicate the $G$ magnitudes associated with various completeness levels using the information from \citet{Arenou2018}, down to the faint magnitude limit of \textit{Gaia} DR2 at $G=21$ (grey line). The bottom panel shows the same contrast curve, converted into mass ratios using the BT-Settl models and adopting an age of 3 Gyr. Since we assumed a primary mass of 1 M$_\odot$, the plot in the bottom panel is also representative of the corresponding mass limits in units of Solar masses. Figure \ref{f:gaia_completeness} clearly demonstrates that when working in mass ratio space the range over which \textit{Gaia} DR2 is not complete for sources with 5-parameter solutions ($<$99\% completeness, below the red dashed lines) is significantly reduced relative to the span of the same incompleteness levels in magnitude space, going from \textit{G}$=$17$-$21 to \textit{q}$=$0.16$-$0.09 M$_\odot$. This implies that in addition to the targets too faint for \textit{Gaia} ($G>21$, below the 0\% completeness grey dashed lines in Figure \ref{f:gaia_completeness}) only the lowest-mass companions have a high chance to of being missed due to survey incompleteness. We note however that the two known companions from our sample that only have a 2-parameter astrometric solution in \textit{Gaia} DR2 ($\tau$ Gem B and HAT-P-20 B) have relatively bright $G$-band magnitudes of 9.42 mag and 12.80 mag, respectively. Their apparent magnitudes fall into a $>$99\% completeness level of sources with full astrometric solutions according to \citet{Arenou2018}. We conclude that it was statistically unlikely to have two bright companions in our sample present in the DR2 catalogue but lacking a 5-parameter solution, and that these two sources are not representative of the completeness of \textit{Gaia} DR2.

For all targets in our sample, we converted the completeness curve obtained from \citet{Arenou2018} into mass ratios as we did for our example target in the bottom panel of Figure \ref{f:gaia_completeness}. We are thus able to assign a completeness factor to each mass ratio value for every detection limit presented in Figure \ref{f:gaia_limits}. Instead of traditional sensitivities, where anything above the mass ratio curves is considered as detectable and anything below is not retrievable, we now associate every point in the separation-mass ratio space to a detection probability, given by the completeness level at any given mass ratio value. The part of the parameter space below the final limits remains at the zero detection probability level regardless of the associated completeness value. We will use these probabilities in the next section to define a 2-dimensional detection probability map for our sample.

\subsection{Detection probability map}
\label{detection_prob_map}

We combined all sensitivity curves obtained in Section \ref{imaging_limits} from imaging data and in Section \ref{gaia_limits} from \textit{Gaia} DR2 to define a single detection probability map for our survey, as was done in \citet{Fontanive2018}. For targets with both Gaia and imaging limits, we started by combining the two sets of contrast curves. Following the approach described in Section \ref{imaging_limits}, we considered the best value available (lowest $q$ value) in the separation ranges where the \textit{Gaia} and imaging limits overlapped, keeping track of the ranges over which the final curves corresponded to the \textit{Gaia} limits. This allowed us to define a unique sensitivity curve for each object.

The mass ratio limits for each target in the sample were then placed on a grid of separations and mass ratios with a resolution of 0.002 in $q$ and steps of 0.01 in log($\rho$). For every cell in the grid, we then identified the number of targets around which a companion of given separation and mass ratio would have been retrieved given the data gathered for this survey. When the considered separation corresponded to the \textit{Gaia} limit of a given target, we then scaled this detection by the \textit{Gaia} completeness level at the associated mass ratio, which we previously computed in Section \ref{gaia_limits}. The number obtained for each cell of the grid was then divided by the total number of objects in our sample, providing a value between 0 and 1 representing the average probability that a companion of projected separation $\rho$ and mass ratio $q$ would have been detected around our 38 targets given the data considered in this work.

\begin{figure}
    \centering
    \includegraphics[width=0.49\textwidth]{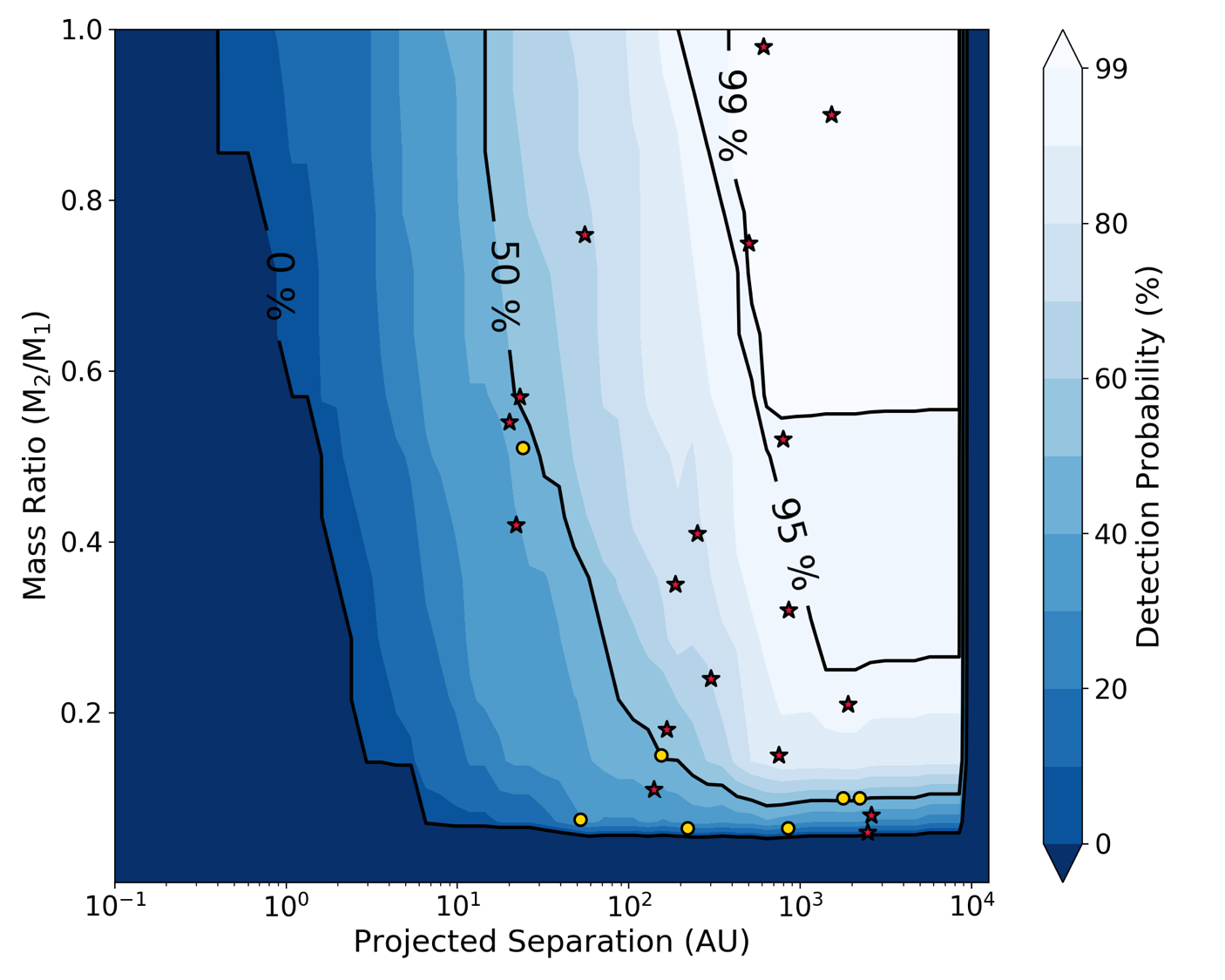}
    \caption{Detection probability map for our sample using the mass ratio sensitivities for our targets from Sections \ref{imaging_limits} and \ref{gaia_limits}, including the completeness of \textit{Gaia} DR2 (see Section \ref{gaia_limits}). Black contours denote the 0\%, 50\% and 95\% and 99\% completeness regions for the full survey. Red stars show the positions of all confirmed companions and yellow circles indicate the positions of candidate companions (see Tables \ref{t:confirmed_binaries} and \ref{t:candidate_binaries}).}
    \label{f:detection_prob_map}
\end{figure}

Figure \ref{f:detection_prob_map} shows the resulting detection probability map for our full sample of 38 objects, considering all available imaging data for the targets in our survey and the \textit{Gaia} DR2 catalogue. Companions inside the 99\% completeness region are essentially detectable around all targets in the sample. We are complete to companions with $q>0.2$ at separations $>$1000 AU around 90\% of our sample, and down to $q\sim0.1$ from separations of $\sim$100 AU around half of our targets (50\% detection probability contour). Confirmed comoving companions were found to be relatively evenly distributed throughout the parameter space (both in separation and mass ratio). In contrast, most candidate companions are concentrated around $q\sim0.1$, which we attribute to the fact that these fainter companions are not detected by \textit{Gaia} and are more likely to lack a second direct imaging epoch, necessary to be astrometrically confirmed.
Interestingly, no companion was found at projected separations $<$20 AU, despite reaching a completeness level up to 50\%, while a number of companions (confirmed and candidates) were retrieved at the same detection probability level at wider separations and low mass ratios.

\section{Statistical analysis}
\label{statistical_analysis}

We used the statistical tool described in \citet{Fontanive2018} to constrain the multiplicity properties of our sample. Examining the binary statistics of these objects will allow us to investigate the possible role of binarity in the formation or evolution of massive, close-in brown dwarfs and planets. The code is based on a Markov Chain Monte Carlo (MCMC) sampling method, using the {\sc emcee} Python package \citep{Foreman-Mackey2013}, and allows us to place robust Bayesian statistical constraints on the binary frequency and companion population distributions for the sample gathered in this study. We add a new capability to the tool in order to account for unconfirmed candidates, which we describe below.

The statistical approach uses the detection limits of the survey, in the form of a detection probability map (see Section \ref{detection_prob_map}, Figure \ref{f:detection_prob_map}), and the properties of detected companions (total number of detections, separations and mass ratios of identified systems) to derive posterior distributions of model parameters (binary fraction and parameters describing the shapes of companion distributions) most compatible with the gathered data.
Based on previous studies of stellar multiplicity in the field \citep{Duquennoy1991, Raghavan2010}, we adopt a lognormal distribution in companion separation $\rho$ (Equation \ref{eq:lognorm}) and a power-law in mass ratio $q$ (Equation \ref{eq:powerlaw}):
\begin{equation} \label{eq:lognorm}
    P(\rho \mid \mu, \sigma) \propto \exp{[-\left(\log_{10}(\rho)\: -\: \mu\right)^2 \:\mathbin{/}\: 2 \sigma^2]}
\end{equation}
\noindent where $\mu$ and $\sigma$ are the mean and standard deviation of the underlying normal distribution in $\log(\rho)$.
The mass ratio distribution ranges from 0 to 1 and is defined by the power-law index $\gamma$:
\begin{equation} \label{eq:powerlaw}
    P(q \mid \gamma) \propto \left\{
                \begin{array}{ll}
                  q^\gamma \; \; \mathrm{for} \: \: \gamma \geqslant 0\\
                  (1-q)^{-\gamma} \; \; \mathrm{for} \: \: \gamma < 0
                  \end{array}
              \right.
\end{equation}
so that negative and positive indices produce symmetric distributions about $q=0.5$ for the same absolute value of $\gamma$.

As was done in \citet{Fontanive2018}, we truncated the model distributions at $\rho=$ 20$-$10,000 AU and $q=$ 0.05$-$1, in order to constrain the binary frequency on those separation and mass ratio ranges. We adopted flat priors for each model parameter, set to unity over the following ranges and to zero elsewhere: 0.5$-$4 for $\mu$, 0.1$-$3 for $\sigma$, -3$-$3 for $\gamma$ and 0$-$1 for the binary fraction $f$.

In order to also take into account the candidate companions identified around the targets in our sample, we used the probabilities of the candidates being physically bound derived in Section \ref{lit_search} and listed in Table \ref{t:candidate_binaries}. At each step in the MCMC chain, we drew for each target with a candidate companion a number between 0 and 1, and counted the candidate as a bonafide companion if the drawn value was below the companionship probability. This ensures that each candidate companion is selected in a fraction of MCMC steps that is representative of its probability of being physically associated to the primary target.
For hierarchical systems (e.g. 30 Ari, WASP-14), we considered the properties of all detected components in the part of the code that constrains the shape of the separation and mass ratio distributions, accounting for candidate companions only when they were selected (e.g. 70 Vir, EPIC 219388192, HD 89744). For systems in which the binary companion is itself a tight binary (AS 205, HD 178911, Kepler-13), we used the combined mass of the binary component, since this total mass would be responsible for any dynamical effect on the close-in planet or brown dwarf. For the section of the tool constraining the binary fraction, we considered the number of multiple systems rather than the total number of companions in order to constrain the multiplicity rate of our sample.

\section{Results}
\label{results}

\subsection{MCMC analysis for the full sample}
\label{results_all}

\begin{figure}
    \centering
    \includegraphics[width=0.49\textwidth]{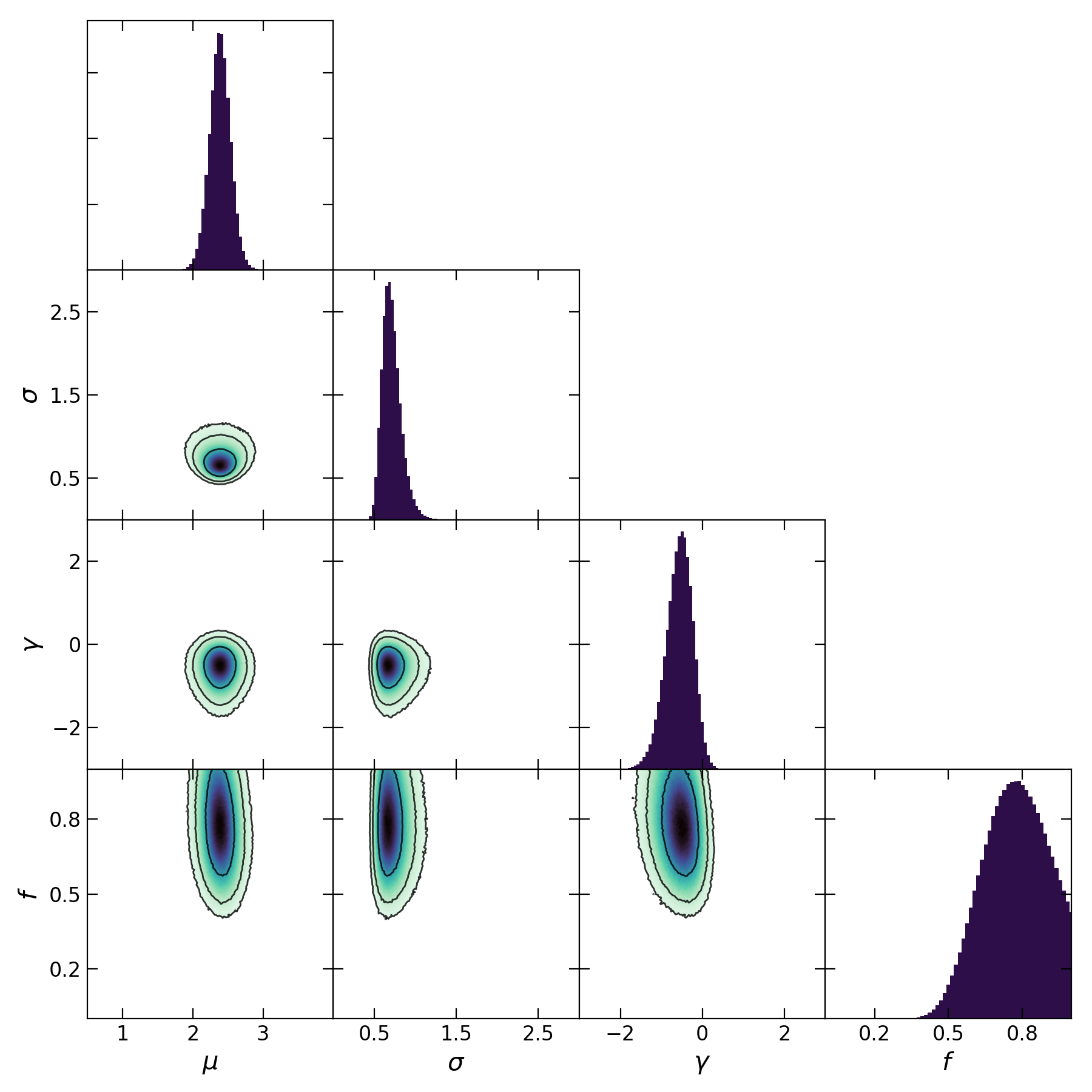}
    \caption{Posterior probability distributions of all model parameters (diagonal) from our MCMC analysis performed on the full sample of 38 objects and correlation among all pairs of parameters (triangle plot). Normalised histograms at the ends of rows are marginalised over all other parameters. Black contour lines in the correlation plots correspond to regions containing 68\%, 95\% and 99\% of the posterior.
}
    \label{f:corner_plot_all}
\end{figure}

\begin{figure*}
\begin{minipage}[c]{0.49\linewidth}
    \centering
    \includegraphics[width=\textwidth]{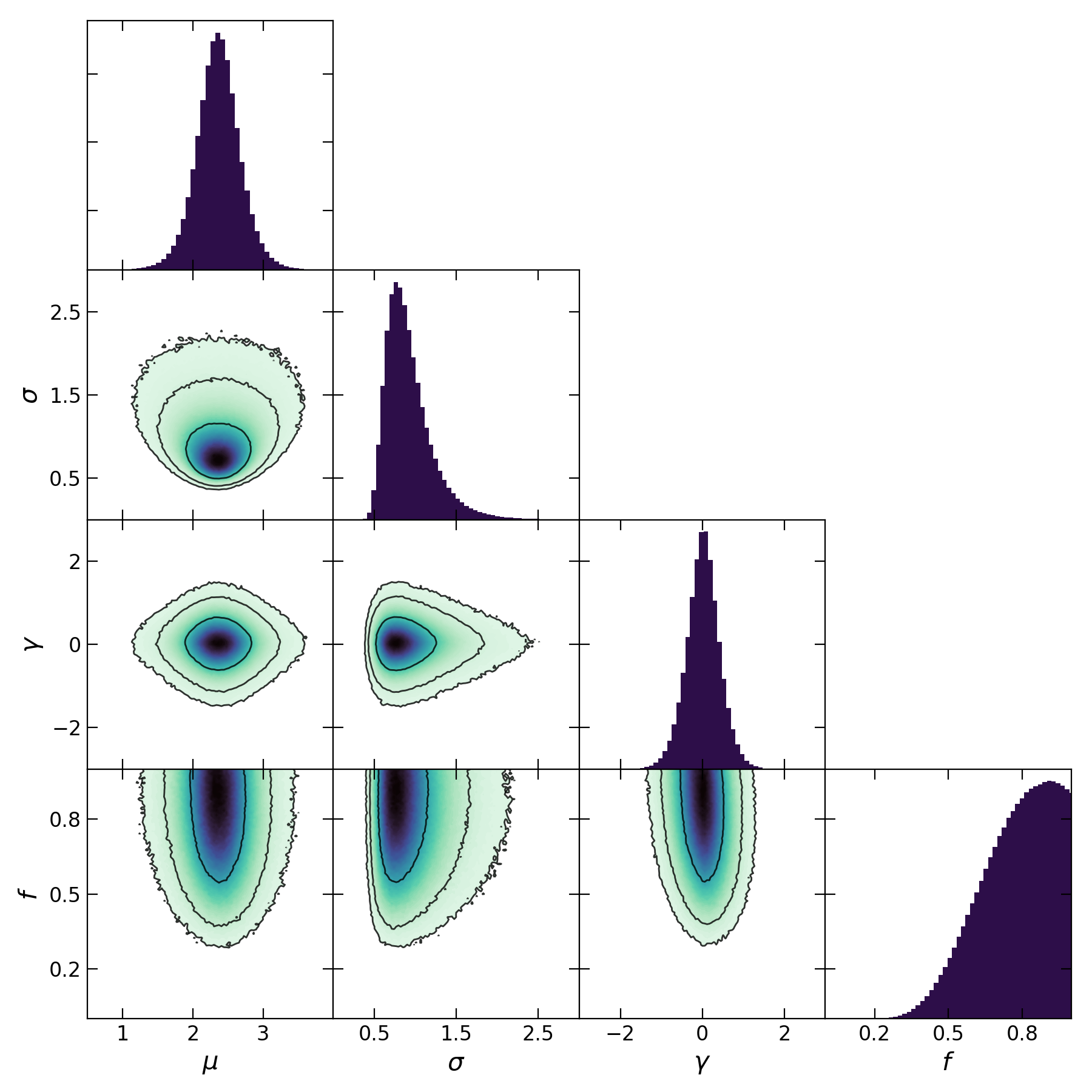}
    \caption{Same as Figure \ref{f:corner_plot_all} for the subset of objects with an inner planet or brown dwarf on an orbit shorter than 10 days.}
    \label{f:corner_plot_Kozai}
\end{minipage}
\hfill
\begin{minipage}[c]{0.49\linewidth}
    \centering
    \includegraphics[width=\textwidth]{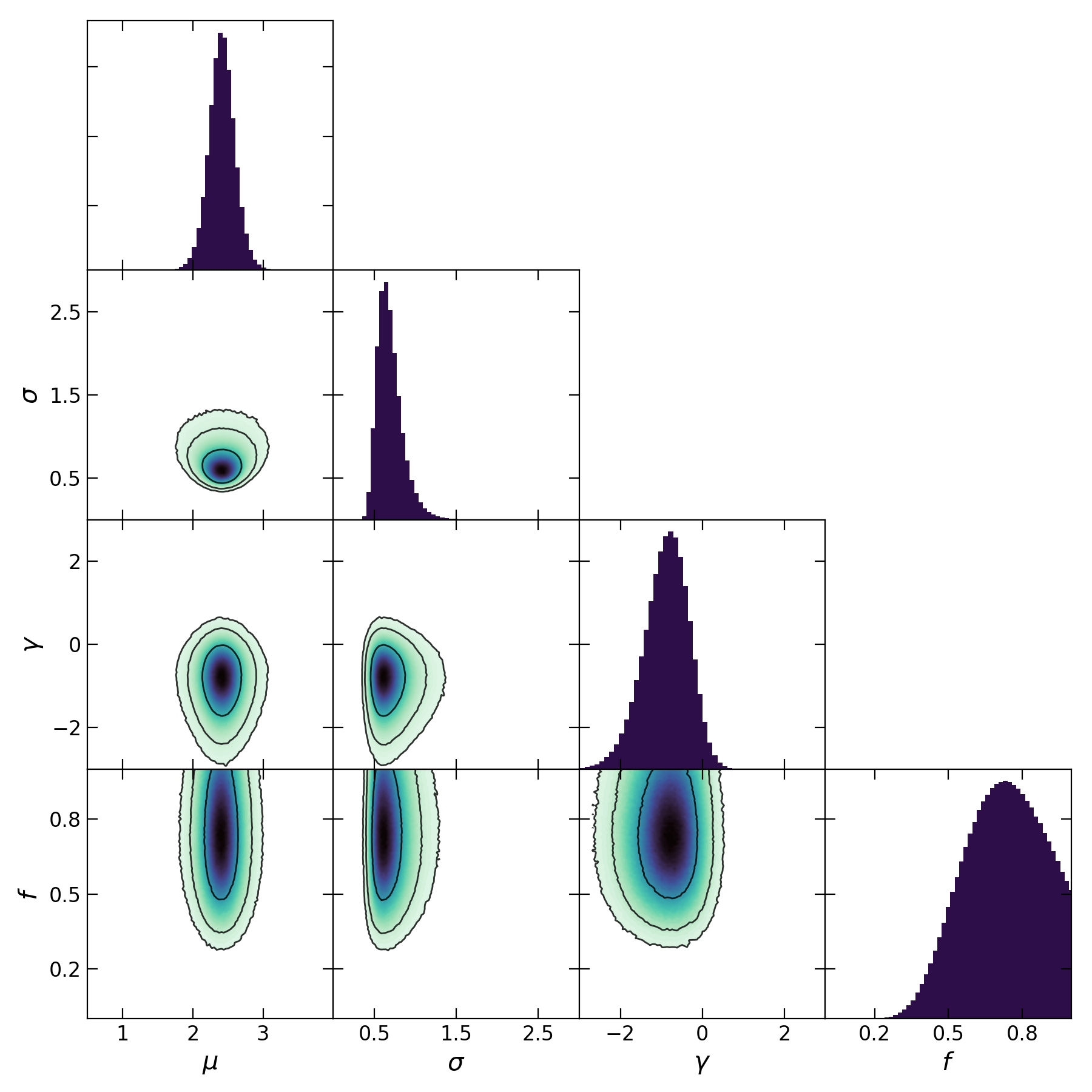}
    \caption{Same as Figure \ref{f:corner_plot_all} for the subsample of targets with an inner companion with a period larger than 10 days.}
    \label{f:corner_plot_notKozai}
\end{minipage}%
\end{figure*}

We ran the MCMC sampler with 2000 walkers taking 5000 steps each on our full sample of 38 objects. We found that walkers were expanding from their initial positions to a reasonable sampling of the parameter space in less than 100 steps, and removed the first 100 steps of this ``burn-in'' phase.
We found a mean fraction of steps accepted for each walker of 0.44, in good agreement with the rule of thumb acceptance fraction suggested by \citet{Foreman-Mackey2013} between 0.2 and 0.5, and trust the obtained value to be a reliable indication of convergence. 

The full output from our MCMC analysis is presented in Figure \ref{f:corner_plot_all}. The best-fit values for the binary parameters of our core sample on separations in the range 20$-$10,000 AU are summarised in Table \ref{t:results}. Errors correspond to 68\% confidence intervals, estimated using a highest posterior density approach to determine the boundaries of Bayesian credible intervals (see \citealp{Fontanive2018}). The highest density region method provides a set of the most probable values of a given parameter. All 4 model parameters were found to be well-constrained, converging to sharply-defined peaks in the posterior distributions, with the binary fraction $f$ showing the broadest posterior distribution.

The obtained posterior distribution for the binary frequency of our sample, $f = 79.0^{+13.2}_{-14.7}$\%, is found to be much higher than the overall multiplicity rate of FGK stars in the field, generally observed to be around 40$-$50\% \citep{Duquennoy1991, Raghavan2010}. The peak of the lognormal in separation corresponds to a value of $\sim$250 AU, also much wider than for the field population \citep{Raghavan2010}. We discuss these features further in Section \ref{discussion} where we provide a detailed assessment of the possible sources and implications of these results.

\begin{table*}
    \centering
    \caption{Summary of multiplicity properties from our study, with a comparison to field stars and hosts to lower-mass planets.}
    \begin{tabular}{l c c c c c c c c l }
    \hline \hline
        Sample & $N$ & $M_2$ & $\mu$ & $\sigma$ & $\gamma$ & $f$ & Binary sep. & Binary $q$ & Reference \\
     &  & (M$_\mathrm{Jup}$) & &  & & (\%) & (AU) &  & \\
        \hline
        Full sample & 38 & 7$-$70 & $2.39^{+0.14}_{-0.15}$ & $0.68^{+0.12}_{-0.10}$ & $-0.52^{+0.31}_{-0.32}$ & $79.0^{+13.2}_{-14.7}$ & 20$-$10,000 & 0.05$-$1 & This work \\ 
        & & & & & & $70\pm10$ & 50$-$2000 & 0.05$-$1 & This work\\ [0.2cm]
        $<$ 10 day subset & 12 & 7$-$70 & $2.36^{+0.28}_{-0.31}$ & $0.76^{+0.29}_{-0.16}$ & $0.03^{+0.38}_{-0.40}$ & $92.0^{+8.0}_{-19.0}$ & 20$-$10,000 & 0.05$-$1 & This work \\ [0.2cm]
        $>$ 10 day subset & 26 & 7$-$70 & $2.40^{+0.18}_{-0.17}$ & $0.63^{+0.15}_{-0.12}$ & $-0.89^{+0.55}_{-0.64}$ & $74.0^{+14.4}_{-15.9}$ & 20$-$10,000 & 0.05$-$1 & This work \\
        \hline
        FGK field stars & 454 & $-$ & 1.70 & 1.68 & $\sim$0 & $44\pm2$ & overall & overall & \citet{Raghavan2010} \\
         & & & &  & & $36\pm2$ & 20$-$10,000 & overall & Scaled in this work \\
         & & & &  & & $16\pm1$ & 50$-$2000 & overall & Scaled in \citet{Ngo2016} \\
        \hline
        Friends of HJs & 77 & $<$ 4* & ... & ... & ... & $47\pm7$ & 50$-$2000 & 0.05$-$1 & \citet{Ngo2016} \\
        \hline \\ [-2.5ex]
    \multicolumn{10}{l}{
    \begin{minipage}{0.9\textwidth}
    *5 objects from the Friends of Hot Jupiters survey have masses between 7$-$12 M$_\mathrm{Jup}$, all of which are part of our studied sample (see text).
    \end{minipage}}
    \end{tabular}
    \label{t:results}
\end{table*}

\subsection{Sample division at 10 days in inner companion period}
\label{results_subsets}

We divided our sample into two subsets, with a cut at 10 days in the period of the inner planets and brown dwarfs, the commonly-accepted threshold for hot Jupiters \citep{Wang2015, Dawson2018}. This allows us to investigate possible differences in the binary properties of the stars with companions on orbits comparable to hot Jupiters, and those with planets or brown dwarfs at slightly wider separations. The hot Jupiter-like subset includes 12 targets, 6 of which are confirmed binaries, with 2 additional targets having at least one high-probability candidate companion. The sample of wider inner companions contains 26 objects, including 10 confirmed multiples and 1 candidate binary. 

Following the approach described above, we created detection probability maps for each subset, considering the available detection limits for all targets from each subsample. We then performed the same statistical analysis as that presented above to constrain the multiplicity rates and binary properties of our samples of objects with periods shorter and longer than 10 days, so as to assess whether statistically significant discrepancies are observed between the two populations.
Figures \ref{f:corner_plot_Kozai} and \ref{f:corner_plot_notKozai} show the output of the MCMC sampler for the shorter and longer-period samples, respectively. As expected from the smaller sample sizes of the two subsets relative to the full sample, the walkers are slightly more widely spread throughout the parameter space than in Figure \ref{f:corner_plot_all}, and this effect is amplified for the smaller sample of $<$10 days companions (Figure \ref{f:corner_plot_Kozai}). Nevertheless, all 4 model parameters are still well-constrained within the explored parameter space in both subsamples. The best-fit values and corresponding 1-$\sigma$ intervals are given in Table \ref{t:results} for each subset.

The model parameters describing the companion separation distribution ($\mu$ and $\sigma$) peak at very similar values for the two subsets, indicating that no significant difference is found in the binary separation of these two populations. The power-law index $\gamma$ describing the mass ratio distribution appears to shift to slightly lower values for the sample of longer-period inner companions, which reflects the generally lower mass ratios of multiple systems found in that subset. 

\begin{figure}
    \centering
    \includegraphics[width=0.42\textwidth]{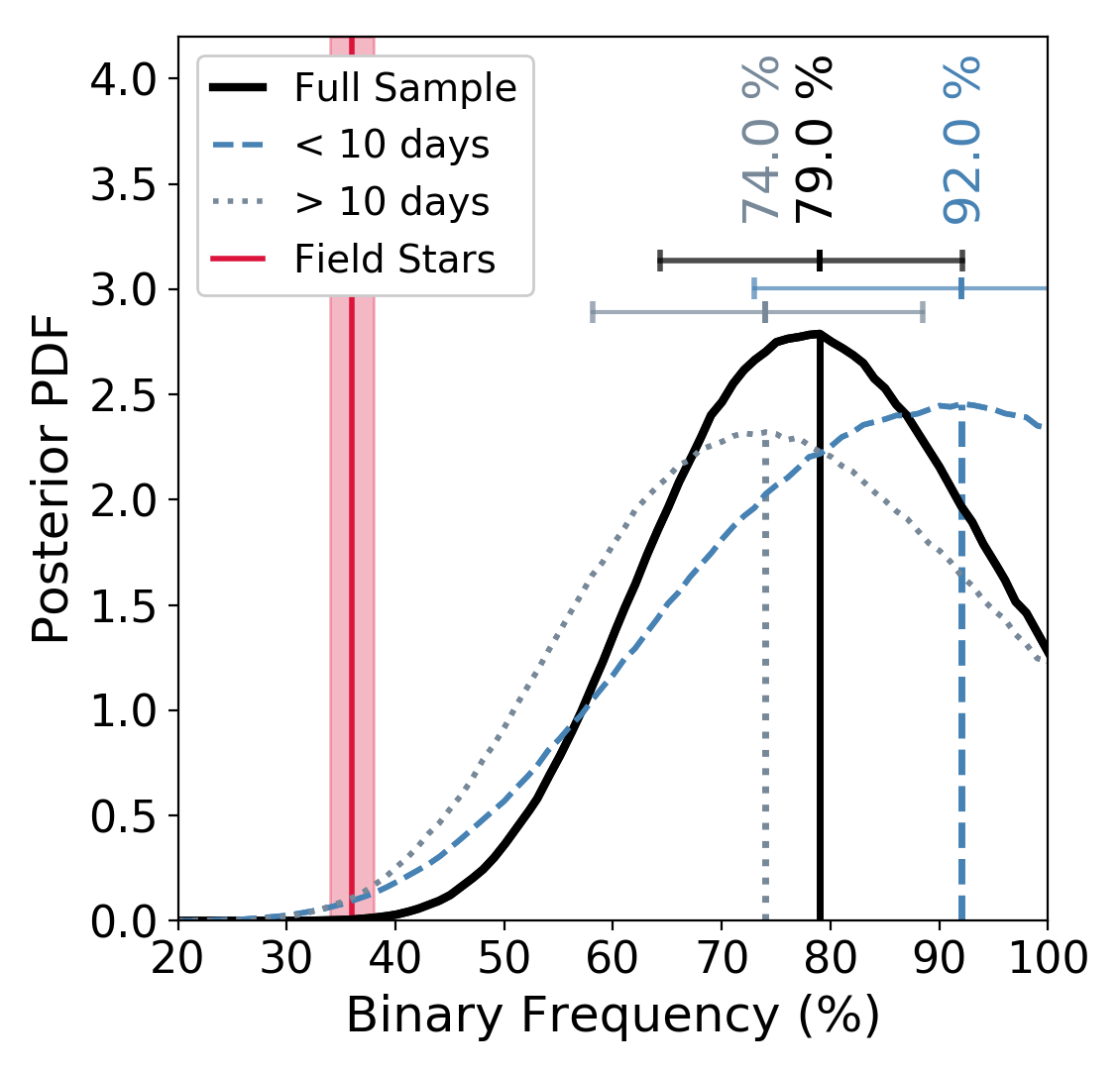}
    \caption{Posterior probability distributions obtained from our MCMC analysis for the binary frequency of our full sample of 38 objects (solid black line), the subset of 12 objects with inner companions on orbits shorter than 10 days (dashed blue line) and the subsample of 26 systems with a wider inner companion (dotted grey line). Binary frequencies are constrained over the separation range 20$-$10,000 AU. The vertical lines show the positions of the most likely value for each distribution and the corresponding values are indicated above. The ranges of the horizontal lines correspond to the 68\% intervals of highest probability. The red line and shaded region show the multiplicity fraction of field stars from \citet{Raghavan2010} which we scaled to the same separation range.}
    \label{f:freq_pdf}
\end{figure}

The binary fraction $f$, on the other hand, shows a larger discrepancy in the output posterior distributions. The obtained probability density function for the sample with inner companions on very short orbits (Figure \ref{f:corner_plot_Kozai}) was found to peak at $92.0^{+8.0}_{-19.0}$\% (68\% confidence), consistent with a binary rate of 100\% at the 1-$\sigma$ level. In contrast, the subset of wider inner companions (Figure \ref{f:corner_plot_notKozai}) has a binary frequency of $74.0^{+14.4}_{-15.9}$\%. We plot these two distributions in Figure \ref{f:freq_pdf}, together with the obtained posterior distribution for $f$ for the full sample. The red line shows the corresponding multiplicity rate of field stars based on results from \citet{Raghavan2010}, scaled to our probed separation range of 20$-$10,000 AU (see Section \ref{discussion} for details). While all 3 distributions are consistent with one another, a clear shift is observed in the peak of the posteriors. In particular, the peak of the binary fraction distribution for the $<$10 day sample is located outside or at the edge of the 68\% confidence interval of the other two posterior distributions. The resulting binary fractions are all much higher than the corresponding value expected for field stars.

\section{Analysis and Discussion}
\label{discussion}

\subsection{Comparison with field stars}

\subsubsection{Multiplicity fraction}

\citet{Raghavan2010} provided a comprehensive assessment of the multiplicity properties of Solar-type stars, searching for companions to 454 F6$-$K3 primaries in the field. Taking into account the completeness limits of their survey, the authors found that about $56\pm2$\% of stars are single, for an overall multiplicity fraction of $44\pm2$\%, in good agreement with previous results from \citet{Duquennoy1991}. Our binary fractions derived in Section \ref{results} were limited to separations in the range 20$-$10,000 AU. We must therefore restrict the overall binary rate from \citet{Raghavan2010} to this separation range in order to compare our findings to the general field population. Taking into account the shape of the distributions obtained by \citet{Raghavan2010} and excluding all companions from that study with separations outside our considered range, the fraction of stars found in binaries or higher-order multiples becomes $36\pm2$\%. This is more than twice as low as the binary rate obtained for our full sample of 38 objects, with a 3-$\sigma$ significance ($f = 79.0^{+13.2}_{-14.7}$\%; see Table \ref{t:results}). We also find the value for field stars to be lower than the binary rates derived for our two separate subsets, although these results have a lower significance ($\sim$2.5-$\sigma$ level) as a result of the smaller number statistics of the individual subsamples. In Figure \ref{f:freq_pdf} we compare the expected fraction of multiples on this separation range to the posterior distributions obtained for the full sample studied in this work and to the two subsets defined in Section \ref{results_subsets}. The plot clearly shows that our derived binary rates are statistically larger than the binary fraction from the overall FGK stellar population.

\citet{ZuckerMahez2002} noted a substantial paucity of high-mass planets with short-period orbits around single stars. In contrast, this feature is not observed around binary star systems, which exhibit a prevalence of short-orbit massive planets \citep{ZuckerMahez2002,Eggenberger2004,Desidera2007,Mugrauer2007}.
Our results are highly consistent with these observations, suggesting that almost all stars with a $>$7 M$_\mathrm{Jup}$ companion within $\sim$1 AU are part of multiple stellar systems.
The statistically higher binary occurrence of hosts to massive planets relative to the general field population indicates that stellar companions may play an important role in the existence of the most massive giant planets and brown dwarfs observed on tight orbits, and that a binary companion may be required to explain their presence. The much higher binary fractions we find for our sample compared to field stars, despite the known biases from transit and radial velocity surveys against close binaries, reinforces the idea of a significant correlation between stellar binarity and the existence of the massive inner companions studied here. While the nature and magnitude of this role are not clear and cannot be established based on this study alone, a number of possibilities have been formulated and explored in the literature to explain the possible influence of binary companions on giant planet formation and evolution. We discuss these theories in Section \ref{formation_evolution}. 

A caveat of this analysis is that \citet{Raghavan2010} studied stars in a volume limited to 25 pc in distance. In order to compare our results to the overall field population, we extrapolated the measurements from \citet{Raghavan2010} out to distances of 500 pc. While the distributions found by \citet{Raghavan2010} are valid for 0.5$-$1.5 M$_\odot$ stars, our sample contains two targets from young star-forming regions (one confirmed binary and one not known to have any companion), as well as a number of giants and subgiants. These populations may have different binary statistics than the main sequence solar-type stars probed by \citet{Raghavan2010} and the assumptions made in our analysis may not be entirely valid. The field study by \citet{Raghavan2010} was also heavily biased towards G stars, while our sample contains a number of more massive A and F stars, which are expected to have a higher binary fraction, as well as some M-dwarfs, expected to have a lower binary frequency. The mass dependence of stellar binarity could therefore be another factor affecting our results. The field stars sample may also be contaminated with planet hosts, and it must be pointed out that the results presented above are not a comparison between planet-free stars and planetary hosts, but rather an assessment of planet hosts multiplicity properties relative to the general stellar population. That being said, the extremely high binary fraction derived for our studied sample is still a robust and significant result by itself, even if the comparison to field stars may not be fully reliable.

\subsubsection{Mass ratio distribution}

\citet{Raghavan2010} found a roughly flat mass ratio distribution for binaries separated by more than 100 days. The value obtained for the power-law index in our full sample indicates a slight preference for lower-mass companions, but is fairly consistent with a flat distribution (i.e. $\gamma = 0$, see Table \ref{t:results}). Our subset of $>$10 day inner companions indicates a moderately larger preference towards low mass ratio companions for these systems, and the subsample of short-period planets was found to exhibit a uniform distribution in wide companion mass ratio. Our probed samples are thus in reasonable agreement with the mass ratios observed around multiple stars in the field, and we find no evidence for distinct populations between our studied targets and the general field population.

\subsubsection{Separation distribution}

\begin{figure}
    \centering
    \includegraphics[width=0.42\textwidth]{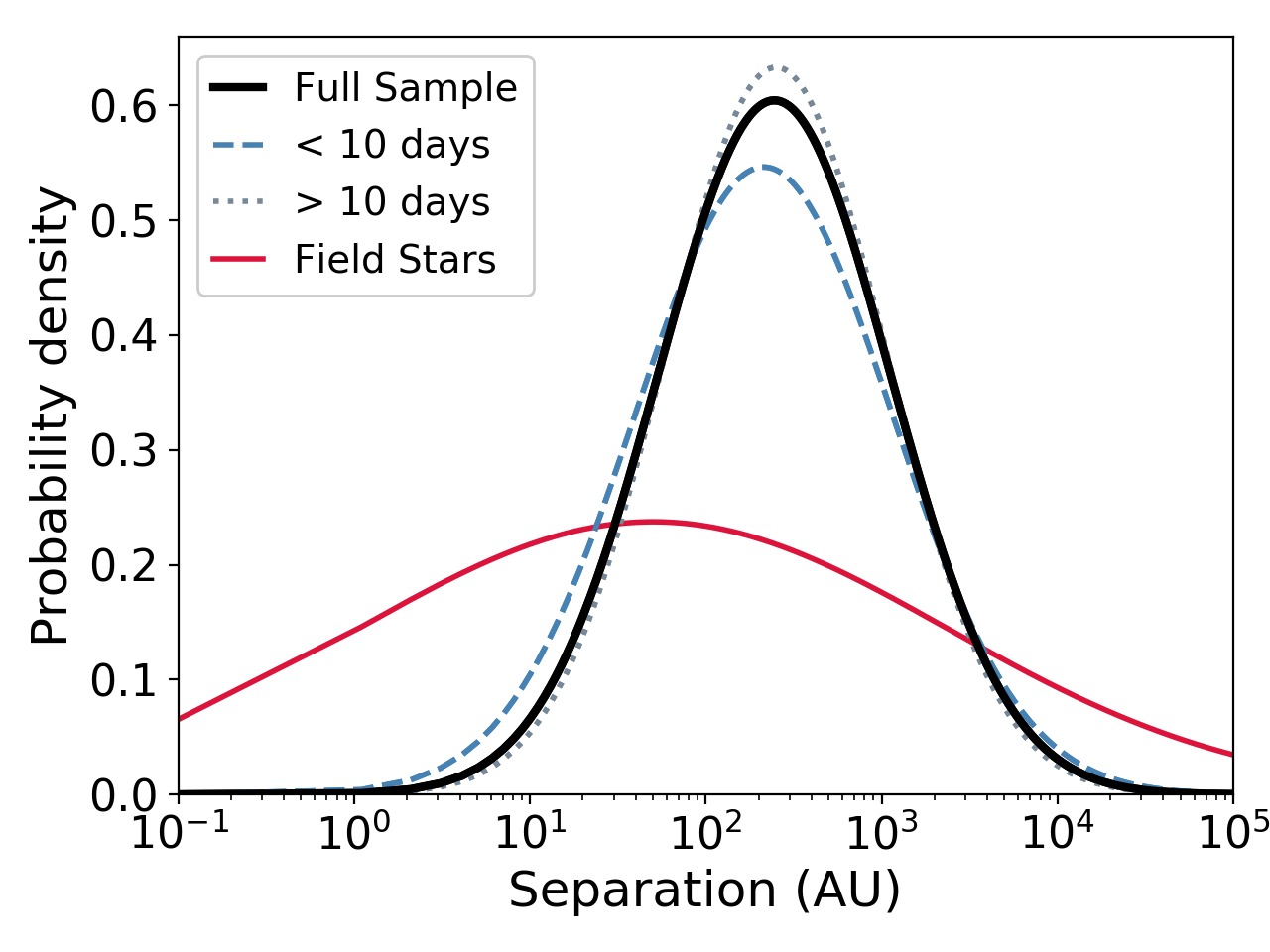}
    \caption{Separation distributions of wide binary or hierarchical companions, comparing the output from our MCMC analysis on the full sample of 38 objects (solid black line) and two subsets (dashed blue line and grey dotted line), to the field population from \citet{Raghavan2010} in red. Our obtained density functions show narrower distributions, peaking at larger separations than for field stars.}
    \label{f:sep_pdf}
\end{figure}

In contrast, we found larger and more significant disparities in binary companion separation between the distributions obtained in Section \ref{results} and the expected distribution from FGK field stars, as shown in Table \ref{t:results}. \citet{Raghavan2010} reported a lognormal distribution in companion separation peaking at 1.70 in $\log_{10}(a)$, with a Gaussian width of 1.68, corresponding to a broad peak around 50 AU. This is significantly smaller than our derived value of $\mu = 2.39^{+0.14}_{-0.15}$ for our full sample, with a mean located at $\sim$250 AU. We also found a much narrower separation distribution, with a Gaussian width of $\sigma = 0.68^{+0.12}_{-0.10}$. The results obtained for our two subsets are in good agreement with each other and with the full sample (Table \ref{t:results}). Figure \ref{f:sep_pdf} shows the constraints obtained from our statistical analysis on the separation distribution of the multiple systems in our core sample and defined subsets. The red distribution represents the results obtained by \citet{Raghavan2010} for solar-type stars in the field, clearly demonstrating the preference for wider binaries among our targets and the strong deficit of closely-separated systems in our studied sample.

While we restricted ourselves to a 20$-$10,000 AU separation range to constrain the binary frequency $f$, the parameters describing the separation distribution ($\mu$ and $\sigma$) were explored over a broad parameter space. The MCMC walkers would have been able to converge to a distribution peaking near or even inside 20 AU had it been compatible with the observed data. As we noted in Section \ref{detection_prob_map}, the lack of tight binaries is unlikely to be the sole result of observational biases and limiting sensitivities. Indeed, we are sensitive to binary companions with separations of 5$-$20 AU around 20$-$60\% of our targets depending on the system mass ratio, and a number of companions (confirmed and candidates) are retrieved at the same detection probability level at larger separations and low-mass ratios (see Figure \ref{f:detection_prob_map}). If the true underlying separation distribution in our sample was comparable to that of field stars, with a broad peak near 50 AU, we would have expected to detect numerous close binaries given the number of wide multiple systems already present in the sample. We thus consider that this observed feature is real and not due to observational limitations.

This deficiency of tight binaries could however be due to selection effects in exoplanet surveys, which are often biased against close binary systems. In addition, it may be harder to detect planets via the radial velocity method in the presence of a close, massive stellar companion. The spurious assumption that a planet host is single when it is in fact an unresolved binary will also lead to erroneous measurements of the planet's physical properties. A population of massive planets and brown dwarfs in closely-separated binaries could hence exist and be underrepresented or misreported among detected exoplanets. If this is the case, the true multiplicity rate of systems hosting massive, close-in planets or brown dwarfs should then be even higher than what we found here.

Nevertheless, our observations are consistent with previous studies. The shortfall of close binaries among planet hosts has indeed been vastly reported in the literature \citep{Roell2012,Bergfors2013,Wang2014a,Wang2014b,Kraus2016} and is generally attributed to a hindrance of planet formation in very tight binaries. While observational constraints remain sparse, the current census is that binarity on scales comparable to the Solar system ($\lesssim$50$-$100 AU) has the potential to affect planet formation and evolution. However, different conclusions have been reached on the theoretical side, and it is not clear whether these processes are altered, inhibited or facilitated by the presence of a binary companion and on what separations these effects may take place. This is further discussed in Section \ref{formation_evolution}.

Another possible explanation for the depletion with tight binary systems among our sample is that the inner planetary and brown dwarf companions to our targets may have formed at much wider separations than their current locations, at radii overlapping with the missing population of binary companions. Such massive planets are indeed expected to form outside at least a few AU for formation by CA \citep{Mordasini2012}, and more likely several tens of AU for GI, in regions of the circumstellar discs that are massive and cool enough to form massive giant planets \citep{Rafikov2005}. If inner companions have formed at such separations, additional, massive binary companions should not exist within a few tens to hundreds of AU around these systems, which would be reconcilable with our observations. However, as this trend is also observed around systems hosting low-mass planets, for which a formation at very wide separations is not required, this may not be the primary phenomenon responsible for this feature.

\subsection{Binarity as a function of planet properties}

\subsubsection{Binary frequency versus inner companion period}

In Section \ref{results_subsets}, we divided our sample into two subsets in order to investigate possible differences in the demographics of stars hosting planetary or brown dwarf companions within and beyond orbital periods of 10 days. While we found no evidence for distinct binary mass ratio or separation distributions between these two populations, our statistical analysis revealed a possibly larger binary frequency for the subset of shorter-period companions, with a peak at 92\%, compared to 74\% for the subsample of more widely-separated systems. These results are marginal due to the less stringent constraints we were able to place on the individual subsets, as shown by the broader posterior distributions in Figure \ref{f:freq_pdf} relative the one obtained for the full sample. Larger sample sizes will be required to confirm this tendency.

This theory is nonetheless supported by the similar trend seen for hosts to lower-mass planets. Surveys searching for wide companions to radial velocity planets from sub-Jupiter masses up to a few M$_\mathrm{Jup}$ and out to 5 AU found that less than $\sim$25\% of these systems were part of binaries or multiple systems (e.g. \citealp{Raghavan2006,Ginski2012}), although these surveys may be biased or incomplete.
In contrast, studies of slightly shorter-period transiting planets (typically $<$100 days) observed rather higher binary fractions, generally around $\sim$50\%, for planets of comparable masses \citep{Adams2012,Adams2013,Ngo2016}. Furthermore, \citet{Tokovinin2006} found that 96\% of spectroscopic binaries with periods $<$3 days have a third component, compared to only 34\% of systems with periods longer than 12 days, albeit some selection biases may affect these results to a limited extent.

The marginal difference in binary occurrence observed in this work between very short-period transiting planets and brown dwarfs, and the somewhat wider population of radial velocity companions, thus indicates that this trend, if real, may also hold for the very massive inner companions studied here. This trend could suggest that binarity greatly helps the formation or migration of massive giant planets and brown dwarfs observed within $\sim$1 AU, and essentially becomes necessary for these companions to reach orbital periods shorter than 10 days.

\subsubsection{Binary frequency versus inner companion mass}

In the final paper of the Friends of hot Jupiters campaign \citep{Knutson2014,Piskorz2015,Ngo2015}, \citet{Ngo2016} found that 47$\pm$7\% of hot Jupiter systems have a stellar companion between 50 and 2000 AU, a binary rate 3 times higher that for field stars in this separation range. The authors concluded that binary companions on these separations facilitate planet formation or help the inward migration of giant planets. Our study probed higher-mass planets than those considered in that survey, allowing us to examine trends in stellar multiplicity as a function of planet mass, including inside the brown dwarf regime. The Friends of hot Jupiters survey looked at systems with planet masses mostly limited to 4 M$_\mathrm{Jup}$. Only five objects with more massive companions were studied in that work, with masses between 7 and 12 M$_\mathrm{Jup}$, all of which are part of our selected targets (3 are confirmed multiples: HAT-P-20, WASP-14 and WASP-18; 2 are apparently single: HAT-P-2 and XO-3). The binary fraction derived here for more massive objects was estimated for separations between 20 and 10,000 AU. The corresponding binary rate restricted to the 50$-$2000 AU separation range becomes 70$\pm$10\% for our core sample, 1.5 times larger than the value from \citet{Ngo2016} at the 2.3-$\sigma$ level. This fraction is 4 times larger for field stars on these separations (see \citealp{Ngo2016}), with a 5-$\sigma$ significance. These results are summarised in Table \ref{t:results}.

Our results for the shorter-period subset are less significant due to the looser constraints we were able to place on the smaller-sized subsample. We thus only compare previous studies with the binary fraction estimated for the full sample, and keep in mind that hosts to shorter-period planets may have an enhanced binary rate, as discussed previously.
Our findings suggest that the trends characterised by \citet{Ngo2016} for hosts to hot Jupiters are also observed and even strengthened for the highest-mass close-in planets and brown dwarfs. These results are in excellent agreement with early observations by \citet{ZuckerMahez2002} and \citet{Eggenberger2004}, who determined that the most massive planets on orbits of a few days are consistently found in binary systems, suggesting that this planetary population does not exist around single stars.

We note that four targets in our sample have notably high mass estimates (40$-$60 M$_\mathrm{Jup}$) relative to the rest of our sample, namely BD+24 4697, HD 77065, HD 134114 and HD 160508. These objects appear somewhat isolated in the period-mass parameter space in Figure \ref{f:planet_binarity}. Given that their mass measurements are lower limits derived from radial velocity information, their true masses are most certainly even higher. Assuming a uniform distribution of inclinations, we may calculate the minimum value for the projected mass that corresponds to a true substellar mass $M_2$ $<$ 70 M$_\mathrm{Jup}$ at a given confidence level. This translates to $M_2 \sin i < 34$ M$_\mathrm{Jup}$ for a 68\% confidence of a true mass $M_2$ below the hydrogen-burning limit. These four targets thus have a large chance of being stellar, and are therefore likely to have formed as tight stellar binary systems, rather than brown dwarf companions forming in a circumstellar disc around the host star. None of these systems were found to have a wide binary companion. Excluding these systems from our survey to focus on substellar companions only would hence have resulted in an even higher binary fraction for the rest of our sample. This further reinforces the idea that the most massive planets and brown dwarfs forming in discs and detected within $\sim$1 AU require a wide stellar companion to form or evolve to their observed orbital configurations.

\subsection{Implications for formation and evolution processes}
\label{formation_evolution}

Our results demonstrate a very robust correlation between binary occurrence rate and the sporadic population of close-in massive giant planets and brown dwarf desert inhabitants. Whatever the underlying processes, this concurrence implies that wide binaries must have an influence on the observed population of short-period planetary and brown dwarf companions, which could occur at the stage of formation or during later evolution.

\citet{ZuckerMahez2002} were the first to raise the possibility that planets in binaries may have a different mass-period distribution, a trend subsequently confirmed by \citet{Eggenberger2004}, \citet{Desidera2007}, \citet{Mugrauer2007} and others. Our results are in very good agreement with these studies, suggesting that the most massive planets observed within $\sim$1 AU are almost exclusively found in binary systems, and that this feature is amplified as planets or brown dwarfs reach shorter periods.
\citet{Desidera2007} concluded that the presence of a stellar companion on separations $<$100$-$300 AU may be able to modify the formation or evolution of giant planets. \citet{Eggenberger2004} also found that massive planets in binary systems with periods shorter than 40 days have very low eccentricities, suggesting that these planets likely underwent some form of migration, possibly induced or driven by outer binary companions, to evolve to their current orbits.
\citet{Duchene2010} investigated these observational trends and concluded that binarity does not affect the formation and growth of planetesimals (see also \citealp{Batygin2011} and \citealp{Rafikov2013}). \citet{Duchene2010} proposed that planet formation in binaries tighter than $\sim$100 AU occurs at a similar rate but through different mechanisms than around wider binaries and single stars, possibly explaining the observed preponderance of very massive, close-in planets found in binaries but rarely seen around isolated stars.

Simulations by \citet{Kley2001} showed that perturbations from a secondary star may alter the formation and evolution of a planet, in particular by enhancing the mass accretion and orbital migration rates. This could explain why the most massive short-period planets are found in multiple systems, the presence of stellar companions enabling massive planets to achieve smaller orbital separations than the corresponding limit for planets orbiting single stars \citep{Eggenberger2004}.
\citet{Jensen2003} found that the distribution of disc mass in $>$200 AU binary systems among T Tauri stars is not always determined by the stellar masses and may be more asymmetric, with the primary retaining a much more substantial disc and the secondary being left with a very low-mass disc. Massive discs around primaries in wide binaries could thus provide larger reservoirs of material for planet formation, which is thought to be favourable to the formation of higher-mass planets, as discussed in \citet{Mordasini2012}.
The shorter lifetime of circumstellar discs in tight binaries (e.g. \citealp{Kraus2012}) argues for a formation via gravitational collapse of the circumstellar disc (thousand year timescale) rather than through core accretion, which requires 1$-$10 million years. A favoured formation by gravitational disc instability is further supported by the very high masses of the giant planets or brown dwarfs considered here. Furthermore, theoretical work by \citet{Boss2006} suggested that a close stellar companion could rapidly induce gravitational perturbations and trigger the instabilities needed for gravitational fragmentation to proceed, even if the disc is not initially unstable to its own gravity.  However, simulations by \citet{Forgan2009} indicate that, rather than promoting fragmentation, perturbations from an outer companion are more likely to make the disc more stable.

The brown dwarf desert is thought to be a natural feature arising from formation around single stars, where massive objects with brown dwarf masses can only form at wide separations and can be challenging to bring inwards through disc migration alone. By modifying the circumstellar disc environment, allowing for different conditions facilitating \textit{in-situ} formation, and/or by triggering migration processes, the presence of a binary companion could help populate the low-mass end of the brown dwarf desert and explain the puzzling existence of the scarce population of very close-in brown dwarf desert inhabitants.

It is worth noting that over half of detected binary companions in our study have projected separations larger than 200 AU. We therefore argue that wider binaries must also be able to impact, almost to the same degree, the formation and/or evolution of these systems.
The processes described above must therefore also be possible from wider separations in order to account for the existence of the planets and brown dwarfs probed in this work. An easy way to facilitate this is to form the inner companions at significantly larger orbital distances (tens of AU), increasing the initial gravitational influence of the outer companion. As mentioned previously, this theory could tentatively explain the shortfall of binaries with separations $<$50$-$100 AU, which would then not be expected around such systems.


\begin{figure}
    \centering
    \includegraphics[width=0.42\textwidth]{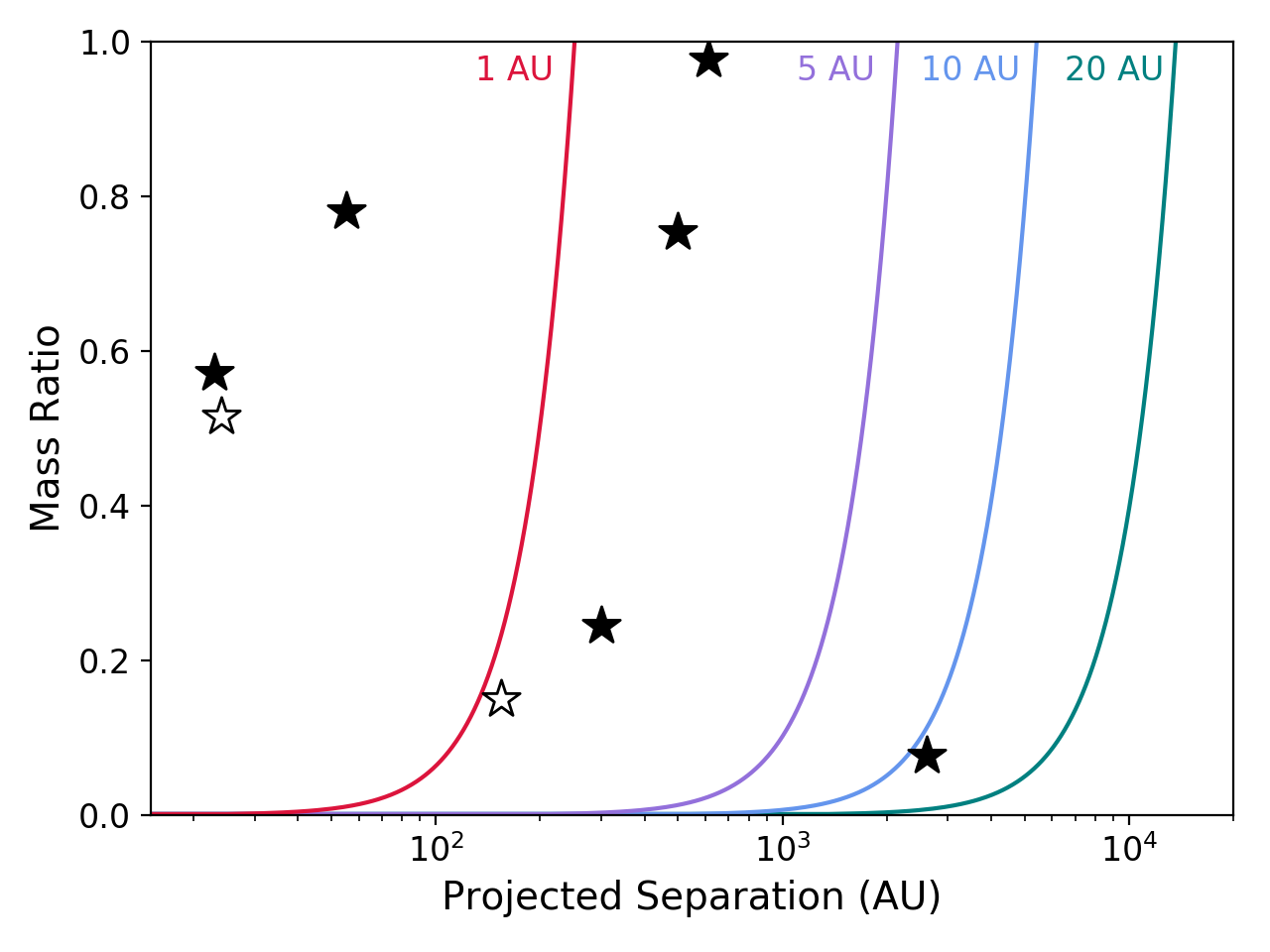}
    \caption{Minimum companion mass ratios necessary to excite Kozai-Lidov oscillations for a 15 M$_\mathrm{Jup}$ planet with an initial semi-major axis of 1 AU, 5 AU, 10 AU and 20 AU around a 1 M$_\odot$ star, as a function of wide companion separation. Companions lying to the left of each line are close and massive enough to induce Kozai-Lidov oscillations and overcome the pericenter precession of inner planets at 1, 5, 10 and 20 AU (see text). We overplot the positions of the confirmed (black symbols) and candidate (white symbols) binaries in our Kozai-consistent subsample.}
    \label{f:Kozai_masses}
\end{figure}

\subsection{Scattering and migration via the Kozai-Lidov mechanism}

One way binary companions could assist the migration of these systems is through the Kozai-Lidov mechanism \citep{Kozai1962,Lidov1962}. In this alternative scenario to produce hot Jupiters, an outer binary companion on an inclined orbit relative to the orbital plane of the planet triggers periodic oscillations of the planet's eccentricity and inclination \citep{Fabrycky2007,Wu2007,Dong2014,Petrovich2015}. Combined with the effects of tidal dissipation, these secular interactions can result in a very short orbit for the inner companion \citep{Rice2015}, possibly with a high spin-orbit misalignment with its host star. The amplitude of these interactions mostly depends on the initial mutual inclination between the inner and outer companions \citep{Fabrycky2007}, allowing Kozai-Lidov cycles to be induced by very distant perturbers.

As noted in Section \ref{sample_selection}, the subset of targets with a planet or brown dwarf within a 10 day orbit corresponds to the systems in our sample with a tidal circularisation timescale shorter than $\sim$15 Gyr, which we consider to be fully consistent with the Kozai-Lidov scenario. In Section \ref{results_subsets}, we showed that this subsample may have a marginally higher binary fraction than the already very high multiplicity rate observed for our full sample. This is thus compatible with the idea that the inner companions from this subsets could have been driven to their current orbital configurations through Kozai-Lidov oscillations under the effect of wide companions. We also note that 4 of our 5 targets also studied in the Friends of hot Jupiters campaign have high obliquities (placed in the ``misaligned'' sample in \citealp{Knutson2014}), a feature often associated with the Kozai-Lidov mechanism.

Unfortunately, full orbital parametrisation, including inclination measurements, is not possible for wide, directly-imaged binaries. Nevertheless, we can determine if the observed wide companions could be responsible for a Kozai-Lidov scattering of the inner planets and brown dwarfs based on their masses and orbital distances. This is done by estimating the minimum companion mass required to excite Kozai-Lidov oscillations on a timescale shorter than general pericenter precession, as was done in \citet{Ngo2016}. We adopted a primary stellar mass of 1 M$_\odot$ and a mass of 15 M$_\mathrm{Jup}$ for the inner companion, close to the median of our Kozai-consistent sample. Equating equations (1) and (23) from \citet{Fabrycky2007} for the Kozai-Lidov oscillation timescale and pericenter precession due to general relativity, respectively, we computed in Figure \ref{f:Kozai_masses} the minimum masses and separations necessary to migrate a 15 M$_\mathrm{Jup}$ companion with initial orbital separations of 1 AU, 5 AU, 10 AU and 20 AU through this scenario. We assumed initially circular orbits for the inner companions and eccentricities of 0.5 for the outer companions, based on the roughly uniform eccentricity distribution between 0 and 1 found by \citet{Raghavan2010} for wide field binaries. We found that almost all detected binary companions could explain the presence of the inner companions in this subset via the Kozai-Lidov mechanism, assuming they formed at separations larger than at least 1$-$10 AU.

These hierarchical systems are hence compatible with a migration of the inner companions through the Kozai-Lidov scenario based on this simple analysis. However, the subset of objects inconsistent the Kozai-Lidov mechanism, based on our tidal circularisation timescale argument (see Section \ref{sample_selection}), was also found to have a particularly large binary frequency. This suggests that these systems do not primarily migrate via Kozai-Lidov oscillations. This is in good agreement with the theoretical study by \citet{Naoz2012} and observational constraints placed by \citet{Ngo2016}, which concurred that only 20 to 30\% of all hot Jupiters can be explained by the Kozai-Lidov migration process. Kozai-Lidov oscillations are therefore unlikely to be the dominant mechanism driving close-in massive planets to their current locations.

\section{Summary and Conclusions}
\label{conclusion}

We have gathered a compilation of 38 planets or brown dwarfs with masses of at least 7 M$_\mathrm{Jup}$ and orbiting within $\sim$1 AU from their host stars, with the aim of examining the multiplicity statistics of these systems. We searched for wide binary companions to these objects using new direct imaging data, observations reported in the literature and the \textit{Gaia} DR2 catalogue. A total of 16 confirmed multiple systems were found, and another 3 targets have at least one high-probability candidate companion. We report here the discovery in \textit{Gaia} DR2 of a new mid-K tertiary component comoving with the WASP-14 binary system, and present an independent confirmation of the wide M7.5 companion to WASP-18.
We used a robust MCMC statistical approach to constrain the binary properties of our sample, correcting for observational biases and incompleteness. Our main results are summarised below.

1. \textit{A very high binary fraction.}
Our analysis revealed a very large binary frequency of $f = 79.0^{+13.2}_{-14.7}$\% for these outer companions on separations between 20 and 10,000 AU, which is more than twice as high as for field stars on the same separation range, with a 3-$\sigma$ significance. These results demonstrate that wide binary companions greatly influence the formation or evolution of these close-in massive planets and brown dwarfs. The presence of a binary companion could allow for different natal environments in circumstellar discs, enabling {\it in-situ} formation at locations where giant planet formation is not normally possible. Stellar companions could also facilitate disc migration beyond the extent normally seen around single stars, or could trigger alternative migration processes through induced secular interactions. 

2. \textit{A deficit of close binaries.}
The output of our statistical analysis also showed a strong preference for wide binaries, with a peak around 250 AU, compared to $\sim$50 AU for the overall field population. The apparent shortfall of $<$50$-$100 AU binaries is consistent with previous studies. It is not clear whether this deficiency indicates that planet formation is inhibited in tight binaries, that our probed planets formed near these separations, or if it is the result of selection biases in exoplanet surveys. Based on these observations, we argue that the mechanisms assisting planet formation or evolution in multiple star systems must be associated with widely-separated binaries, on distances larger than several hundreds of AU. However, we did find that the Kozai-Lidov mechanism is unlikely to be the dominant underlying process.

3. \textit{A higher binary rate for higher-mass planets.}
A comparison with the population of lower-mass planets suggests that binary occurrence increases with planet mass for these close-in objects. This is in good agreement with prior studies that found the most massive planets to be almost exclusively observed in binary systems, and indicates that the role played by binary companions in the existence of these systems becomes more critical for higher-mass planets.

4. \textit{A higher binary rate for shorter-period planets.}
Dividing our sample into two subsets, we found that stars hosting planets or brown dwarfs with orbital periods $<$10 days have a marginally larger binary rate than systems with longer-period inner companions, consistent with a 100\% multiplicity fraction at the 1-$\sigma$ level. If confirmed, this trend could suggest that the influence of binarity on the formation/evolution of the most massive planetary companions is enhanced for shorter-period planets, and may even become a requirement for the very closest planets and brown dwarfs.

We conclude that wide binary companions have a crucial influence on planet formation and/or evolution and may be responsible for the sporadic population of high-mass planets and brown dwarf desert members observed on very tight orbital configurations, which seem to rarely exist around isolated stars.

\section*{Acknowledgements}
We thank our anonymous referee for their insightful comments and suggestions. This paper comes from work undertaken during a visit funded by the Scottish Universties Physics Alliance (SUPA) Postgraduate, Postdoctoral and Early Career Researcher Short-Term Visits funding.
This work benefited from the Exoplanet Summer Program in the Other Worlds Laboratory (OWL) at the University of California, Santa Cruz, a program funded by the Heising-Simons Foundation.
KM acknowledges funding by the Science and Technology Foundation of Portugal (FCT), grants No. IF/00194/2015 and PTDC/FIS-AST/28731/2017.
This work has made use of data from the European Space Agency (ESA) mission
{\it Gaia} (\url{https://www.cosmos.esa.int/gaia}), processed by the {\it Gaia} Data Processing and Analysis Consortium (DPAC, \url{https://www.cosmos.esa.int/web/gaia/dpac/consortium}).
This study is based on observations obtained at the Gemini Observatory, which is operated by the Association of Universities for Research in Astronomy, Inc., under a cooperative agreement with the NSF on behalf of the Gemini partnership: the National Science Foundation (United States), National Research Council (Canada), CONICYT (Chile), Ministerio de Ciencia, Tecnolog\'{i}a e Innovaci\'{o}n Productiva (Argentina), Minist\'{e}rio da Ci\^{e}ncia, Tecnologia e Inova\c{c}\~{a}o (Brazil), and Korea Astronomy and Space Science Institute (Republic of Korea).
This work is based on observations collected at the European Southern Observatory under ESO programme 099.C-0728.
Some of the observations in the paper made use of the NN-EXPLORE Exoplanet and Stellar Speckle Imager (NESSI). NESSI was funded by the NASA Exoplanet Exploration Program and the NASA Ames Research Center. NESSI was built at the Ames Research Center by Steve B. Howell, Nic Scott, Elliott P. Horch, and Emmett Quigley.
This research has made use of the NASA Exoplanet Archive, which is operated by the California Institute of Technology, under contract with the National Aeronautics and Space Administration under the Exoplanet Exploration Program. This research has made use of the Exoplanet Orbit Database, and the Exoplanet Data Explorer at exoplanets.org, as well as the Extrasolar Planets Encyclopaedia at exoplanet.eu. 
This publication makes use of data products from the Two Micron All Sky Survey, which is a joint project of the Uni- versity of Massachusetts and the Infrared Processing and Analysis Center/California Institute of Technology, funded by the National Aeronautics and Space Administration and the National Science Foundation.
This research has made use of the SIMBAD database, operated at CDS, Strasbourg, France.

\bibliographystyle{mnras}

\begin{thebibliography}{}
\makeatletter
\relax
\def\mn@urlcharsother{\let\do\@makeother \do\$\do\&\do\#\do\^\do\_\do\%\do\~}
\def\mn@doi{\begingroup\mn@urlcharsother \@ifnextchar [ {\mn@doi@}
  {\mn@doi@[]}}
\def\mn@doi@[#1]#2{\def\@tempa{#1}\ifx\@tempa\@empty \href
  {http://dx.doi.org/#2} {doi:#2}\else \href {http://dx.doi.org/#2} {#1}\fi
  \endgroup}
\def\mn@eprint#1#2{\mn@eprint@#1:#2::\@nil}
\def\mn@eprint@arXiv#1{\href {http://arxiv.org/abs/#1} {{\tt arXiv:#1}}}
\def\mn@eprint@dblp#1{\href {http://dblp.uni-trier.de/rec/bibtex/#1.xml}
  {dblp:#1}}
\def\mn@eprint@#1:#2:#3:#4\@nil{\def\@tempa {#1}\def\@tempb {#2}\def\@tempc
  {#3}\ifx \@tempc \@empty \let \@tempc \@tempb \let \@tempb \@tempa \fi \ifx
  \@tempb \@empty \def\@tempb {arXiv}\fi \@ifundefined
  {mn@eprint@\@tempb}{\@tempb:\@tempc}{\expandafter \expandafter \csname
  mn@eprint@\@tempb\endcsname \expandafter{\@tempc}}}

\bibitem[\protect\citeauthoryear{{Adams}, {Ciardi}, {Dupree}, {Gautier},
  {Kulesa}  \& {McCarthy}}{{Adams} et~al.}{2012}]{Adams2012}
{Adams} E.~R.,  {Ciardi} D.~R.,  {Dupree} A.~K.,  {Gautier} T.~N. I.,  {Kulesa}
  C.,   {McCarthy} D.,  2012, \mn@doi [\aj] {10.1088/0004-6256/144/2/42}, \href
  {https://ui.adsabs.harvard.edu/#abs/2012AJ....144...42A} {144, 42}

\bibitem[\protect\citeauthoryear{{Adams}, {Dupree}, {Kulesa}  \&
  {McCarthy}}{{Adams} et~al.}{2013}]{Adams2013}
{Adams} E.~R.,  {Dupree} A.~K.,  {Kulesa} C.,   {McCarthy} D.,  2013, \mn@doi
  [\aj] {10.1088/0004-6256/146/1/9}, \href
  {http://adsabs.harvard.edu/abs/2013AJ....146....9A} {146, 9}

\bibitem[\protect\citeauthoryear{{Aguilera-G{\'o}mez}, {Ram{\'\i}rez}  \&
  {Chanam{\'e}}}{{Aguilera-G{\'o}mez} et~al.}{2018}]{Aguilera-Gomez2018}
{Aguilera-G{\'o}mez} C.,  {Ram{\'\i}rez} I.,   {Chanam{\'e}} J.,  2018, \mn@doi
  [\aap] {10.1051/0004-6361/201732209}, \href
  {https://ui.adsabs.harvard.edu/#abs/2018A&A...614A..55A} {614, A55}

\bibitem[\protect\citeauthoryear{{Allard}, {Homeier}, {Freytag}  \&
  {Sharp}}{{Allard} et~al.}{2012}]{Allard2012}
{Allard} F.,  {Homeier} D.,  {Freytag} B.,   {Sharp} C.~M.,  2012, in
  {Reyl{\'e}} C.,  {Charbonnel} C.,   {Schultheis} M.,  eds,  EAS Publications
  Series Vol. 57, EAS Publications Series. pp 3--43 (\mn@eprint {arXiv}
  {1206.1021}), \mn@doi{10.1051/eas/1257001}

\bibitem[\protect\citeauthoryear{{Almeida}, {Gameiro}, {Petrov}, {Melo},
  {Santos}, {Figueira}  \& {Alencar}}{{Almeida} et~al.}{2017}]{Almeida2017}
{Almeida} P.~V.,  {Gameiro} J.~F.,  {Petrov} P.~P.,  {Melo} C.,  {Santos}
  N.~C.,  {Figueira} P.,   {Alencar} S.~H.~P.,  2017, \mn@doi [\aap]
  {10.1051/0004-6361/201629749}, \href
  {https://ui.adsabs.harvard.edu/#abs/2017A&A...600A..84A} {600, A84}

\bibitem[\protect\citeauthoryear{{Andrews} \& {Williams}}{{Andrews} \&
  {Williams}}{2007}]{Andrews2007}
{Andrews} S.~M.,  {Williams} J.~P.,  2007, \mn@doi [\apj] {10.1086/511741},
  \href {http://adsabs.harvard.edu/abs/2007ApJ...659..705A} {659, 705}

\bibitem[\protect\citeauthoryear{{Arenou} et~al.,}{{Arenou}
  et~al.}{2018}]{Arenou2018}
{Arenou} F.,  et~al., 2018, \mn@doi [\aap] {10.1051/0004-6361/201833234}, \href
  {http://adsabs.harvard.edu/abs/2018A\%26A...616A..17A} {616, A17}

\bibitem[\protect\citeauthoryear{{Augereau} \& {Papaloizou}}{{Augereau} \&
  {Papaloizou}}{2004}]{Augereau2004}
{Augereau} J.~C.,  {Papaloizou} J.~C.~B.,  2004, \mn@doi [\aap]
  {10.1051/0004-6361:20031622}, \href
  {http://adsabs.harvard.edu/abs/2004A%26A...414.1153A} {414, 1153}

\bibitem[\protect\citeauthoryear{{Bailer-Jones}, {Rybizki}, {Fouesneau},
  {Mantelet}  \& {Andrae}}{{Bailer-Jones} et~al.}{2018}]{Bailer-Jones2018}
{Bailer-Jones} C.~A.~L.,  {Rybizki} J.,  {Fouesneau} M.,  {Mantelet} G.,
  {Andrae} R.,  2018, \mn@doi [\aj] {10.3847/1538-3881/aacb21}, \href
  {http://adsabs.harvard.edu/abs/2018AJ....156...58B} {156, 58}

\bibitem[\protect\citeauthoryear{{Bakos} et~al.,}{{Bakos}
  et~al.}{2011}]{Bakos2011}
{Bakos} G.~{\'A}.,  et~al., 2011, \mn@doi [\apj] {10.1088/0004-637X/742/2/116},
  \href {http://adsabs.harvard.edu/abs/2011ApJ...742..116B} {742, 116}

\bibitem[\protect\citeauthoryear{{Baranec}, {Ziegler}, {Law}, {Morton},
  {Riddle}, {Atkinson}, {Schonhut}  \& {Crepp}}{{Baranec}
  et~al.}{2016}]{Baranec2016}
{Baranec} C.,  {Ziegler} C.,  {Law} N.~M.,  {Morton} T.,  {Riddle} R.,
  {Atkinson} D.,  {Schonhut} J.,   {Crepp} J.,  2016, \mn@doi [\aj]
  {10.3847/0004-6256/152/1/18}, \href
  {http://adsabs.harvard.edu/abs/2016AJ....152...18B} {152, 18}

\bibitem[\protect\citeauthoryear{{Batygin}, {Morbidelli}  \&
  {Tsiganis}}{{Batygin} et~al.}{2011}]{Batygin2011}
{Batygin} K.,  {Morbidelli} A.,   {Tsiganis} K.,  2011, \mn@doi [\aap]
  {10.1051/0004-6361/201117193}, \href
  {http://adsabs.harvard.edu/abs/2011A%26A...533A...7B} {533, A7}

\bibitem[\protect\citeauthoryear{{Bell}, {Cassen}, {Klahr}  \&
  {Henning}}{{Bell} et~al.}{1997}]{Bell1997}
{Bell} K.~R.,  {Cassen} P.~M.,  {Klahr} H.~H.,   {Henning} T.,  1997, \mn@doi
  [\apj] {10.1086/304514}, \href
  {http://adsabs.harvard.edu/abs/1997ApJ...486..372B} {486, 372}

\bibitem[\protect\citeauthoryear{{Bergfors} et~al.,}{{Bergfors}
  et~al.}{2013}]{Bergfors2013}
{Bergfors} C.,  et~al., 2013, \mn@doi [\mnras] {10.1093/mnras/sts019}, \href
  {http://adsabs.harvard.edu/abs/2013MNRAS.428..182B} {428, 182}

\bibitem[\protect\citeauthoryear{{Bonfanti}, {Ortolani}  \&
  {Nascimbeni}}{{Bonfanti} et~al.}{2016}]{Bonfanti2016}
{Bonfanti} A.,  {Ortolani} S.,   {Nascimbeni} V.,  2016, \mn@doi [\aap]
  {10.1051/0004-6361/201527297}, \href
  {http://adsabs.harvard.edu/abs/2016A\%26A...585A...5B} {585, A5}

\bibitem[\protect\citeauthoryear{{Bonomo} et~al.,}{{Bonomo}
  et~al.}{2017}]{Bonomo2017}
{Bonomo} A.~S.,  et~al., 2017, \mn@doi [\aap] {10.1051/0004-6361/201629882},
  \href {http://adsabs.harvard.edu/abs/2017A\%26A...602A.107B} {602, A107}

\bibitem[\protect\citeauthoryear{{Boss}}{{Boss}}{1998}]{Boss1998}
{Boss} A.~P.,  1998, \mn@doi [\apj] {10.1086/306036}, \href
  {http://adsabs.harvard.edu/abs/1998ApJ...503..923B} {503, 923}

\bibitem[\protect\citeauthoryear{{Boss}}{{Boss}}{2006}]{Boss2006}
{Boss} A.~P.,  2006, \mn@doi [\apj] {10.1086/500530}, \href
  {http://adsabs.harvard.edu/abs/2006ApJ...641.1148B} {641, 1148}

\bibitem[\protect\citeauthoryear{{Bowler}, {Liu}  \& {Cushing}}{{Bowler}
  et~al.}{2009}]{Bowler2009}
{Bowler} B.~P.,  {Liu} M.~C.,   {Cushing} M.~C.,  2009, \mn@doi [\apj]
  {10.1088/0004-637X/706/2/1114}, \href
  {http://adsabs.harvard.edu/abs/2009ApJ...706.1114B} {706, 1114}

\bibitem[\protect\citeauthoryear{{Brandner} et~al.,}{{Brandner}
  et~al.}{2000}]{Brandner2000}
{Brandner} W.,  et~al., 2000, \mn@doi [\aj] {10.1086/301483}, \href
  {http://adsabs.harvard.edu/abs/2000AJ....120..950B} {120, 950}

\bibitem[\protect\citeauthoryear{{Butler}, {Marcy}, {Williams}, {Hauser}  \&
  {Shirts}}{{Butler} et~al.}{1997}]{Butler1997}
{Butler} R.~P.,  {Marcy} G.~W.,  {Williams} E.,  {Hauser} H.,   {Shirts} P.,
  1997, \mn@doi [\apjl] {10.1086/310444}, \href
  {http://adsabs.harvard.edu/abs/1997ApJ...474L.115B} {474, L115}

\bibitem[\protect\citeauthoryear{{Butler}, {Marcy}, {Fischer}, {Brown},
  {Contos}, {Korzennik}, {Nisenson}  \& {Noyes}}{{Butler}
  et~al.}{1999}]{Butler1999}
{Butler} R.~P.,  {Marcy} G.~W.,  {Fischer} D.~A.,  {Brown} T.~M.,  {Contos}
  A.~R.,  {Korzennik} S.~G.,  {Nisenson} P.,   {Noyes} R.~W.,  1999, \mn@doi
  [\apj] {10.1086/308035}, \href
  {http://adsabs.harvard.edu/abs/1999ApJ...526..916B} {526, 916}

\bibitem[\protect\citeauthoryear{{Butler} et~al.,}{{Butler}
  et~al.}{2006}]{Butler2006}
{Butler} R.~P.,  et~al., 2006, \mn@doi [\apj] {10.1086/504701}, \href
  {http://adsabs.harvard.edu/abs/2006ApJ...646..505B} {646, 505}

\bibitem[\protect\citeauthoryear{{Chauvin}, {Lagrange}, {Udry}, {Fusco},
  {Galland}, {Naef}, {Beuzit}  \& {Mayor}}{{Chauvin}
  et~al.}{2006}]{Chauvin2006}
{Chauvin} G.,  {Lagrange} A.-M.,  {Udry} S.,  {Fusco} T.,  {Galland} F.,
  {Naef} D.,  {Beuzit} J.-L.,   {Mayor} M.,  2006, \mn@doi [\aap]
  {10.1051/0004-6361:20054709}, \href
  {http://adsabs.harvard.edu/abs/2006A\%26A...456.1165C} {456, 1165}

\bibitem[\protect\citeauthoryear{{Chauvin} et~al.,}{{Chauvin}
  et~al.}{2012}]{Chauvin2012}
{Chauvin} G.,  et~al., 2012, \mn@doi [\aap] {10.1051/0004-6361/201118346},
  \href {http://adsabs.harvard.edu/abs/2012A%26A...542A..41C} {542, A41}

\bibitem[\protect\citeauthoryear{{Chen}, {Henning}, {van Boekel}  \&
  {Grady}}{{Chen} et~al.}{2006}]{Chen2006}
{Chen} X.~P.,  {Henning} T.,  {van Boekel} R.,   {Grady} C.~A.,  2006, \mn@doi
  [\aap] {10.1051/0004-6361:20054122}, \href
  {http://adsabs.harvard.edu/abs/2006A%26A...445..331C} {445, 331}

\bibitem[\protect\citeauthoryear{{Clarke} \& {Lodato}}{{Clarke} \&
  {Lodato}}{2009}]{Clarke2009}
{Clarke} C.~J.,  {Lodato} G.,  2009, \mn@doi [\mnras]
  {10.1111/j.1745-3933.2009.00695.x}, \href
  {http://adsabs.harvard.edu/abs/2009MNRAS.398L...6C} {398, L6}

\bibitem[\protect\citeauthoryear{{Coker}, {Gaudi}, {Pogge}  \& {Horch}}{{Coker}
  et~al.}{2018}]{Coker2018}
{Coker} C.~T.,  {Gaudi} B.~S.,  {Pogge} R.~W.,   {Horch} E.,  2018, \aj, \href
  {http://adsabs.harvard.edu/abs/2018AJ....155...27C} {155, 27}

\bibitem[\protect\citeauthoryear{{Csizmadia}, {Hellard}  \&
  {Smith}}{{Csizmadia} et~al.}{2019}]{Csizmadia2019}
{Csizmadia} S.,  {Hellard} H.,   {Smith} A.~M.~S.,  2019, \mn@doi [\aap]
  {10.1051/0004-6361/201834376}, \href
  {http://adsabs.harvard.edu/abs/2019A%26A...623A..45C} {623, A45}

\bibitem[\protect\citeauthoryear{{Curiel}, {Cant{\'o}}, {Georgiev},
  {Ch{\'a}vez}  \& {Poveda}}{{Curiel} et~al.}{2011}]{Curiel2011}
{Curiel} S.,  {Cant{\'o}} J.,  {Georgiev} L.,  {Ch{\'a}vez} C.~E.,   {Poveda}
  A.,  2011, \mn@doi [\aap] {10.1051/0004-6361/201015693}, \href
  {http://adsabs.harvard.edu/abs/2011A\%26A...525A..78C} {525, A78}

\bibitem[\protect\citeauthoryear{{Curtis}, {Wolfgang}, {Wright}, {Brewer}  \&
  {Johnson}}{{Curtis} et~al.}{2013}]{Curtis2013}
{Curtis} J.~L.,  {Wolfgang} A.,  {Wright} J.~T.,  {Brewer} J.~M.,   {Johnson}
  J.~A.,  2013, \mn@doi [\aj] {10.1088/0004-6256/145/5/134}, \href
  {http://adsabs.harvard.edu/abs/2013AJ....145..134C} {145, 134}

\bibitem[\protect\citeauthoryear{{Daemgen}, {Hormuth}, {Brandner}, {Bergfors},
  {Janson}, {Hippler}  \& {Henning}}{{Daemgen} et~al.}{2009}]{Daemgen2009}
{Daemgen} S.,  {Hormuth} F.,  {Brandner} W.,  {Bergfors} C.,  {Janson} M.,
  {Hippler} S.,   {Henning} T.,  2009, \mn@doi [\aap]
  {10.1051/0004-6361/200810988}, \href
  {http://adsabs.harvard.edu/abs/2009A%26A...498..567D} {498, 567}

\bibitem[\protect\citeauthoryear{{Daemgen}, {Bonavita}, {Jayawardhana},
  {Lafreni{\`e}re}  \& {Janson}}{{Daemgen} et~al.}{2015}]{Daemgen2015}
{Daemgen} S.,  {Bonavita} M.,  {Jayawardhana} R.,  {Lafreni{\`e}re} D.,
  {Janson} M.,  2015, \mn@doi [\apj] {10.1088/0004-637X/799/2/155}, \href
  {http://adsabs.harvard.edu/abs/2015ApJ...799..155D} {799, 155}

\bibitem[\protect\citeauthoryear{Dawson \& Johnson}{Dawson \&
  Johnson}{2018}]{Dawson2018}
Dawson R.~I.,  Johnson J.~A.,  2018, \mn@doi [Annual Review of Astronomy and
  Astrophysics] {10.1146/annurev-astro-081817-051853}, 56, 175

\bibitem[\protect\citeauthoryear{{Desidera} \& {Barbieri}}{{Desidera} \&
  {Barbieri}}{2007}]{Desidera2007}
{Desidera} S.,  {Barbieri} M.,  2007, \mn@doi [\aap]
  {10.1051/0004-6361:20066319}, \href
  {http://adsabs.harvard.edu/abs/2007A%26A...462..345D} {462, 345}

\bibitem[\protect\citeauthoryear{{Desidera} et~al.,}{{Desidera}
  et~al.}{2004}]{Desidera2004}
{Desidera} S.,  et~al., 2004, in {Beaulieu} J.,  {Lecavelier Des Etangs} A.,
  {Terquem} C.,  eds,  Astronomical Society of the Pacific Conference Series
  Vol. 321, Extrasolar Planets: Today and Tomorrow. p.~103

\bibitem[\protect\citeauthoryear{{D{\'{\i}}az} et~al.,}{{D{\'{\i}}az}
  et~al.}{2012}]{Diaz2012}
{D{\'{\i}}az} R.~F.,  et~al., 2012, \mn@doi [\aap]
  {10.1051/0004-6361/201117935}, \href
  {http://adsabs.harvard.edu/abs/2012A\%26A...538A.113D} {538, A113}

\bibitem[\protect\citeauthoryear{{Diolaiti}, {Bendinelli}, {Bonaccini},
  {Close}, {Currie}  \& {Parmeggiani}}{{Diolaiti} et~al.}{2000}]{Diolaiti2000}
{Diolaiti} E.,  {Bendinelli} O.,  {Bonaccini} D.,  {Close} L.,  {Currie} D.,
  {Parmeggiani} G.,  2000, \mn@doi [\aaps] {10.1051/aas:2000305}, \href
  {http://adsabs.harvard.edu/abs/2000A%26AS..147..335D} {147, 335}

\bibitem[\protect\citeauthoryear{{Dobbs-Dixon}, {Lin}  \&
  {Mardling}}{{Dobbs-Dixon} et~al.}{2004}]{Dobbs-Dixon2004}
{Dobbs-Dixon} I.,  {Lin} D.~N.~C.,   {Mardling} R.~A.,  2004, \mn@doi [\apj]
  {10.1086/421510}, \href
  {https://ui.adsabs.harvard.edu/\#abs/2004ApJ...610..464D} {610, 464}

\bibitem[\protect\citeauthoryear{{D{\"o}llinger}, {Hatzes}, {Pasquini},
  {Guenther}, {Hartmann}, {Girardi}  \& {Esposito}}{{D{\"o}llinger}
  et~al.}{2007}]{Dollinger2007}
{D{\"o}llinger} M.~P.,  {Hatzes} A.~P.,  {Pasquini} L.,  {Guenther} E.~W.,
  {Hartmann} M.,  {Girardi} L.,   {Esposito} M.,  2007, \mn@doi [\aap]
  {10.1051/0004-6361:20066987}, \href
  {http://adsabs.harvard.edu/abs/2007A%26A...472..649D} {472, 649}

\bibitem[\protect\citeauthoryear{{Dommanget} \& {Nys}}{{Dommanget} \&
  {Nys}}{2000}]{Dommanget2000}
{Dommanget} J.,  {Nys} O.,  2000, \aap, \href
  {http://adsabs.harvard.edu/abs/2000A\%26A...363..991D} {363, 991}

\bibitem[\protect\citeauthoryear{{Dong}, {Katz}  \& {Socrates}}{{Dong}
  et~al.}{2014}]{Dong2014}
{Dong} S.,  {Katz} B.,   {Socrates} A.,  2014, \mn@doi [\apjl]
  {10.1088/2041-8205/781/1/L5}, \href
  {http://adsabs.harvard.edu/abs/2014ApJ...781L...5D} {781, L5}

\bibitem[\protect\citeauthoryear{{Dong}, {Zhu}, {Fung}, {Rafikov}, {Chiang}  \&
  {Wagner}}{{Dong} et~al.}{2016}]{dong2016}
{Dong} R.,  {Zhu} Z.,  {Fung} J.,  {Rafikov} R.,  {Chiang} E.,   {Wagner} K.,
  2016, \mn@doi [\apjl] {10.3847/2041-8205/816/1/L12}, \href
  {http://adsabs.harvard.edu/abs/2016ApJ...816L..12D} {816, L12}

\bibitem[\protect\citeauthoryear{{Duch{\^e}ne}}{{Duch{\^e}ne}}{2010}]{Duchene2010}
{Duch{\^e}ne} G.,  2010, \mn@doi [\apjl] {10.1088/2041-8205/709/2/L114}, \href
  {http://adsabs.harvard.edu/abs/2010ApJ...709L.114D} {709, L114}

\bibitem[\protect\citeauthoryear{{Duquennoy} \& {Mayor}}{{Duquennoy} \&
  {Mayor}}{1991}]{Duquennoy1991}
{Duquennoy} A.,  {Mayor} M.,  1991, \aap, \href
  {http://adsabs.harvard.edu/abs/1991A\%26A...248..485D} {248, 485}

\bibitem[\protect\citeauthoryear{{ESA}}{{ESA}}{1997}]{ESA1997}
{ESA} ed. 1997, {The HIPPARCOS and TYCHO catalogues. Astrometric and
  photometric star catalogues derived from the ESA HIPPARCOS Space Astrometry
  Mission}  ESA Special Publication Vol. 1200

\bibitem[\protect\citeauthoryear{{Eggenberger}, {Udry}  \&
  {Mayor}}{{Eggenberger} et~al.}{2004}]{Eggenberger2004}
{Eggenberger} A.,  {Udry} S.,   {Mayor} M.,  2004, \mn@doi [\aap]
  {10.1051/0004-6361:20034164}, \href
  {http://adsabs.harvard.edu/abs/2004A\%26A...417..353E} {417, 353}

\bibitem[\protect\citeauthoryear{{Eggenberger}, {Udry}, {Chauvin}, {Beuzit},
  {Lagrange}, {S{\'e}gransan}  \& {Mayor}}{{Eggenberger}
  et~al.}{2007}]{Eggenberger2007}
{Eggenberger} A.,  {Udry} S.,  {Chauvin} G.,  {Beuzit} J.-L.,  {Lagrange}
  A.-M.,  {S{\'e}gransan} D.,   {Mayor} M.,  2007, \mn@doi [\aap]
  {10.1051/0004-6361:20077447}, \href
  {http://adsabs.harvard.edu/abs/2007A\%26A...474..273E} {474, 273}

\bibitem[\protect\citeauthoryear{{Eggenberger}, {Udry}, {Chauvin}, {Forveille},
  {Beuzit}, {Lagrange}  \& {Mayor}}{{Eggenberger}
  et~al.}{2011}]{Eggenberger2011}
{Eggenberger} A.,  {Udry} S.,  {Chauvin} G.,  {Forveille} T.,  {Beuzit} J.-L.,
  {Lagrange} A.-M.,   {Mayor} M.,  2011, in {Sozzetti} A.,  {Lattanzi} M.~G.,
  {Boss} A.~P.,  eds,  IAU Symposium Vol. 276, The Astrophysics of Planetary
  Systems: Formation, Structure, and Dynamical Evolution. pp 409--410
  (\mn@eprint {arXiv} {1101.0432}), \mn@doi{10.1017/S1743921311020564}

\bibitem[\protect\citeauthoryear{{Eggleton} \& {Tokovinin}}{{Eggleton} \&
  {Tokovinin}}{2008}]{Eggleton2008}
{Eggleton} P.~P.,  {Tokovinin} A.~A.,  2008, \mn@doi [\mnras]
  {10.1111/j.1365-2966.2008.13596.x}, \href
  {http://adsabs.harvard.edu/abs/2008MNRAS.389..869E} {389, 869}

\bibitem[\protect\citeauthoryear{{Eggleton}, {Kiseleva}  \& {Hut}}{{Eggleton}
  et~al.}{1998}]{Eggleton1998}
{Eggleton} P.~P.,  {Kiseleva} L.~G.,   {Hut} P.,  1998, \mn@doi [\apj]
  {10.1086/305670}, \href {http://adsabs.harvard.edu/abs/1998ApJ...499..853E}
  {499, 853}

\bibitem[\protect\citeauthoryear{{Eisner}, {Hillenbrand}, {White}, {Akeson}  \&
  {Sargent}}{{Eisner} et~al.}{2005}]{Eisner2005}
{Eisner} J.~A.,  {Hillenbrand} L.~A.,  {White} R.~J.,  {Akeson} R.~L.,
  {Sargent} A.~I.,  2005, \mn@doi [\apj] {10.1086/428828}, \href
  {http://adsabs.harvard.edu/abs/2005ApJ...623..952E} {623, 952}

\bibitem[\protect\citeauthoryear{{Esteves}, {De Mooij}  \&
  {Jayawardhana}}{{Esteves} et~al.}{2015}]{Esteves2015}
{Esteves} L.~J.,  {De Mooij} E. J.~W.,   {Jayawardhana} R.,  2015, \mn@doi
  [\apj] {10.1088/0004-637X/804/2/150}, \href
  {https://ui.adsabs.harvard.edu/#abs/2015ApJ...804..150E} {804, 150}

\bibitem[\protect\citeauthoryear{{Fabrycky} \& {Tremaine}}{{Fabrycky} \&
  {Tremaine}}{2007}]{Fabrycky2007}
{Fabrycky} D.,  {Tremaine} S.,  2007, \mn@doi [\apj] {10.1086/521702}, \href
  {http://adsabs.harvard.edu/abs/2007ApJ...669.1298F} {669, 1298}

\bibitem[\protect\citeauthoryear{{Filippazzo}, {Rice}, {Faherty}, {Cruz}, {Van
  Gordon}  \& {Looper}}{{Filippazzo} et~al.}{2015}]{Filippazzo2015}
{Filippazzo} J.~C.,  {Rice} E.~L.,  {Faherty} J.,  {Cruz} K.~L.,  {Van Gordon}
  M.~M.,   {Looper} D.~L.,  2015, \mn@doi [\apj] {10.1088/0004-637X/810/2/158},
  \href {http://adsabs.harvard.edu/abs/2015ApJ...810..158F} {810, 158}

\bibitem[\protect\citeauthoryear{{Fontanive}, {Biller}, {Bonavita}  \&
  {Allers}}{{Fontanive} et~al.}{2018}]{Fontanive2018}
{Fontanive} C.,  {Biller} B.,  {Bonavita} M.,   {Allers} K.,  2018, \mn@doi
  [\mnras] {10.1093/mnras/sty1682}, \href
  {http://adsabs.harvard.edu/abs/2018MNRAS.479.2702F} {479, 2702}

\bibitem[\protect\citeauthoryear{{Foreman-Mackey}, {Hogg}, {Lang}  \&
  {Goodman}}{{Foreman-Mackey} et~al.}{2013}]{Foreman-Mackey2013}
{Foreman-Mackey} D.,  {Hogg} D.~W.,  {Lang} D.,   {Goodman} J.,  2013, \mn@doi
  [\pasp] {10.1086/670067}, \href
  {http://adsabs.harvard.edu/abs/2013PASP..125..306F} {125, 306}

\bibitem[\protect\citeauthoryear{{Forgan} \& {Rice}}{{Forgan} \&
  {Rice}}{2009}]{Forgan2009}
{Forgan} D.,  {Rice} K.,  2009, \mn@doi [\mnras]
  {10.1111/j.1365-2966.2009.15596.x}, \href
  {http://adsabs.harvard.edu/abs/2009MNRAS.400.2022F} {400, 2022}

\bibitem[\protect\citeauthoryear{{Forgan} \& {Rice}}{{Forgan} \&
  {Rice}}{2011}]{Forgan2011}
{Forgan} D.,  {Rice} K.,  2011, \mn@doi [\mnras]
  {10.1111/j.1365-2966.2011.19380.x}, \href
  {http://adsabs.harvard.edu/abs/2011MNRAS.417.1928F} {417, 1928}

\bibitem[\protect\citeauthoryear{{Gaia Collaboration} et~al.,}{{Gaia
  Collaboration} et~al.}{2016}]{GaiaCollaboration2016}
{Gaia Collaboration} et~al., 2016, \mn@doi [\aap]
  {10.1051/0004-6361/201629272}, \href
  {http://adsabs.harvard.edu/abs/2016A\%26A...595A...1G} {595, A1}

\bibitem[\protect\citeauthoryear{{Gaia Collaboration} et~al.,}{{Gaia
  Collaboration} et~al.}{2018}]{GaiaCollaboration2018}
{Gaia Collaboration} et~al., 2018, \mn@doi [\aap]
  {10.1051/0004-6361/201833051}, \href
  {http://adsabs.harvard.edu/abs/2018A\%26A...616A...1G} {616, A1}

\bibitem[\protect\citeauthoryear{{Galland}, {Lagrange}, {Udry}, {Chelli},
  {Pepe}, {Beuzit}  \& {Mayor}}{{Galland} et~al.}{2005}]{Galland2005}
{Galland} F.,  {Lagrange} A.-M.,  {Udry} S.,  {Chelli} A.,  {Pepe} F.,
  {Beuzit} J.-L.,   {Mayor} M.,  2005, \mn@doi [\aap]
  {10.1051/0004-6361:200500176}, \href
  {http://adsabs.harvard.edu/abs/2005A\%26A...444L..21G} {444, L21}

\bibitem[\protect\citeauthoryear{{Galland}, {Lagrange}, {Udry}, {Beuzit},
  {Pepe}  \& {Mayor}}{{Galland} et~al.}{2006}]{Galland2006}
{Galland} F.,  {Lagrange} A.-M.,  {Udry} S.,  {Beuzit} J.-L.,  {Pepe} F.,
  {Mayor} M.,  2006, \mn@doi [\aap] {10.1051/0004-6361:20054079}, \href
  {http://adsabs.harvard.edu/abs/2006A\%26A...452..709G} {452, 709}

\bibitem[\protect\citeauthoryear{{Ghez}, {Neugebauer}  \& {Matthews}}{{Ghez}
  et~al.}{1993}]{Ghez1993}
{Ghez} A.~M.,  {Neugebauer} G.,   {Matthews} K.,  1993, \mn@doi [\aj]
  {10.1086/116782}, \href {http://adsabs.harvard.edu/abs/1993AJ....106.2005G}
  {106, 2005}

\bibitem[\protect\citeauthoryear{{Ginski}, {Mugrauer}, {Seeliger}  \&
  {Eisenbeiss}}{{Ginski} et~al.}{2012}]{Ginski2012}
{Ginski} C.,  {Mugrauer} M.,  {Seeliger} M.,   {Eisenbeiss} T.,  2012, \mn@doi
  [\mnras] {10.1111/j.1365-2966.2012.20485.x}, \href
  {http://adsabs.harvard.edu/abs/2012MNRAS.421.2498G} {421, 2498}

\bibitem[\protect\citeauthoryear{{Ginski} et~al.,}{{Ginski}
  et~al.}{2016}]{Ginski2016}
{Ginski} C.,  et~al., 2016, \mn@doi [\mnras] {10.1093/mnras/stw049}, \href
  {http://adsabs.harvard.edu/abs/2016MNRAS.457.2173G} {457, 2173}

\bibitem[\protect\citeauthoryear{Grether \& Lineweaver}{Grether \&
  Lineweaver}{2006}]{Grether2006}
Grether D.,  Lineweaver C.~H.,  2006, The Astrophysical Journal, 640, 1051

\bibitem[\protect\citeauthoryear{{Guenther}, {Hartmann}, {Esposito}, {Hatzes},
  {Cusano}  \& {Gandolfi}}{{Guenther} et~al.}{2009}]{Guenther2009}
{Guenther} E.~W.,  {Hartmann} M.,  {Esposito} M.,  {Hatzes} A.~P.,  {Cusano}
  F.,   {Gandolfi} D.,  2009, \mn@doi [\aap] {10.1051/0004-6361/200912112},
  \href {http://adsabs.harvard.edu/abs/2009A\%26A...507.1659G} {507, 1659}

\bibitem[\protect\citeauthoryear{{Guilloteau}, {Simon}, {Pi{\'e}tu}, {Di
  Folco}, {Dutrey}, {Prato}  \& {Chapillon}}{{Guilloteau}
  et~al.}{2014}]{Guilloteau2014}
{Guilloteau} S.,  {Simon} M.,  {Pi{\'e}tu} V.,  {Di Folco} E.,  {Dutrey} A.,
  {Prato} L.,   {Chapillon} E.,  2014, \mn@doi [\aap]
  {10.1051/0004-6361/201423765}, \href
  {http://adsabs.harvard.edu/abs/2014A%26A...567A.117G} {567, A117}

\bibitem[\protect\citeauthoryear{{Hall}, {Forgan}  \& {Rice}}{{Hall}
  et~al.}{2017}]{Hall2017}
{Hall} C.,  {Forgan} D.,   {Rice} K.,  2017, \mn@doi [\mnras]
  {10.1093/mnras/stx1244}, \href
  {http://adsabs.harvard.edu/abs/2017MNRAS.470.2517H} {470, 2517}

\bibitem[\protect\citeauthoryear{{Hellier} et~al.,}{{Hellier}
  et~al.}{2009}]{Hellier2009}
{Hellier} C.,  et~al., 2009, \mn@doi [\nat] {10.1038/nature08245}, \href
  {https://ui.adsabs.harvard.edu/#abs/2009Natur.460.1098H} {460, 1098}

\bibitem[\protect\citeauthoryear{{Herriot} et~al.,}{{Herriot}
  et~al.}{2000}]{Herriot2000}
{Herriot} G.,  et~al., 2000, in {Wizinowich} P.~L.,  ed.,  \procspie Vol. 4007,
  Adaptive Optical Systems Technology. pp 115--125, \mn@doi{10.1117/12.390288}

\bibitem[\protect\citeauthoryear{{Hodapp} et~al.,}{{Hodapp}
  et~al.}{2003}]{Hodapp2003}
{Hodapp} K.~W.,  et~al., 2003, \mn@doi [\pasp] {10.1086/379669}, \href
  {http://adsabs.harvard.edu/abs/2003PASP..115.1388H} {115, 1388}

\bibitem[\protect\citeauthoryear{{Horch}, {Veillette}, {Baena Gall{\'e}},
  {Shah}, {O'Rielly}  \& {van Altena}}{{Horch} et~al.}{2009}]{Horch2009}
{Horch} E.~P.,  {Veillette} D.~R.,  {Baena Gall{\'e}} R.,  {Shah} S.~C.,
  {O'Rielly} G.~V.,   {van Altena} W.~F.,  2009, \mn@doi [\aj]
  {10.1088/0004-6256/137/6/5057}, \href
  {http://adsabs.harvard.edu/abs/2009AJ....137.5057H} {137, 5057}

\bibitem[\protect\citeauthoryear{{Horch}, {Howell}, {Everett}  \&
  {Ciardi}}{{Horch} et~al.}{2012}]{Horch2012}
{Horch} E.~P.,  {Howell} S.~B.,  {Everett} M.~E.,   {Ciardi} D.~R.,  2012,
  \mn@doi [\aj] {10.1088/0004-6256/144/6/165}, \href
  {http://adsabs.harvard.edu/abs/2012AJ....144..165H} {144, 165}

\bibitem[\protect\citeauthoryear{{Hu{\ss}mann}, {Stolte}, {Brandner}, {Gennaro}
   \& {Liermann}}{{Hu{\ss}mann} et~al.}{2012}]{Hussmann12}
{Hu{\ss}mann} B.,  {Stolte} A.,  {Brandner} W.,  {Gennaro} M.,   {Liermann} A.,
   2012, \mn@doi [\aap] {10.1051/0004-6361/201117637}, \href
  {http://adsabs.harvard.edu/abs/2012A%26A...540A..57H} {540, A57}

\bibitem[\protect\citeauthoryear{{Irwin} et~al.,}{{Irwin}
  et~al.}{2010}]{Irwin2010}
{Irwin} J.,  et~al., 2010, \mn@doi [\apj] {10.1088/0004-637X/718/2/1353}, \href
  {https://ui.adsabs.harvard.edu/#abs/2010ApJ...718.1353I} {718, 1353}

\bibitem[\protect\citeauthoryear{{Jang-Condell}, {Mugrauer}  \&
  {Schmidt}}{{Jang-Condell} et~al.}{2008}]{Jang-Condell2008}
{Jang-Condell} H.,  {Mugrauer} M.,   {Schmidt} T.,  2008, \mn@doi [\apjl]
  {10.1086/591791}, \href {http://adsabs.harvard.edu/abs/2008ApJ...683L.191J}
  {683, L191}

\bibitem[\protect\citeauthoryear{{Jenkins} et~al.,}{{Jenkins}
  et~al.}{2017}]{Jenkins2017}
{Jenkins} J.~S.,  et~al., 2017, \mn@doi [\mnras] {10.1093/mnras/stw2811}, \href
  {http://adsabs.harvard.edu/abs/2017MNRAS.466..443J} {466, 443}

\bibitem[\protect\citeauthoryear{{Jensen} \& {Akeson}}{{Jensen} \&
  {Akeson}}{2003}]{Jensen2003}
{Jensen} E.~L.~N.,  {Akeson} R.~L.,  2003, \mn@doi [\apj] {10.1086/345719},
  \href {http://adsabs.harvard.edu/abs/2003ApJ...584..875J} {584, 875}

\bibitem[\protect\citeauthoryear{{Jofr{\'e}}, {Petrucci}, {Saffe}, {Saker}, {de
  la Villarmois}, {Chavero}, {G{\'o}mez}  \& {Mauas}}{{Jofr{\'e}}
  et~al.}{2015}]{Jofre2015}
{Jofr{\'e}} E.,  {Petrucci} R.,  {Saffe} C.,  {Saker} L.,  {de la Villarmois}
  E.~A.,  {Chavero} C.,  {G{\'o}mez} M.,   {Mauas} P.~J.~D.,  2015, \mn@doi
  [\aap] {10.1051/0004-6361/201424474}, \href
  {https://ui.adsabs.harvard.edu/#abs/2015A&A...574A..50J} {574, A50}

\bibitem[\protect\citeauthoryear{{Johansen}, {Oishi}, {Mac Low}, {Klahr},
  {Henning}  \& {Youdin}}{{Johansen} et~al.}{2007}]{Johansen2007}
{Johansen} A.,  {Oishi} J.~S.,  {Mac Low} M.-M.,  {Klahr} H.,  {Henning} T.,
  {Youdin} A.,  2007, \mn@doi [\nat] {10.1038/nature06086}, \href
  {http://adsabs.harvard.edu/abs/2007Natur.448.1022J} {448, 1022}

\bibitem[\protect\citeauthoryear{{Johns-Krull} et~al.,}{{Johns-Krull}
  et~al.}{2016}]{Johns-Krull2016}
{Johns-Krull} C.~M.,  et~al., 2016, \mn@doi [\apj]
  {10.3847/0004-637X/826/2/206}, \href
  {http://adsabs.harvard.edu/abs/2016ApJ...826..206J} {826, 206}

\bibitem[\protect\citeauthoryear{{Johnson} et~al.,}{{Johnson}
  et~al.}{2011}]{Johnson2011}
{Johnson} J.~A.,  et~al., 2011, \mn@doi [\apj] {10.1088/0004-637X/735/1/24},
  \href {http://adsabs.harvard.edu/abs/2011ApJ...735...24J} {735, 24}

\bibitem[\protect\citeauthoryear{{Johnson}, {Cochran}, {Albrecht},
  {Dodson-Robinson}, {Winn}  \& {Gullikson}}{{Johnson}
  et~al.}{2014}]{Johnson2014}
{Johnson} M.~C.,  {Cochran} W.~D.,  {Albrecht} S.,  {Dodson-Robinson} S.~E.,
  {Winn} J.~N.,   {Gullikson} K.,  2014, \mn@doi [\apj]
  {10.1088/0004-637X/790/1/30}, \href
  {https://ui.adsabs.harvard.edu/#abs/2014ApJ...790...30J} {790, 30}

\bibitem[\protect\citeauthoryear{{Jones}, {Jenkins}, {Rojo}, {Melo}  \&
  {Bluhm}}{{Jones} et~al.}{2013}]{Jones2013}
{Jones} M.~I.,  {Jenkins} J.~S.,  {Rojo} P.,  {Melo} C.~H.~F.,   {Bluhm} P.,
  2013, \mn@doi [\aap] {10.1051/0004-6361/201321660}, \href
  {https://ui.adsabs.harvard.edu/#abs/2013A&A...556A..78J} {556, A78}

\bibitem[\protect\citeauthoryear{{Jones}, {Jenkins}, {Bluhm}, {Rojo}  \&
  {Melo}}{{Jones} et~al.}{2014}]{Jones2014}
{Jones} M.~I.,  {Jenkins} J.~S.,  {Bluhm} P.,  {Rojo} P.,   {Melo} C.~H.~F.,
  2014, \mn@doi [\aap] {10.1051/0004-6361/201323345}, \href
  {https://ui.adsabs.harvard.edu/#abs/2014A&A...566A.113J} {566, A113}

\bibitem[\protect\citeauthoryear{{Jones} et~al.,}{{Jones}
  et~al.}{2015}]{Jones2015}
{Jones} J.,  et~al., 2015, \mn@doi [\apj] {10.1088/0004-637X/813/1/58}, \href
  {https://ui.adsabs.harvard.edu/#abs/2015ApJ...813...58J} {813, 58}

\bibitem[\protect\citeauthoryear{{Kaib}, {Raymond}  \& {Duncan}}{{Kaib}
  et~al.}{2013}]{Kaib2013}
{Kaib} N.~A.,  {Raymond} S.~N.,   {Duncan} M.,  2013, \mn@doi [\nat]
  {10.1038/nature11780}, \href
  {http://adsabs.harvard.edu/abs/2013Natur.493..381K} {493, 381}

\bibitem[\protect\citeauthoryear{{Kane} et~al.,}{{Kane}
  et~al.}{2011a}]{Kane2011a}
{Kane} S.~R.,  et~al., 2011a, \mn@doi [\apj] {10.1088/0004-637X/733/1/28},
  \href {https://ui.adsabs.harvard.edu/#abs/2011ApJ...733...28K} {733, 28}

\bibitem[\protect\citeauthoryear{{Kane}, {Henry}, {Dragomir}, {Fischer},
  {Howard}, {Wang}  \& {Wright}}{{Kane} et~al.}{2011b}]{Kane2011b}
{Kane} S.~R.,  {Henry} G.~W.,  {Dragomir} D.,  {Fischer} D.~A.,  {Howard}
  A.~W.,  {Wang} X.,   {Wright} J.~T.,  2011b, \mn@doi [\apj]
  {10.1088/2041-8205/735/2/L41}, \href
  {https://ui.adsabs.harvard.edu/#abs/2011ApJ...735L..41K} {735, L41}

\bibitem[\protect\citeauthoryear{{Kane} et~al.,}{{Kane}
  et~al.}{2015}]{Kane2015}
{Kane} S.~R.,  et~al., 2015, \mn@doi [\apj] {10.1088/0004-637X/815/1/32}, \href
  {https://ui.adsabs.harvard.edu/#abs/2015ApJ...815...32K} {815, 32}

\bibitem[\protect\citeauthoryear{{Kley}}{{Kley}}{2001}]{Kley2001}
{Kley} W.,  2001, in {Zinnecker} H.,  {Mathieu} R.,  eds,  IAU Symposium Vol.
  200, The Formation of Binary Stars. p.~511

\bibitem[\protect\citeauthoryear{{Knutson} et~al.,}{{Knutson}
  et~al.}{2014}]{Knutson2014}
{Knutson} H.~A.,  et~al., 2014, \mn@doi [\apj] {10.1088/0004-637X/785/2/126},
  \href {https://ui.adsabs.harvard.edu/#abs/2014ApJ...785..126K} {785, 126}

\bibitem[\protect\citeauthoryear{{Kozai}}{{Kozai}}{1962}]{Kozai1962}
{Kozai} Y.,  1962, \mn@doi [\aj] {10.1086/108790}, \href
  {http://adsabs.harvard.edu/abs/1962AJ.....67..591K} {67, 591}

\bibitem[\protect\citeauthoryear{{Kratter}, {Murray-Clay}  \&
  {Youdin}}{{Kratter} et~al.}{2010}]{Kratter2010}
{Kratter} K.~M.,  {Murray-Clay} R.~A.,   {Youdin} A.~N.,  2010, \mn@doi [\apj]
  {10.1088/0004-637X/710/2/1375}, \href
  {http://adsabs.harvard.edu/abs/2010ApJ...710.1375K} {710, 1375}

\bibitem[\protect\citeauthoryear{{Kraus}, {Ireland}, {Hillenbrand}  \&
  {Martinache}}{{Kraus} et~al.}{2012}]{Kraus2012}
{Kraus} A.~L.,  {Ireland} M.~J.,  {Hillenbrand} L.~A.,   {Martinache} F.,
  2012, \mn@doi [\apj] {10.1088/0004-637X/745/1/19}, \href
  {http://adsabs.harvard.edu/abs/2012ApJ...745...19K} {745, 19}

\bibitem[\protect\citeauthoryear{{Kraus}, {Ireland}, {Huber}, {Mann}  \&
  {Dupuy}}{{Kraus} et~al.}{2016}]{Kraus2016}
{Kraus} A.~L.,  {Ireland} M.~J.,  {Huber} D.,  {Mann} A.~W.,   {Dupuy} T.~J.,
  2016, \mn@doi [\aj] {10.3847/0004-6256/152/1/8}, \href
  {https://ui.adsabs.harvard.edu/#abs/2016AJ....152....8K} {152, 8}

\bibitem[\protect\citeauthoryear{{Lafreni{\`e}re}, {Jayawardhana}  \& {van
  Kerkwijk}}{{Lafreni{\`e}re} et~al.}{2008}]{Lafreniere2008}
{Lafreni{\`e}re} D.,  {Jayawardhana} R.,   {van Kerkwijk} M.~H.,  2008, \mn@doi
  [\apjl] {10.1086/595870}, \href
  {http://adsabs.harvard.edu/abs/2008ApJ...689L.153L} {689, L153}

\bibitem[\protect\citeauthoryear{{Lafreni{\`e}re}, {Jayawardhana}, {van
  Kerkwijk}, {Brandeker}  \& {Janson}}{{Lafreni{\`e}re}
  et~al.}{2014}]{Lafreniere2014}
{Lafreni{\`e}re} D.,  {Jayawardhana} R.,  {van Kerkwijk} M.~H.,  {Brandeker}
  A.,   {Janson} M.,  2014, \mn@doi [\apj] {10.1088/0004-637X/785/1/47}, \href
  {http://adsabs.harvard.edu/abs/2014ApJ...785...47L} {785, 47}

\bibitem[\protect\citeauthoryear{{Lambrechts} \& {Johansen}}{{Lambrechts} \&
  {Johansen}}{2014}]{Lambrechts2014}
{Lambrechts} M.,  {Johansen} A.,  2014, \mn@doi [\aap]
  {10.1051/0004-6361/201424343}, \href
  {http://adsabs.harvard.edu/abs/2014A%26A...572A.107L} {572, A107}

\bibitem[\protect\citeauthoryear{{Law} et~al.,}{{Law} et~al.}{2014}]{Law2014}
{Law} N.~M.,  et~al., 2014, \mn@doi [\apj] {10.1088/0004-637X/791/1/35}, \href
  {http://adsabs.harvard.edu/abs/2014ApJ...791...35L} {791, 35}

\bibitem[\protect\citeauthoryear{{Lenzen} et~al.,}{{Lenzen}
  et~al.}{2003}]{Lenzen2003}
{Lenzen} R.,  et~al., 2003, in {Iye} M.,  {Moorwood} A.~F.~M.,  eds,  \procspie
  Vol. 4841, Instrument Design and Performance for Optical/Infrared
  Ground-based Telescopes. pp 944--952, \mn@doi{10.1117/12.460044}

\bibitem[\protect\citeauthoryear{{Lewis} et~al.,}{{Lewis}
  et~al.}{2013}]{Lewis2013}
{Lewis} N.~K.,  et~al., 2013, \mn@doi [\apj] {10.1088/0004-637X/766/2/95},
  \href {https://ui.adsabs.harvard.edu/#abs/2013ApJ...766...95L} {766, 95}

\bibitem[\protect\citeauthoryear{{Lidov}}{{Lidov}}{1962}]{Lidov1962}
{Lidov} M.~L.,  1962, \mn@doi [\planss] {10.1016/0032-0633(62)90129-0}, \href
  {http://adsabs.harvard.edu/abs/1962P%26SS....9..719L} {9, 719}

\bibitem[\protect\citeauthoryear{{Lindegren} et~al.,}{{Lindegren}
  et~al.}{2018}]{Lindegren2018}
{Lindegren} L.,  et~al., 2018, \mn@doi [\aap] {10.1051/0004-6361/201832727},
  \href {http://adsabs.harvard.edu/abs/2018A\%26A...616A...2L} {616, A2}

\bibitem[\protect\citeauthoryear{{Liu} et~al.,}{{Liu} et~al.}{2008}]{Liu2008}
{Liu} Y.~J.,  et~al., 2008, \mn@doi [\apj] {10.1086/523297}, \href
  {https://ui.adsabs.harvard.edu/#abs/2008ApJ...672..553L} {672, 553}

\bibitem[\protect\citeauthoryear{{Lowrance}, {Kirkpatrick}  \&
  {Beichman}}{{Lowrance} et~al.}{2002}]{Lowrance2002}
{Lowrance} P.~J.,  {Kirkpatrick} J.~D.,   {Beichman} C.~A.,  2002, \mn@doi
  [\apjl] {10.1086/341554}, \href
  {http://adsabs.harvard.edu/abs/2002ApJ...572L..79L} {572, L79}

\bibitem[\protect\citeauthoryear{Ma \& Ge}{Ma \& Ge}{2014}]{Ma2014}
Ma B.,  Ge J.,  2014, \mn@doi [Monthly Notices of the Royal Astronomical
  Society] {10.1093/mnras/stu134}, 439, 2781

\bibitem[\protect\citeauthoryear{{Ma} et~al.,}{{Ma} et~al.}{2016}]{Ma2016}
{Ma} B.,  et~al., 2016, \mn@doi [\aj] {10.3847/0004-6256/152/5/112}, \href
  {http://adsabs.harvard.edu/abs/2016AJ....152..112M} {152, 112}

\bibitem[\protect\citeauthoryear{{Makarov} \& {Kaplan}}{{Makarov} \&
  {Kaplan}}{2005}]{Makarov2005}
{Makarov} V.~V.,  {Kaplan} G.~H.,  2005, \mn@doi [\aj] {10.1086/429590}, \href
  {http://adsabs.harvard.edu/abs/2005AJ....129.2420M} {129, 2420}

\bibitem[\protect\citeauthoryear{{Maldonado} \& {Villaver}}{{Maldonado} \&
  {Villaver}}{2017}]{Manaldo2017}
{Maldonado} J.,  {Villaver} E.,  2017, \mn@doi [\aap]
  {10.1051/0004-6361/201630120}, \href
  {https://ui.adsabs.harvard.edu/#abs/2017A&A...602A..38M} {602, A38}

\bibitem[\protect\citeauthoryear{{Marcy} \& {Butler}}{{Marcy} \&
  {Butler}}{2000}]{Marcy2000}
{Marcy} G.~W.,  {Butler} R.~P.,  2000, \mn@doi [Publications of the
  Astronomical Society of the Pacific] {10.1086/316516}, \href
  {https://ui.adsabs.harvard.edu/\#abs/2000PASP..112..137M} {112, 137}

\bibitem[\protect\citeauthoryear{{Mardling} \& {Lin}}{{Mardling} \&
  {Lin}}{2002}]{Mardling2002}
{Mardling} R.~A.,  {Lin} D.~N.~C.,  2002, \mn@doi [\apj] {10.1086/340752},
  \href {http://adsabs.harvard.edu/abs/2002ApJ...573..829M} {573, 829}

\bibitem[\protect\citeauthoryear{{Masana}, {Jordi}  \& {Ribas}}{{Masana}
  et~al.}{2006}]{Masana2006}
{Masana} E.,  {Jordi} C.,   {Ribas} I.,  2006, \mn@doi [\aap]
  {10.1051/0004-6361:20054021}, \href
  {http://adsabs.harvard.edu/abs/2006A\%26A...450..735M} {450, 735}

\bibitem[\protect\citeauthoryear{{Mason}, {Wycoff}, {Hartkopf}, {Douglass}  \&
  {Worley}}{{Mason} et~al.}{2001}]{Mason2001}
{Mason} B.~D.,  {Wycoff} G.~L.,  {Hartkopf} W.~I.,  {Douglass} G.~G.,
  {Worley} C.~E.,  2001, \mn@doi [\aj] {10.1086/323920}, \href
  {http://adsabs.harvard.edu/abs/2001AJ....122.3466M} {122, 3466}

\bibitem[\protect\citeauthoryear{{Matsuo}, {Shibai}, {Ootsubo}  \&
  {Tamura}}{{Matsuo} et~al.}{2007}]{Matsuo2007}
{Matsuo} T.,  {Shibai} H.,  {Ootsubo} T.,   {Tamura} M.,  2007, \mn@doi [\apj]
  {10.1086/517964}, \href {http://adsabs.harvard.edu/abs/2007ApJ...662.1282M}
  {662, 1282}

\bibitem[\protect\citeauthoryear{{Mayer}, {Wadsley}, {Quinn}  \&
  {Stadel}}{{Mayer} et~al.}{2005}]{Mayer2005}
{Mayer} L.,  {Wadsley} J.,  {Quinn} T.,   {Stadel} J.,  2005, \mn@doi [\mnras]
  {10.1111/j.1365-2966.2005.09468.x}, \href
  {http://adsabs.harvard.edu/abs/2005MNRAS.363..641M} {363, 641}

\bibitem[\protect\citeauthoryear{{McAlister}, {Hartkopf}, {Hutter}  \&
  {Franz}}{{McAlister} et~al.}{1987}]{McAlister1987}
{McAlister} H.~A.,  {Hartkopf} W.~I.,  {Hutter} D.~J.,   {Franz} O.~G.,  1987,
  \mn@doi [\aj] {10.1086/114353}, \href
  {http://adsabs.harvard.edu/abs/1987AJ.....93..688M} {93, 688}

\bibitem[\protect\citeauthoryear{{McArthur}, {Benedict}, {Barnes}, {Martioli},
  {Korzennik}, {Nelan}  \& {Butler}}{{McArthur} et~al.}{2010}]{McArthur2010}
{McArthur} B.~E.,  {Benedict} G.~F.,  {Barnes} R.,  {Martioli} E.,  {Korzennik}
  S.,  {Nelan} E.,   {Butler} R.~P.,  2010, \mn@doi [\apj]
  {10.1088/0004-637X/715/2/1203}, \href
  {http://adsabs.harvard.edu/abs/2010ApJ...715.1203M} {715, 1203}

\bibitem[\protect\citeauthoryear{{Mitchell}, {Reffert}, {Trifonov},
  {Quirrenbach}  \& {Fischer}}{{Mitchell} et~al.}{2013}]{Mitchell2013}
{Mitchell} D.~S.,  {Reffert} S.,  {Trifonov} T.,  {Quirrenbach} A.,   {Fischer}
  D.~A.,  2013, \mn@doi [\aap] {10.1051/0004-6361/201321714}, \href
  {http://adsabs.harvard.edu/abs/2013A\%26A...555A..87M} {555, A87}

\bibitem[\protect\citeauthoryear{{Morbey} \& {Brosterhus}}{{Morbey} \&
  {Brosterhus}}{1974}]{Morbey1974}
{Morbey} C.~L.,  {Brosterhus} E.~B.,  1974, \mn@doi [\pasp] {10.1086/129630},
  \href {http://adsabs.harvard.edu/abs/1974PASP...86..455M} {86, 455}

\bibitem[\protect\citeauthoryear{{Mordasini}}{{Mordasini}}{2018}]{Mordasini2018}
{Mordasini} C.,  2018, {Planetary Population Synthesis}.
p.~143, \mn@doi{10.1007/978-3-319-55333-7_143}

\bibitem[\protect\citeauthoryear{{Mordasini}, {Alibert}, {Benz}, {Klahr}  \&
  {Henning}}{{Mordasini} et~al.}{2012}]{Mordasini2012}
{Mordasini} C.,  {Alibert} Y.,  {Benz} W.,  {Klahr} H.,   {Henning} T.,  2012,
  \mn@doi [\aap] {10.1051/0004-6361/201117350}, \href
  {http://adsabs.harvard.edu/abs/2012A\%26A...541A..97M} {541, A97}

\bibitem[\protect\citeauthoryear{{Morton}, {Bryson}, {Coughlin}, {Rowe},
  {Ravichandran}, {Petigura}, {Haas}  \& {Batalha}}{{Morton}
  et~al.}{2016}]{Morton2016}
{Morton} T.~D.,  {Bryson} S.~T.,  {Coughlin} J.~L.,  {Rowe} J.~F.,
  {Ravichandran} G.,  {Petigura} E.~A.,  {Haas} M.~R.,   {Batalha} N.~M.,
  2016, \mn@doi [\apj] {10.3847/0004-637X/822/2/86}, \href
  {https://ui.adsabs.harvard.edu/#abs/2016ApJ...822...86M} {822, 86}

\bibitem[\protect\citeauthoryear{{Moutou}, {Vigan}, {Mesa}, {Desidera},
  {Th{\'e}bault}, {Zurlo}  \& {Salter}}{{Moutou} et~al.}{2017}]{Moutou2017}
{Moutou} C.,  {Vigan} A.,  {Mesa} D.,  {Desidera} S.,  {Th{\'e}bault} P.,
  {Zurlo} A.,   {Salter} G.,  2017, \mn@doi [\aap]
  {10.1051/0004-6361/201630173}, \href
  {http://adsabs.harvard.edu/abs/2017A\%26A...602A..87M} {602, A87}

\bibitem[\protect\citeauthoryear{{Mugrauer} \& {Neuh{\"a}user}}{{Mugrauer} \&
  {Neuh{\"a}user}}{2009}]{Mugrauer2009}
{Mugrauer} M.,  {Neuh{\"a}user} R.,  2009, \mn@doi [\aap]
  {10.1051/0004-6361:200810639}, \href
  {http://adsabs.harvard.edu/abs/2009A%26A...494..373M} {494, 373}

\bibitem[\protect\citeauthoryear{{Mugrauer}, {Neuh{\"a}user}, {Mazeh},
  {Guenther}  \& {Fern{\'a}ndez}}{{Mugrauer} et~al.}{2004}]{Mugrauer2004}
{Mugrauer} M.,  {Neuh{\"a}user} R.,  {Mazeh} T.,  {Guenther} E.,
  {Fern{\'a}ndez} M.,  2004, \mn@doi [Astronomische Nachrichten]
  {10.1002/asna.200410252}, \href
  {https://ui.adsabs.harvard.edu/\#abs/2004AN....325..718M} {325, 718}

\bibitem[\protect\citeauthoryear{{Mugrauer}, {Neuh{\"a}user}, {Mazeh},
  {Guenther}, {Fern{\'a}ndez}  \& {Broeg}}{{Mugrauer}
  et~al.}{2006}]{Mugrauer2006}
{Mugrauer} M.,  {Neuh{\"a}user} R.,  {Mazeh} T.,  {Guenther} E.,
  {Fern{\'a}ndez} M.,   {Broeg} C.,  2006, \mn@doi [Astronomische Nachrichten]
  {10.1002/asna.200510528}, \href
  {http://adsabs.harvard.edu/abs/2006AN....327..321M} {327, 321}

\bibitem[\protect\citeauthoryear{{Mugrauer}, {Neuh{\"a}user}  \&
  {Mazeh}}{{Mugrauer} et~al.}{2007}]{Mugrauer2007}
{Mugrauer} M.,  {Neuh{\"a}user} R.,   {Mazeh} T.,  2007, \mn@doi [\aap]
  {10.1051/0004-6361:20065883}, \href
  {http://adsabs.harvard.edu/abs/2007A\%26A...469..755M} {469, 755}

\bibitem[\protect\citeauthoryear{{Naoz}, {Farr}  \& {Rasio}}{{Naoz}
  et~al.}{2012}]{Naoz2012}
{Naoz} S.,  {Farr} W.~M.,   {Rasio} F.~A.,  2012, \mn@doi [\apjl]
  {10.1088/2041-8205/754/2/L36}, \href
  {http://adsabs.harvard.edu/abs/2012ApJ...754L..36N} {754, L36}

\bibitem[\protect\citeauthoryear{{Nayakshin}}{{Nayakshin}}{2010}]{Nayakshin2010}
{Nayakshin} S.,  2010, \mn@doi [\mnras] {10.1111/j.1745-3933.2010.00923.x},
  \href {http://adsabs.harvard.edu/abs/2010MNRAS.408L..36N} {408, L36}

\bibitem[\protect\citeauthoryear{{Ngo} et~al.,}{{Ngo} et~al.}{2015}]{Ngo2015}
{Ngo} H.,  et~al., 2015, \mn@doi [\apj] {10.1088/0004-637X/800/2/138}, \href
  {http://adsabs.harvard.edu/abs/2015ApJ...800..138N} {800, 138}

\bibitem[\protect\citeauthoryear{{Ngo} et~al.,}{{Ngo} et~al.}{2016}]{Ngo2016}
{Ngo} H.,  et~al., 2016, \mn@doi [\apj] {10.3847/0004-637X/827/1/8}, \href
  {http://adsabs.harvard.edu/abs/2016ApJ...827....8N} {827, 8}

\bibitem[\protect\citeauthoryear{{Nowak} et~al.,}{{Nowak}
  et~al.}{2017}]{Nowak2017}
{Nowak} G.,  et~al., 2017, \mn@doi [\aj] {10.3847/1538-3881/aa5cb6}, \href
  {http://adsabs.harvard.edu/abs/2017AJ....153..131N} {153, 131}

\bibitem[\protect\citeauthoryear{{P{\'a}l} et~al.,}{{P{\'a}l}
  et~al.}{2010}]{Pal2010}
{P{\'a}l} A.,  et~al., 2010, \mn@doi [\mnras]
  {10.1111/j.1365-2966.2009.15849.x}, \href
  {https://ui.adsabs.harvard.edu/#abs/2010MNRAS.401.2665P} {401, 2665}

\bibitem[\protect\citeauthoryear{{Patience} et~al.,}{{Patience}
  et~al.}{2002}]{Patience2002}
{Patience} J.,  et~al., 2002, \mn@doi [\apj] {10.1086/342982}, \href
  {http://adsabs.harvard.edu/abs/2002ApJ...581..654P} {581, 654}

\bibitem[\protect\citeauthoryear{{Petrovich}}{{Petrovich}}{2015}]{Petrovich2015}
{Petrovich} C.,  2015, \mn@doi [\apj] {10.1088/0004-637X/799/1/27}, \href
  {http://adsabs.harvard.edu/abs/2015ApJ...799...27P} {799, 27}

\bibitem[\protect\citeauthoryear{{Pichardo}, {Sparke}  \& {Aguilar}}{{Pichardo}
  et~al.}{2005}]{Pichardo2005}
{Pichardo} B.,  {Sparke} L.~S.,   {Aguilar} L.~A.,  2005, \mn@doi [\mnras]
  {10.1111/j.1365-2966.2005.08905.x}, \href
  {http://adsabs.harvard.edu/abs/2005MNRAS.359..521P} {359, 521}

\bibitem[\protect\citeauthoryear{{Pinfield}, {Jones}, {Lucas}, {Kendall},
  {Folkes}, {Day-Jones}, {Chappelle}  \& {Steele}}{{Pinfield}
  et~al.}{2006}]{Pinfield2006}
{Pinfield} D.~J.,  {Jones} H.~R.~A.,  {Lucas} P.~W.,  {Kendall} T.~R.,
  {Folkes} S.~L.,  {Day-Jones} A.~C.,  {Chappelle} R.~J.,   {Steele} I.~A.,
  2006, \mn@doi [\mnras] {10.1111/j.1365-2966.2006.10213.x}, \href
  {http://adsabs.harvard.edu/abs/2006MNRAS.368.1281P} {368, 1281}

\bibitem[\protect\citeauthoryear{{Piskorz}, {Knutson}, {Ngo}, {Muirhead},
  {Batygin}, {Crepp}, {Hinkley}  \& {Morton}}{{Piskorz}
  et~al.}{2015}]{Piskorz2015}
{Piskorz} D.,  {Knutson} H.~A.,  {Ngo} H.,  {Muirhead} P.~S.,  {Batygin} K.,
  {Crepp} J.~R.,  {Hinkley} S.,   {Morton} T.~D.,  2015, \mn@doi [\apj]
  {10.1088/0004-637X/814/2/148}, \href
  {http://adsabs.harvard.edu/abs/2015ApJ...814..148P} {814, 148}

\bibitem[\protect\citeauthoryear{{Pollack}, {Hubickyj}, {Bodenheimer},
  {Lissauer}, {Podolak}  \& {Greenzweig}}{{Pollack} et~al.}{1996}]{Pollack1996}
{Pollack} J.~B.,  {Hubickyj} O.,  {Bodenheimer} P.,  {Lissauer} J.~J.,
  {Podolak} M.,   {Greenzweig} Y.,  1996, \mn@doi [\icarus]
  {10.1006/icar.1996.0190}, \href
  {http://adsabs.harvard.edu/abs/1996Icar..124...62P} {124, 62}

\bibitem[\protect\citeauthoryear{{Prato}, {Greene}  \& {Simon}}{{Prato}
  et~al.}{2003}]{Prato2003}
{Prato} L.,  {Greene} T.~P.,   {Simon} M.,  2003, \mn@doi [\apj]
  {10.1086/345828}, \href {http://adsabs.harvard.edu/abs/2003ApJ...584..853P}
  {584, 853}

\bibitem[\protect\citeauthoryear{{Rafikov}}{{Rafikov}}{2005}]{Rafikov2005}
{Rafikov} R.~R.,  2005, \mn@doi [\apjl] {10.1086/428899}, \href
  {http://adsabs.harvard.edu/abs/2005ApJ...621L..69R} {621, L69}

\bibitem[\protect\citeauthoryear{{Rafikov}}{{Rafikov}}{2013}]{Rafikov2013}
{Rafikov} R.~R.,  2013, \mn@doi [\apjl] {10.1088/2041-8205/765/1/L8}, \href
  {http://adsabs.harvard.edu/abs/2013ApJ...765L...8R} {765, L8}

\bibitem[\protect\citeauthoryear{{Raghavan}, {Henry}, {Mason}, {Subasavage},
  {Jao}, {Beaulieu}  \& {Hambly}}{{Raghavan} et~al.}{2006}]{Raghavan2006}
{Raghavan} D.,  {Henry} T.~J.,  {Mason} B.~D.,  {Subasavage} J.~P.,  {Jao}
  W.-C.,  {Beaulieu} T.~D.,   {Hambly} N.~C.,  2006, \mn@doi [\apj]
  {10.1086/504823}, \href {http://adsabs.harvard.edu/abs/2006ApJ...646..523R}
  {646, 523}

\bibitem[\protect\citeauthoryear{{Raghavan} et~al.,}{{Raghavan}
  et~al.}{2010}]{Raghavan2010}
{Raghavan} D.,  et~al., 2010, \mn@doi [\apjs] {10.1088/0067-0049/190/1/1},
  \href {http://adsabs.harvard.edu/abs/2010ApJS..190....1R} {190, 1}

\bibitem[\protect\citeauthoryear{{Reboussin}, {Guilloteau}, {Simon}, {Grosso},
  {Wakelam}, {Di Folco}, {Dutrey}  \& {Pi{\'e}tu}}{{Reboussin}
  et~al.}{2015}]{Reboussin2015}
{Reboussin} L.,  {Guilloteau} S.,  {Simon} M.,  {Grosso} N.,  {Wakelam} V.,
  {Di Folco} E.,  {Dutrey} A.,   {Pi{\'e}tu} V.,  2015, \mn@doi [\aap]
  {10.1051/0004-6361/201525705}, \href
  {http://adsabs.harvard.edu/abs/2015A\%26A...578A..31R} {578, A31}

\bibitem[\protect\citeauthoryear{{Rice} \& {Armitage}}{{Rice} \&
  {Armitage}}{2009}]{Rice2009}
{Rice} W.~K.~M.,  {Armitage} P.~J.,  2009, \mn@doi [\mnras]
  {10.1111/j.1365-2966.2009.14879.x}, \href
  {http://adsabs.harvard.edu/abs/2009MNRAS.396.2228R} {396, 2228}

\bibitem[\protect\citeauthoryear{{Rice}, {Veljanoski}  \& {Collier
  Cameron}}{{Rice} et~al.}{2012}]{Rice2012}
{Rice} W.~K.~M.,  {Veljanoski} J.,   {Collier Cameron} A.,  2012, \mn@doi
  [\mnras] {10.1111/j.1365-2966.2012.21728.x}, \href
  {http://adsabs.harvard.edu/abs/2012MNRAS.425.2567R} {425, 2567}

\bibitem[\protect\citeauthoryear{{Rice}, {Lopez}, {Forgan}  \& {Biller}}{{Rice}
  et~al.}{2015}]{Rice2015}
{Rice} K.,  {Lopez} E.,  {Forgan} D.,   {Biller} B.,  2015, \mn@doi [\mnras]
  {10.1093/mnras/stv1997}, \href
  {http://adsabs.harvard.edu/abs/2015MNRAS.454.1940R} {454, 1940}

\bibitem[\protect\citeauthoryear{{Riddle} et~al.,}{{Riddle}
  et~al.}{2015}]{Riddle2015}
{Riddle} R.~L.,  et~al., 2015, \mn@doi [\apj] {10.1088/0004-637X/799/1/4},
  \href {http://adsabs.harvard.edu/abs/2015ApJ...799....4R} {799, 4}

\bibitem[\protect\citeauthoryear{{Roberts} \& {Mason}}{{Roberts} \&
  {Mason}}{2018}]{Roberts2018}
{Roberts} L.~C.,  {Mason} B.~D.,  2018, \mn@doi [\mnras]
  {10.1093/mnras/stx2559}, \href
  {http://adsabs.harvard.edu/abs/2018MNRAS.473.4497R} {473, 4497}

\bibitem[\protect\citeauthoryear{{Roberts}, {Turner}, {ten Brummelaar}, {Mason}
   \& {Hartkopf}}{{Roberts} et~al.}{2011}]{Roberts2011}
{Roberts} Jr. L.~C.,  {Turner} N.~H.,  {ten Brummelaar} T.~A.,  {Mason} B.~D.,
   {Hartkopf} W.~I.,  2011, \mn@doi [\aj] {10.1088/0004-6256/142/5/175}, \href
  {http://adsabs.harvard.edu/abs/2011AJ....142..175R} {142, 175}

\bibitem[\protect\citeauthoryear{{Roberts}, {Tokovinin}, {Mason}, {Riddle},
  {Hartkopf}, {Law}  \& {Baranec}}{{Roberts} et~al.}{2015}]{Roberts2015}
{Roberts} Jr. L.~C.,  {Tokovinin} A.,  {Mason} B.~D.,  {Riddle} R.~L.,
  {Hartkopf} W.~I.,  {Law} N.~M.,   {Baranec} C.,  2015, \mn@doi [\aj]
  {10.1088/0004-6256/149/4/118}, \href
  {http://adsabs.harvard.edu/abs/2015AJ....149..118R} {149, 118}

\bibitem[\protect\citeauthoryear{{Roell}, {Neuh{\"a}user}, {Seifahrt}  \&
  {Mugrauer}}{{Roell} et~al.}{2012}]{Roell2012}
{Roell} T.,  {Neuh{\"a}user} R.,  {Seifahrt} A.,   {Mugrauer} M.,  2012,
  \mn@doi [\aap] {10.1051/0004-6361/201118051}, \href
  {http://adsabs.harvard.edu/abs/2012A\%26A...542A..92R} {542, A92}

\bibitem[\protect\citeauthoryear{{Rousset} et~al.,}{{Rousset}
  et~al.}{2003}]{Rousset2003}
{Rousset} G.,  et~al., 2003, in {Wizinowich} P.~L.,  {Bonaccini} D.,  eds,
  \procspie Vol. 4839, Adaptive Optical System Technologies II. pp 140--149,
  \mn@doi{10.1117/12.459332}

\bibitem[\protect\citeauthoryear{{Santerne} et~al.,}{{Santerne}
  et~al.}{2012}]{Santerne2012}
{Santerne} A.,  et~al., 2012, \mn@doi [\aap] {10.1051/0004-6361/201219899},
  \href {https://ui.adsabs.harvard.edu/#abs/2012A&A...544L..12S} {544, L12}

\bibitem[\protect\citeauthoryear{{Santos} et~al.,}{{Santos}
  et~al.}{2002}]{Santos2002}
{Santos} N.~C.,  et~al., 2002, \mn@doi [\aap] {10.1051/0004-6361:20020876},
  \href {https://ui.adsabs.harvard.edu/#abs/2002A&A...392..215S} {392, 215}

\bibitem[\protect\citeauthoryear{{Sato} et~al.,}{{Sato}
  et~al.}{2008}]{Sato2008}
{Sato} B.,  et~al., 2008, \mn@doi [Publications of the Astronomical Society of
  Japan] {10.1093/pasj/60.3.539}, \href
  {https://ui.adsabs.harvard.edu/#abs/2008PASJ...60..539S} {60, 539}

\bibitem[\protect\citeauthoryear{{Sato} et~al.,}{{Sato}
  et~al.}{2010}]{Sato2010}
{Sato} B.,  et~al., 2010, \mn@doi [Publications of the Astronomical Society of
  Japan] {10.1093/pasj/62.4.1063}, \href
  {https://ui.adsabs.harvard.edu/#abs/2010PASJ...62.1063S} {62, 1063}

\bibitem[\protect\citeauthoryear{{Schlaufman}}{{Schlaufman}}{2018}]{Schlaufman2018}
{Schlaufman} K.~C.,  2018, \mn@doi [\apj] {10.3847/1538-4357/aa961c}, \href
  {http://adsabs.harvard.edu/abs/2018ApJ...853...37S} {853, 37}

\bibitem[\protect\citeauthoryear{{Schmidt-Kaler}}{{Schmidt-Kaler}}{1982}]{Schmidt-Kaler1982}
{Schmidt-Kaler} T.,  1982, Bulletin d'Information du Centre de Donnees
  Stellaires, \href {http://adsabs.harvard.edu/abs/1982BICDS..23....2S} {23, 2}

\bibitem[\protect\citeauthoryear{{Scott}, {Howell}, {Horch}  \&
  {Everett}}{{Scott} et~al.}{2018}]{Scott2018}
{Scott} N.~J.,  {Howell} S.~B.,  {Horch} E.~P.,   {Everett} M.~E.,  2018,
  \mn@doi [\pasp] {10.1088/1538-3873/aab484}, \href
  {http://adsabs.harvard.edu/abs/2018PASP..130e4502S} {130, 054502}

\bibitem[\protect\citeauthoryear{{Shatsky}}{{Shatsky}}{2001}]{Shatsky2001}
{Shatsky} N.,  2001, \mn@doi [\aap] {10.1051/0004-6361:20011401}, \href
  {https://ui.adsabs.harvard.edu/#abs/2001A&A...380..238S} {380, 238}

\bibitem[\protect\citeauthoryear{{Shaya} \& {Olling}}{{Shaya} \&
  {Olling}}{2011}]{Shaya2011}
{Shaya} E.~J.,  {Olling} R.~P.,  2011, \mn@doi [\apjs]
  {10.1088/0067-0049/192/1/2}, \href
  {http://adsabs.harvard.edu/abs/2011ApJS..192....2S} {192, 2}

\bibitem[\protect\citeauthoryear{{Shporer} et~al.,}{{Shporer}
  et~al.}{2014}]{Shporer2014}
{Shporer} A.,  et~al., 2014, \mn@doi [\apj] {10.1088/0004-637X/788/1/92}, \href
  {https://ui.adsabs.harvard.edu/#abs/2014ApJ...788...92S} {788, 92}

\bibitem[\protect\citeauthoryear{{Siverd} et~al.,}{{Siverd}
  et~al.}{2012}]{Siverd2012}
{Siverd} R.~J.,  et~al., 2012, \mn@doi [\apj] {10.1088/0004-637X/761/2/123},
  \href {http://adsabs.harvard.edu/abs/2012ApJ...761..123S} {761, 123}

\bibitem[\protect\citeauthoryear{{Skrutskie} et~al.,}{{Skrutskie}
  et~al.}{2006}]{Skrutskie2006}
{Skrutskie} M.~F.,  et~al., 2006, \mn@doi [\aj] {10.1086/498708}, \href
  {http://adsabs.harvard.edu/abs/2006AJ....131.1163S} {131, 1163}

\bibitem[\protect\citeauthoryear{{Southworth}}{{Southworth}}{2012}]{Southworth2012}
{Southworth} J.,  2012, \mn@doi [\mnras] {10.1111/j.1365-2966.2012.21756.x},
  \href {http://adsabs.harvard.edu/abs/2012MNRAS.426.1291S} {426, 1291}

\bibitem[\protect\citeauthoryear{{Stamatellos}}{{Stamatellos}}{2013}]{Stamatellos2013}
{Stamatellos} D.,  2013, in European Physical Journal Web of Conferences. p.
  08001 (\mn@eprint {arXiv} {1302.3955}), \mn@doi{10.1051/epjconf/20134708001}

\bibitem[\protect\citeauthoryear{{Stamatellos} \& {Herczeg}}{{Stamatellos} \&
  {Herczeg}}{2015}]{Stamatellos2015}
{Stamatellos} D.,  {Herczeg} G.~J.,  2015, \mn@doi [\mnras]
  {10.1093/mnras/stv526}, \href
  {http://adsabs.harvard.edu/abs/2015MNRAS.449.3432S} {449, 3432}

\bibitem[\protect\citeauthoryear{{Stamatellos} \& {Whitworth}}{{Stamatellos} \&
  {Whitworth}}{2008}]{Stamatellos2008}
{Stamatellos} D.,  {Whitworth} A.~P.,  2008, \mn@doi [\aap]
  {10.1051/0004-6361:20078628}, \href
  {http://adsabs.harvard.edu/abs/2008A%26A...480..879S} {480, 879}

\bibitem[\protect\citeauthoryear{{Szab{\'o}} et~al.,}{{Szab{\'o}}
  et~al.}{2011}]{Szabo2011}
{Szab{\'o}} G.~M.,  et~al., 2011, \mn@doi [\apj] {10.1088/2041-8205/736/1/L4},
  \href {https://ui.adsabs.harvard.edu/#abs/2011ApJ...736L...4S} {736, L4}

\bibitem[\protect\citeauthoryear{{Tamuz} et~al.,}{{Tamuz}
  et~al.}{2008}]{Tamuz2008}
{Tamuz} O.,  et~al., 2008, \mn@doi [\aap] {10.1051/0004-6361:20078737}, \href
  {http://adsabs.harvard.edu/abs/2008A\%26A...480L..33T} {480, L33}

\bibitem[\protect\citeauthoryear{{Thebault} \& {Haghighipour}}{{Thebault} \&
  {Haghighipour}}{2015}]{Thebault2015}
{Thebault} P.,  {Haghighipour} N.,  2015, {Planet Formation in Binaries}.
pp 309--340, \mn@doi{10.1007/978-3-662-45052-9_13}

\bibitem[\protect\citeauthoryear{{Tokovinin}, {Griffin}, {Balega}, {Pluzhnik}
  \& {Udry}}{{Tokovinin} et~al.}{2000}]{Tokovinin2000}
{Tokovinin} A.~A.,  {Griffin} R.~F.,  {Balega} Y.~Y.,  {Pluzhnik} E.~A.,
  {Udry} S.,  2000, \mn@doi [Astronomy Letters] {10.1134/1.20374}, \href
  {https://ui.adsabs.harvard.edu/#abs/2000AstL...26..116T} {26, 116}

\bibitem[\protect\citeauthoryear{{Tokovinin}, {Thomas}, {Sterzik}  \&
  {Udry}}{{Tokovinin} et~al.}{2006}]{Tokovinin2006}
{Tokovinin} A.,  {Thomas} S.,  {Sterzik} M.,   {Udry} S.,  2006, \mn@doi [\aap]
  {10.1051/0004-6361:20054427}, \href
  {http://adsabs.harvard.edu/abs/2006A%26A...450..681T} {450, 681}

\bibitem[\protect\citeauthoryear{{Turner}, {ten Brummelaar}, {Roberts}, {Gies},
  {Mason}  \& {Hartkopf}}{{Turner} et~al.}{2006}]{Turner2006}
{Turner} N.,  {ten Brummelaar} T.,  {Roberts} Jr. L.,  {Gies} D.,  {Mason} B.,
   {Hartkopf} W.,  2006, in The Advanced Maui Optical and Space Surveillance
  Technologies Conference. p.~E104

\bibitem[\protect\citeauthoryear{{Udry}, {Mayor}, {Naef}, {Pepe}, {Queloz},
  {Santos}  \& {Burnet}}{{Udry} et~al.}{2002}]{Udry2002}
{Udry} S.,  {Mayor} M.,  {Naef} D.,  {Pepe} F.,  {Queloz} D.,  {Santos} N.~C.,
   {Burnet} M.,  2002, \mn@doi [\aap] {10.1051/0004-6361:20020685}, \href
  {http://adsabs.harvard.edu/abs/2002A\%26A...390..267U} {390, 267}

\bibitem[\protect\citeauthoryear{{Uyama} et~al.,}{{Uyama}
  et~al.}{2017}]{Uyama2017}
{Uyama} T.,  et~al., 2017, \mn@doi [\aj] {10.3847/1538-3881/153/3/106}, \href
  {http://adsabs.harvard.edu/abs/2017AJ....153..106U} {153, 106}

\bibitem[\protect\citeauthoryear{{Vanhollebeke}, {Groenewegen}  \&
  {Girardi}}{{Vanhollebeke} et~al.}{2009}]{Vanhollebeke2009}
{Vanhollebeke} E.,  {Groenewegen} M.~A.~T.,   {Girardi} L.,  2009, \mn@doi
  [\aap] {10.1051/0004-6361/20078472}, \href
  {http://adsabs.harvard.edu/abs/2009A\%26A...498...95V} {498, 95}

\bibitem[\protect\citeauthoryear{{Wagner}, {Apai}, {Kasper}  \&
  {Robberto}}{{Wagner} et~al.}{2015}]{Wagner2015}
{Wagner} K.,  {Apai} D.,  {Kasper} M.,   {Robberto} M.,  2015, \mn@doi [\apjl]
  {10.1088/2041-8205/813/1/L2}, \href
  {http://adsabs.harvard.edu/abs/2015ApJ...813L...2W} {813, L2}

\bibitem[\protect\citeauthoryear{{Wang}, {Xie}, {Barclay}  \& {Fischer}}{{Wang}
  et~al.}{2014a}]{Wang2014a}
{Wang} J.,  {Xie} J.-W.,  {Barclay} T.,   {Fischer} D.~A.,  2014a, \mn@doi
  [\apj] {10.1088/0004-637X/783/1/4}, \href
  {http://adsabs.harvard.edu/abs/2014ApJ...783....4W} {783, 4}

\bibitem[\protect\citeauthoryear{{Wang}, {Fischer}, {Xie}  \& {Ciardi}}{{Wang}
  et~al.}{2014b}]{Wang2014b}
{Wang} J.,  {Fischer} D.~A.,  {Xie} J.-W.,   {Ciardi} D.~R.,  2014b, \mn@doi
  [\apj] {10.1088/0004-637X/791/2/111}, \href
  {http://adsabs.harvard.edu/abs/2014ApJ...791..111W} {791, 111}

\bibitem[\protect\citeauthoryear{{Wang}, {Fischer}, {Horch}  \& {Huang}}{{Wang}
  et~al.}{2015}]{Wang2015}
{Wang} J.,  {Fischer} D.~A.,  {Horch} E.~P.,   {Huang} X.,  2015, \mn@doi
  [\apj] {10.1088/0004-637X/799/2/229}, \href
  {http://adsabs.harvard.edu/abs/2015ApJ...799..229W} {799, 229}

\bibitem[\protect\citeauthoryear{{White} \& {Ghez}}{{White} \&
  {Ghez}}{2001}]{White2001}
{White} R.~J.,  {Ghez} A.~M.,  2001, \mn@doi [\apj] {10.1086/321542}, \href
  {https://ui.adsabs.harvard.edu/#abs/2001ApJ...556..265W} {556, 265}

\bibitem[\protect\citeauthoryear{{Wilson}, {Kirkpatrick}, {Gizis}, {Skrutskie},
  {Monet}  \& {Houck}}{{Wilson} et~al.}{2001}]{Wilson2001}
{Wilson} J.~C.,  {Kirkpatrick} J.~D.,  {Gizis} J.~E.,  {Skrutskie} M.~F.,
  {Monet} D.~G.,   {Houck} J.~R.,  2001, \mn@doi [\aj] {10.1086/323134}, \href
  {https://ui.adsabs.harvard.edu/#abs/2001AJ....122.1989W} {122, 1989}

\bibitem[\protect\citeauthoryear{{Wilson} et~al.,}{{Wilson}
  et~al.}{2016}]{Wilson2016}
{Wilson} P.~A.,  et~al., 2016, \mn@doi [\aap] {10.1051/0004-6361/201527581},
  \href {http://adsabs.harvard.edu/abs/2016A\%26A...588A.144W} {588, A144}

\bibitem[\protect\citeauthoryear{{Winn} et~al.,}{{Winn}
  et~al.}{2008}]{Winn2008}
{Winn} J.~N.,  et~al., 2008, \mn@doi [\apj] {10.1086/589737}, \href
  {https://ui.adsabs.harvard.edu/#abs/2008ApJ...683.1076W} {683, 1076}

\bibitem[\protect\citeauthoryear{{Wittenmyer}, {Endl}  \&
  {Cochran}}{{Wittenmyer} et~al.}{2007}]{Wittenmyer2007}
{Wittenmyer} R.~A.,  {Endl} M.,   {Cochran} W.~D.,  2007, \mn@doi [\apj]
  {10.1086/509110}, \href {http://adsabs.harvard.edu/abs/2007ApJ...654..625W}
  {654, 625}

\bibitem[\protect\citeauthoryear{{Wittenmyer}, {Endl}, {Cochran}, {Levison}  \&
  {Henry}}{{Wittenmyer} et~al.}{2009}]{Wittenmyer2009}
{Wittenmyer} R.~A.,  {Endl} M.,  {Cochran} W.~D.,  {Levison} H.~F.,   {Henry}
  G.~W.,  2009, \mn@doi [\apjs] {10.1088/0067-0049/182/1/97}, \href
  {http://adsabs.harvard.edu/abs/2009ApJS..182...97W} {182, 97}

\bibitem[\protect\citeauthoryear{{W{\"o}llert} \& {Brandner}}{{W{\"o}llert} \&
  {Brandner}}{2015}]{WollertBrandner2015}
{W{\"o}llert} M.,  {Brandner} W.,  2015, \mn@doi [\aap]
  {10.1051/0004-6361/201526525}, \href
  {http://adsabs.harvard.edu/abs/2015A%26A...579A.129W} {579, A129}

\bibitem[\protect\citeauthoryear{{W{\"o}llert}, {Brandner}, {Bergfors}  \&
  {Henning}}{{W{\"o}llert} et~al.}{2015}]{Wollert2015}
{W{\"o}llert} M.,  {Brandner} W.,  {Bergfors} C.,   {Henning} T.,  2015,
  \mn@doi [\aap] {10.1051/0004-6361/201424091}, \href
  {http://adsabs.harvard.edu/abs/2015A\%26A...575A..23W} {575, A23}

\bibitem[\protect\citeauthoryear{{Wong} et~al.,}{{Wong}
  et~al.}{2015}]{Wong2015}
{Wong} I.,  et~al., 2015, \mn@doi [\apj] {10.1088/0004-637X/811/2/122}, \href
  {https://ui.adsabs.harvard.edu/#abs/2015ApJ...811..122W} {811, 122}

\bibitem[\protect\citeauthoryear{{Wu}, {Murray}  \& {Ramsahai}}{{Wu}
  et~al.}{2007}]{Wu2007}
{Wu} Y.,  {Murray} N.~W.,   {Ramsahai} J.~M.,  2007, \mn@doi [\apj]
  {10.1086/521996}, \href {http://adsabs.harvard.edu/abs/2007ApJ...670..820W}
  {670, 820}

\bibitem[\protect\citeauthoryear{{Youdin} \& {Goodman}}{{Youdin} \&
  {Goodman}}{2005}]{Youdin2005}
{Youdin} A.~N.,  {Goodman} J.,  2005, \mn@doi [\apj] {10.1086/426895}, \href
  {http://adsabs.harvard.edu/abs/2005ApJ...620..459Y} {620, 459}

\bibitem[\protect\citeauthoryear{{Zacharias}, {Monet}, {Levine}, {Urban},
  {Gaume}  \& {Wycoff}}{{Zacharias} et~al.}{2004}]{Zacharias2004}
{Zacharias} N.,  {Monet} D.~G.,  {Levine} S.~E.,  {Urban} S.~E.,  {Gaume} R.,
  {Wycoff} G.~L.,  2004, in American Astronomical Society Meeting Abstracts.
  p.~1418

\bibitem[\protect\citeauthoryear{{Ziegler} et~al.,}{{Ziegler}
  et~al.}{2017}]{Ziegler2017}
{Ziegler} C.,  et~al., 2017, \mn@doi [\aj] {10.3847/1538-3881/153/2/66}, \href
  {http://adsabs.harvard.edu/abs/2017AJ....153...66Z} {153, 66}

\bibitem[\protect\citeauthoryear{{Ziegler} et~al.,}{{Ziegler}
  et~al.}{2018}]{Ziegler2018}
{Ziegler} C.,  et~al., 2018, \mn@doi [\aj] {10.3847/1538-3881/aad80a}, \href
  {http://adsabs.harvard.edu/abs/2018AJ....156..259Z} {156, 259}

\bibitem[\protect\citeauthoryear{{Zucker} \& {Mazeh}}{{Zucker} \&
  {Mazeh}}{2002}]{ZuckerMahez2002}
{Zucker} S.,  {Mazeh} T.,  2002, \mn@doi [\apjl] {10.1086/340373}, \href
  {http://adsabs.harvard.edu/abs/2002ApJ...568L.113Z} {568, L113}

\bibitem[\protect\citeauthoryear{{Zucker}, {Mazeh}, {Santos}, {Udry}  \&
  {Mayor}}{{Zucker} et~al.}{2003}]{Zucker2003}
{Zucker} S.,  {Mazeh} T.,  {Santos} N.~C.,  {Udry} S.,   {Mayor} M.,  2003,
  \mn@doi [\aap] {10.1051/0004-6361:20030499}, \href
  {https://ui.adsabs.harvard.edu/#abs/2003A&A...404..775Z} {404, 775}

\bibitem[\protect\citeauthoryear{{Zucker}, {Mazeh}, {Santos}, {Udry}  \&
  {Mayor}}{{Zucker} et~al.}{2004}]{Zucker2004}
{Zucker} S.,  {Mazeh} T.,  {Santos} N.~C.,  {Udry} S.,   {Mayor} M.,  2004,
  \mn@doi [\aap] {10.1051/0004-6361:20040384}, \href
  {https://ui.adsabs.harvard.edu/#abs/2004A&A...426..695Z} {426, 695}

\bibitem[\protect\citeauthoryear{{Zuckerman}}{{Zuckerman}}{2014}]{Zuckerman2014}
{Zuckerman} B.,  2014, \mn@doi [\apjl] {10.1088/2041-8205/791/2/L27}, \href
  {http://adsabs.harvard.edu/abs/2014ApJ...791L..27Z} {791, L27}

\makeatother
\end{thebibliography}
\input{ms.bbl}


\appendix

\section{Notes on individual targets}
\label{A:binary_search}

\subsection{Bound systems}
\label{A:binaries}

\textbf{11 Com} (HD 107383, HIP 60202) is a common proper motion binary found in the Catalog of Components of Double and Multiple Stars (CCDM; \citealp{Dommanget2000}). The system has a magnitude difference $\Delta V = 8.0$ and an angular separation of 9\farcs1, corresponding to a projected separation of 850 AU at the distance of 11 Com. From the reported magnitude difference, we infer a mass of 0.7 M$_\odot$ for the secondary using the BT-Settl models \citep{Allard2012} and stellar parameters given in Table \ref{t:stellar_properties} for the primary.

\textbf{30 Ari B} (HD 16232, HIP 12184) is part of a hierarchical system. Along with 30 Ari A (HD 16246, HIP 12189), it forms a physical pair with a projected separation of 38\farcs2 or 1520 AU \citep{Shatsky2001}. Both components of the F5V+F8V 30 Ari system are in turn close binaries. In addition to the 9.88 M$_\mathrm{Jup}$ planet orbiting 30 Ari B with a period of $335.1\pm2.5$ days \citep{Guenther2009}, \citet{Riddle2015} found that 30 Ari B is also orbited by another companion, 30 Ari C, with a separation of 22 AU (0\farcs536). \citet{Roberts2015} subsequently demonstrated that the B-C pair is indeed comoving and inferred a mass of 0.5 M$_\odot$ for the C component, classified as an M1V dwarf (see also \citealp{Kane2015}). Moreover, the primary component 30 Ari A is itself a spectroscopic binary with a 1.1 day period \citep{Morbey1974} and a total mass of 1.32 M$_\odot$ \citep{Guenther2009}.

\begin{figure}
    \centering
    \includegraphics[width=0.45\textwidth]{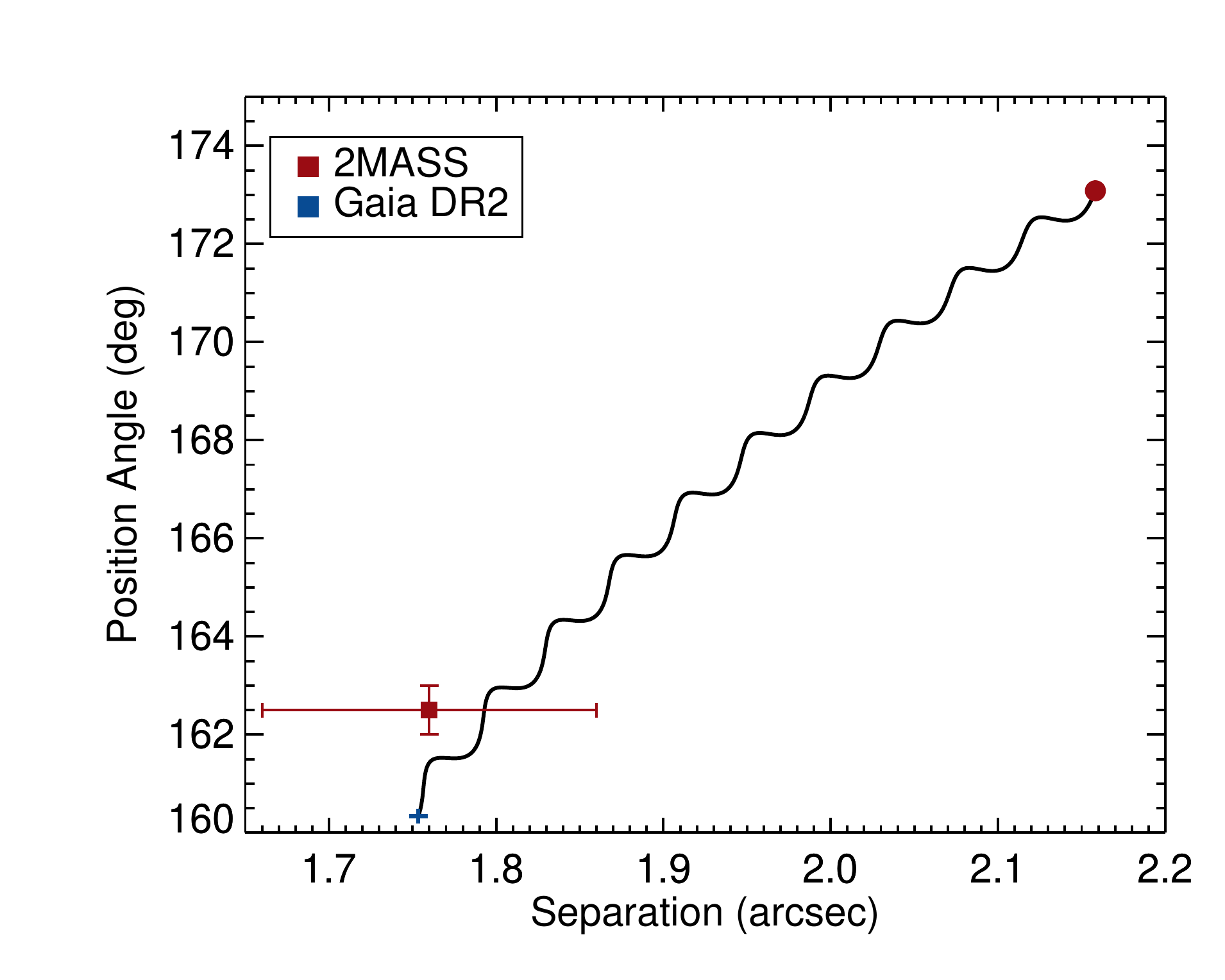}
    \caption{Common proper motion analysis of $\tau$ Gem and its companion over the $\sim$10 year baseline between \textit{Gaia} DR2 (red cross) and the astrometry from 2004 data provided by \citet{Roberts2018} (red square). The black line shows the motion of a background object relative to $\tau$ Gem based on the \textit{Gaia} DR2 parallax and proper motion of the primary, and the red dot indicates the expected position of a background object at the epoch of the 2MASS observations. The close companion is clearly found to be comoving with our target.}
    \label{f:tau_Gem_cpm}
\end{figure}

\textbf{$\tau$ Gem} (HD 54719, HIP 34693) is reported in the CCDM and Washington Double Star (WDS; \citealp{Mason2001}) catalogues to have a candidate companion at a separation of 1\farcs9 and a magnitude $V = 11$ mag. The system was determined to be most likely bound in \citep{Mitchell2013}, who estimated the companion to be a K0 dwarf with a mass of 0.8 M$_\odot$ separated by 187 AU, if real. \citet{Roberts2018} recently provided astrometry for this candidate using data obtained in 2004. They found a separation of 1\farcs76 at a position angle of 162.5 deg. This source is also found in \textit{Gaia} DR2, although it only has a 2-parameter astrometric solution (position only) and therefore does not have parallax and proper motion measurements. From the relative positions of the primary target and candidate in \textit{Gaia} we are able to confirm the bound nature of this system based on the 10 year baseline between the 2004 observations and \textit{Gaia} measurements. Figure \ref{f:tau_Gem_cpm} shows the relative positions of of $\tau$ Gem and its companion at the two epochs, clearly demonstrating that the two objects share common proper motion and thus confirming that they form a physical pair.
The CCDM also reports another candidate to $\tau$ Gem at 59\arcsec. However this latter source is found in the Naval Observatory Merged Astrometric Dataset (Nomad-I; \citealp{Zacharias2004}) to have a proper motion inconsistent with that of the primary (see \citealp{Roell2012}) and we therefore discard this candidate in our survey.

\textbf{$\upsilon$ And} (HD 9826, HIP 7513) was found by \citet{Lowrance2002} to be a wide common proper motion pair on a 55\arcsec separation (750 AU). The secondary stellar component $\upsilon$ And B has a $J$-band magnitude of $9.39\pm0.03$ mag and was estimated by \citet{Lowrance2002} to have an M4.5 V spectral type and a mass of 0.2 M$_\odot$. The primary is the host to 4 close-in planets and substellar companions \citep{Butler1997,Butler1999, McArthur2010, Curiel2011}. This binary system is also mentioned as a physical pair in \citet{Raghavan2006, Raghavan2010}. \citet{Patience2002} and \citet{Roberts2011} also observed the target but did not have a sufficiently wide field of view to detect the distant companion. Neither studies report any additional, more closely-separated candidates around $\upsilon$ And A. 

\textbf{AS 205} (V866 Sco, EPIC 205249328) is an extremely young ($\sim$0.5 Gyr) T Tauri star part of a hierarchical system in Upper sco \citep{Reboussin2015}. The K5 dwarf, and brightest component of the system, was found by \citet{Ghez1993} to form a common proper motion system separated by 1\farcs3 (corresponding to 166 AU at the distance of the system) with a low-mass spectroscopic binary (K7+M0; \citealp{Eisner2005}). \citet{Prato2003} estimated a mass of a mass ratio of \textit{q}$\sim$0.2 between the A and BC components, suggesting a mass of $\sim$0.22 M$_\odot$ for the binary secondary.

\textbf{HAT-P-20} has a red M-dwarf companion at a separation of 6\farcs86 (500 AU) fainter by $\sim$2 mag (WDS catalogue), which was confirmed by \citet{Bakos2011} to form a common proper motion pair using Palomar sky survey archival data. The companion was successfully imaged in the Lucky Imaging survey by \citet{WollertBrandner2015} but was missed in observations from \citet{Ngo2015} due to the restricted field of view of their data. From the reported photometry and adopting the stellar parameters in Table \ref{t:stellar_properties}, we derive a mass of 0.57 M$_\odot$ for this companion using the BT-Settl models at the age of the system. 

\textbf{HD 41004 B} (HIP 28393) was identified in \citet{Santos2002} and \citet{Zucker2003} as the lowest-mass component of a K1V+M2V visual binary with a 0\farcs54 separation, corresponding to 23 AU. The system has a $V$-magnitude difference of 3.7 mag and is catalogued as a physical pair in the WDS, CCDM and Tycho-Hipparcos catalogues (see \citealp{Roell2012}). Both components are hosts to close-in substellar companions: the 0.7 M$_\odot$ primary, HD 41004 A, is orbited by a giant planet at 1.33 AU with a projected mass of $2.54\pm0.74$ M$_\mathrm{Jup}$ \citep{Zucker2004}, while HD 41004 B (0.4 M$_\odot$) hosts a brown dwarf companion at 0.017 AU with a minimum mass of $18.37\pm0.22$ M$_\mathrm{Jup}$ \citep{Zucker2003}.

\textbf{HD 87646 A} (HIP 49522) is flagged as a binary in the Tycho and Hipparcos catalogues with a magnitude difference in the Hipparcos $V$-band of $2.66\pm0.97$ mag. \citet{Ma2016} acquired high-resolution images of the system and found a separation 0\farcs26 ($\sim$20 AU) between the G-dwarf primary and K-dwarf secondary. The authors derive a mass of $1.12\pm0.09$ M$_\odot$ for the A component and estimate a mass ratio of $q\sim0.5$ for the system. In addition to the 12.4 M$_\mathrm{Jup}$ giant planet found at 0.117 AU around HD 87646 A, \citet{Ma2016} also report a 57 M$_\mathrm{Jup}$ brown dwarf candidate companion on an eccentric 1.6 AU orbit around the primary star.

\textbf{HD 89744} (HIP 50786) is a wide binary on a 63\arcsec separation first identified spectroscopically by \citet{Wilson2001} and subsequently confirmed astrometrically to form a common proper motion pair by \citet{Mugrauer2004}. The large angular separation of the binary corresponds to a projected separation of $\sim$2460 AU. \citet{Mugrauer2004} estimated a mass of 0.08 M$_\odot$ for the secondary, near the hydrogen-burning limit. \citet{Raghavan2006} estimated an L0V spectral type for the companion. HD 89744 was also observed by \citet{Roberts2011} with Adaptive Optics on the AEOS telescope, who found a faint candidate companion at 5\farcs62 with a magnitude difference of $\Delta I = 13\pm2$ mag that is yet to be confirmed. Other sets of observations with PUEO-KIR at CFHT by \citet{Chauvin2006} or the UFTI data obtained by \citet{Mugrauer2004} do not go deep enough at that separation to retrieve this candidate. Given the observed magnitude difference, we infer a mass of 0.08 M$_\odot$ for this candidate from the BT-Settl isochrones.

\textbf{HD 114762} (HIP 64426) is a WDS 3\farcs2 (140 AU) binary pair confirmed to be comoving by \citet{Patience2002} using data from Keck/NIRC and Shane/IRCAL. \citet{Bowler2009} further characterised the system, estimating an M9 spectral type and inferring a mass of 0.09 M$_\odot$ for the companion. The companion is also reported in the Adaptive Optics survey by \citet{Roberts2011}.

\textbf{HD 156846} (HIP 84856) is reported as a wide, bound binary in the WDS catalogue, with a separation of 5\farcs1 (250 AU). \citet{Tamuz2008} characterised the companion to the G0 planet host as an M4 dwarf of mass 0.59 M$_\odot$.

\textbf{HD 178911 B} (HIP 94075) is the fainter component of a 16\farcs1 (790 AU) physical pair found in the Tycho-Hipparcos catalogue. The primary component HD 178911 AC is itself a 4.9 AU spectroscopic binary discovered by \citet{McAlister1987}. The triple system was established to be comoving by \citet{Tokovinin2000} and subsequently confirmed by \citet{Raghavan2006}. \citet{Tokovinin2000} estimated a combined mass of 1.9 M$_\odot$ for the AC component, consistent with the value reported in \citet{Mugrauer2007}, while the planet host HD 178911 B has a mass of 1 M$_\odot$ \citep{Mugrauer2007, Bonfanti2016}.

\textbf{Kepler-13 A} (KOI-13) has been extensively targeted with direct imaging \citep{Adams2012, Law2014, Shporer2014, WollertBrandner2015, Kraus2016}. \citet{Szabo2011} reported and confirmed Kepler-13 as a common proper motion system composed of two massive A stars, also found in the CCDM catalogue. \citet{Santerne2012} found the secondary component to be a spectroscopic binary. \citet{Johnson2014} later constrained the mass of Kepler-13 C to be between $0.4-0.75$ M$_\odot$, for a total mass of $1.68\pm0.10$ M$_\odot$ for the BC component, and $1.72\pm0.10$ M$_\odot$ for Kepler-13 A, respectively \citep{Shporer2014}. The A-BC system has a projected angular separation of 1\farcs15, corresponding to a physical projected separation of 610 AU \citep{Szabo2011, Adams2012, Law2014}.

\textbf{NLTT 41135} was identified by \citet{Irwin2010} as a physically associated companion to NLTT 41136 at 2\farcs4 separation (55 AU). From their characterisation of the system, the authors inferred masses of 0.16 M$_\odot$ for NLTT 41135 and 0.21 M$_\odot$ for NLTT 41136, respectively.

\textbf{WASP-14} was found in \citet{Wollert2015} to have a candidate companion at 1\farcs4, 5.4 magnitudes fainter than the primary in AstraLux Norte observations at the Calar Alto 2.2 m telescope. \citet{Ngo2015} independently identified the same candidate and were able to confirm the source to be a common proper motion companion to WASP-14 with a mass of $0.33\pm0.04$ M$_\odot$. We also found in this work a distant companion to the system at 1900 AU, identified in the \textit{Gaia} DR2 catalogue. We characterise WASP-14 C as an 0.28 M$_\odot$ K5 dwarf (see Section \ref{WASP-14}).

\subsection{Unconfirmed Candidate Companions}
\label{A:good_candidates}

\textbf{70 Vir} (HD 117176, HIP 65721) was observed by \citet{Roberts2011} using the Advance Electro-Optical System (AEOS) telescope. The authors report a candidate companion at a separation of 2\farcs86 (52 AU) around 70 Vir, which they classify as an M5 dwarf or later. With a magnitude difference of $\Delta I = 11.4\pm1.2$, we estimate a mass of 0.08 M$_\odot$ for this candidate using the BT-Settl models.
\citet{Pinfield2006} reported an L-dwarf candidate at 43\arcsec (848 AU) based on data from the 2MASS All Sky Catalogue. Common proper motion with the primary has yet to be determined for both candidates. We did not find the latter candidate as a \textit{Gaia} DR2 source, most likely too faint for \textit{Gaia}. 70 Vir had also previously been observed by \citet{Patience2002}. Observations from this survey are not deep enough to retrieve the candidate found by \citet{Roberts2011} and do not have a large enough field of view to detect the wide source from \citet{Patience2002}. Given the faint infrared 2MASS magnitude of the wide source ($J = 15.84\pm0.16$ mag), we estimate a mass of 0.07 M$_\odot$ for the candidate adopting the age of the primary and the BT-Settl isochrones.

\textbf{EPIC 219388192} is solar twin in the old Ruprecht 147 star cluster \citep{Curtis2013, Nowak2017} which was found by \citet{Nowak2017} to host an eccentric transiting brown dwarf companion. The team acquired Subaru/IRCS+AO188 images of the target to search for nearby companions and found two point sources at 6\arcsec and 7\farcs5 with contrasts of $\Delta H = 7.1$ mag $\Delta H = 7.7$ mag, respectively. \citet{Nowak2017} estimated that the candidates, if found to be bound, would be late-type M dwarfs with masses less than $\sim$0.1 M$_\odot$. Both sources are found in the \textit{Gaia} DR2 catalogue with separations and position angles consistent with those reported by \citet{Nowak2017}. However, given the relatively small proper motion of the target and the short time baseline between \textit{Gaia} DR2 and the direct imaging data ($\sim$1 year), we are not able to confirm or refute either of those candidates.
Curtis et al. independently studied the same object and found 4 km s$^{-1}$ offset between the center-of-mass radial velocity of the star and Ruprecht 147's bulk velocity (announced at the Cool Stars 19 workshop \footnote{\url{https://doi.org/10.5281/zenodo.58758}}). As the star's proper motion supports its cluster membership, Curtis et al. also obtained aperture-masking interferometry with Keck II and uncovered a 0.52 M$_\odot$ companion at 82 mas (24 AU) with a magnitude contrast $\Delta K = 2.24$, explaining the observed offset (Curtis et al. private communication).

\textbf{KELT-1} was found by \citet{Siverd2012} to have a faint candidate companion at $588\pm1$ mas ($154\pm8$ AU) based on Keck/NIRC2 AO data. The relative brightness of the candidate was found to be $\Delta H = 5.90\pm0.10$ maf and $\Delta K' = 5.59\pm0.12$ mag. The reported photometry suggests a mass of 0.2 M$_\odot$ based on the BT-Settl models. \citet{Siverd2012} estimated an M4$-$5 spectral type and concluded that the companion is physically associated to the primary, with a $\sim$0.05\% probability of being an unrelated background star based on Galatic models, in excellent agreement with our estimates (see Table \ref{t:candidate_binaries}).
This target was more recently observed by \citet{Coker2018} with the WIYN 3.5 m telescope and by \citet{WollertBrandner2015} with the Calar Alto 2.2 m telescope, although neither of these sets of observations were deep enough the retrieve the candidate identified in \citet{Siverd2012}.

\subsection{Rejected Candidates}
\label{A:rejected_candidates}

\textbf{HD 162020} (HIP 87330) had previously been observed with NACO by \citet{Eggenberger2007}, who found two point sources within 5\arcsec from the star. The first, closer candidate was found by \citet{Eggenberger2007} to be background, while the nature of the second source was inconclusive. With new NACO data for this target, we were able to refute the bound nature of this companion based on the \textit{Gaia} DR2 astrometry of the primary and a decade-long baseline between the archival and new observations (see Figure \ref{f:CPM_HD_162020}). Our proper motion analysis of HD 162020 and this companion is presented in Section \ref{HD_162020}.

\textbf{HD 168443} (HIP 89844) was observed with SPHERE at VLT in the survey conducted in \citet{Moutou2017}. Three point sources are reported within 2\farcs5 of the primary in that paper. \citet{Moutou2017} found that given the galactic latitude of the target and the crowded field of view at wider separations around this object, the three identified sources are likely background contaminants due to the local environment of HD 168443. Using the Trilegal galactic models \citep{Vanhollebeke2009} and following the approach described in Section \ref{lit_search}, we infer probabilities $<$ 1\% for any of these three sources to be physical associated to the primary and do not consider them as bonafide candidates for the purpose of our study.

\textbf{XO-3} has a faint candidate companion ($i$ = 18.43 mag) first reported in \citet{Bergfors2013}. The widely-separated candidate (6\arcsec or 1500 AU projected separation) was found by \citet{Bergfors2013} to likely be a physically unrelated background object if it is a main-sequence star based on a colour analysis, although the authors mention the possibility of a coeval white dwarf. This target was observed with Keck/NIRC2 in \citet{Ngo2015} but was not retrieved in the field of view of the images acquired for that survey. \citet{WollertBrandner2015} also imaged XO-3 in a search for wide companions and detected the same source in AstraLux Norte data. A faint source is found in \textit{Gaia} DR2 ($G$ = 18.45 mag) at the same angular separation and position angle as the detected candidate. Given the comparable photometry and the short timespan between the \textit{Gaia} DR2 epoch (2015.5) and the observations from \citet{WollertBrandner2015} (March 2015), we conclude that this is indeed the same source. The \textit{Gaia} DR2 source (GDR2 470650457698311296) has a full 5-parameter astrometic solution and has parallax and proper motion measurements highly inconsistent with those of XO-3 in \textit{Gaia} DR2. We thus conclude that it is an unrelated background object and rule out this candidate.

\subsection{Null-detections}
\label{A:null_detecitons}

\textbf{BD+24 4697} (HIP 113698) was observed with Gemini North/NIRI as part of this survey. Our data did not reveal the presence of any candidate companion within the field of view and detection limits of our observations.

\textbf{CI Tau} (EPIC 247584113) is a $\sim$2 Myr T Tauri star located in the Taurus star-forming region with an infrared excess in its SED and a dics resolved by \citet{Andrews2007}. It was observed by \citet{Uyama2017} with the Subaru Telescope, using the NIR camera HiCIAO together with the AO188 adaptive optics system, in quad PDI and standard ADI modes. The authors did not find any candidate companion within the 20\arcsec$\times$20\arcsec field of view of their observations. This targets is also reported to be single in \citet{Kraus2012} based on analyses of 2MASS images and in the HST young binary survey by \citet{White2001}. 

\textbf{HAT-P-2} (HD 147506, HIP 80076) was found by \citet{Lewis2013} to have a long-term radial velocity trend, suggesting the presence of an outer companion in addition to its known 9 $M_\mathrm{Jup}$ planet on an eccentric 5 days orbit. \citet{Bonomo2017} subsequently placed lower limits on the period and mass of this possible outer companion of $\geqslant$49.2 yrs ($\sim$13 AU) and $\geqslant$39.5 M$_\mathrm{Jup}$ based on radial velocity data. This is consistent with results from \citet{Knutson2014} who constrained the companion properties to $M_2\sin{i}=8-200$ M$_\mathrm{Jup}$ and $a=4-31$ AU. Observations with NIRC2 on Keck II \citep{Lewis2013, Ngo2015} and with AstraLux Norte \citep{Bergfors2013} did not reveal any companion but only excluded the presence of an $\sim$equal-mass binary from $\sim$10 AU and companions near the hydrogen-burning limit from $\sim$50$-$100 AU. A companion responsible for the observed RV trend could therefore still remain undetected.

\textbf{HD 5891} (HIP 4715) was observed by \citet{Ginski2016} with the Lucky Imaging instrument AstraLux at the Calar Alto 2.2 m telescope and did not find any companion, achieving contrasts of 4 mag at 1\arcsec and 9.5 mag at 5\arcsec.

\textbf{HD 33564} (HIP 25220) is listed in the CCDM as a 25\arcsec binary although the 2 components display inconsistent proper motions and do not form a physical pair \citep{Roell2012}. \citet{Ginski2012} acquired observations of HD 33564 and excluded the presence of companions down to the substellar limit on separations of 20$-$100 AU. This star is also reported as a single object in \citet{Eggleton2008}.

\textbf{HD 77065} (HIP 44259) is one of the two targets we observed with NIRI on Gemini North. We did not find any candidates around this target in our images, ruling out companions at the hydrogen-burning limit from separations of 5 AU and substellar companions with masses $> 40$ M$_\mathrm{Jup}$ from 70 AU.

\textbf{HD 104985} (HIP 58952) was observed with the lucky imaging camera AstraLux on the Calar Alto 2.2-m telescope by \citet{Ginski2012}. The team did not find any candidate around this target.

\textbf{HD 134113} (HIP 74033) is part of the Arcturus moving group. We observed this target with the WIYN telescope and did not find any companions within our detection limits. HD 134113 has no previous direct imaging observations reported in the literature.

\textbf{HD 156279} (HIP 84171) was observed by \citet{Ginski2016} with the AstraLux instrument on the Calar Alto 2.2-m telescope. No companion was detected in the obtained lucky imaging data.

\textbf{HD 160508} (HIP 86394) was observed as part in this work using the WIYN imaging facilities. We did not detect any companions around this object within the field of view of our images.

\textbf{HD 180314} (HIP 94576) was targeted by \citet{Ginski2016} with lucky imaging at Calar Alto. No source was uncovered in the obtained data within 12\arcsec, down to low-mass stellar companions.

\textbf{HD 203949} (HIP 105854) was observed with VLT/SPHERE in \citet{Moutou2017}. That survey does not report the detection of any candidates around this target.

\textbf{WASP-18} (HD 10069, HIP 7562) was part of our observed sample and no source was detected in the field of view of our images. This object had already been observed with Keck II/NIRC2 in \citet{Ngo2015}. No candidate was reported around WASP-18 in that survey. We achieved a better contrast than that reported in \citet{Ngo2015} both at diffraction and background-limited separations and our observations allowed us to rule out the presence of lower-mass companions around WASP-18. A comoving object was however found in this work in GDR2 at 3300 AU, outside the field of view of the direct imaging data (26\farcs7), for which we estimated a spectral type of M7.5 and a mass of 0.092 M$_\odot$ (see Section \ref{WASP-18}).

\section{\textit{Gaia} DR2 Analysis}
\label{A:Gaia_analysis}

In Section \ref{gaia_dr2} we searched for sources in the \textit{Gaia} DR2 catalogue with fractional differences of less than 20\% in parallax and at least one proper motion component relative to the \textit{Gaia} astrometry of our targets. Using these selection constraints, we identified a total of 11 binaries in \textit{Gaia} DR2 among the targets in our sample, 9 of which were previously known. We now examine those systems more carefully as well as the remaining systems from Table \ref{t:confirmed_binaries} in order to evaluate and refine our selection criteria, if needed.

\subsection{Binary Completeness}

For completeness, we first searched for other known binaries in our sample that may have been missed by our chosen constraints. A total of 7 known, comoving systems are missing from our identified \textit{Gaia} binaries, corresponding to the companions with no parallax or proper motion listed in Table \ref{t:confirmed_binaries}. From those, 30 Ari BC, HD 41004 AB and HD 87646 AB have angular separations $<$1\arcsec, the resolving limit of \textit{Gaia} DR2, and were therefore missed because of angular resolution limitations.

While near-equal brightness binaries ($\Delta G < 1$ mag) are typically resolved with \textit{Gaia} from separations of $\sim$1\arcsec (e.g. Kepler-13, 1\farcs15 separation, $\Delta$mag = 0.2 mag; AS 205 AB, 1\farcs3, $\Delta G = 1$ mag), larger separations are required to resolve lower mass ratio binaries. \citet{Ziegler2018} estimated that companions with $\Delta G$ down to $\sim$6 mag are consistently recovered at separations of 3\arcsec$-$5\arcsec, with a roughly linear decrease in the recoverable magnitude difference between 1\arcsec$-$3\arcsec. Based on these results, it is not surprising that systems such as WASP-14 AB (1\farcs45, $\Delta$\textit{J}$=$5.2 mag) and HD 114762 AB (3\farcs2, $\Delta$\textit{J}$=$7.6 mag) are not retrieved in \textit{Gaia} DR2. We thus conclude that these companions are missing from our \textit{Gaia} binary sample because they are fainter than the completeness level of \textit{Gaia} DR2.

Finally, the last missing binaries are $\tau$ Gem AB and HAT-P-20 AB. In both cases, the two binary components were found to be \textit{Gaia} DR2 sources, but the fainter component only had a two-parameter astrometric solution (position only) rather than the full 5-parameter solution (position, parallax and proper motion). With no parallax and proper motion measurements, we were not able to select these systems in our analysis and we attribute the fact that we missed them to the remaining incompleteness of \textit{Gaia} DR2 and not to our selection criteria. We thus conclude that our selection method successfully identified all known binaries that were recoverable.

\subsection{Binaries with excessive astrometric disparities}

Table \ref{t:Gaia_frac_diff} reports the relative differences in parallax and proper motion, together with their associated uncertainties, obtained for all identified \textit{Gaia} binaries (see also Figure \ref{f:Gaia_binaries}). While the majority of the errors are comparable in size to the calculated values themselves, all systems remain fully within our arbitrary cuts at 20\% at the 1-$\sigma$ level (with the exception of the newly discovered WASP-18 AB system which is discussed in Section \ref{WASP-18}).

A number of binaries in Table \ref{t:confirmed_binaries} are part of hierarchical systems and we find that 4 of the 9 previously known \textit{Gaia} systems have an unresolved component in \textit{Gaia} DR2 (30 Ari BC, AS 205 BC, HD 178911 AC and Kepler-13 BC), which correspond to the blue stars in Figure \ref{f:Gaia_binaries}. Looking at the positions of these specific systems in the parameter-space from Figure \ref{f:Gaia_binaries}, we find that they have the largest relative offsets in parallax and/or proper motion, and are the only systems for which the relative difference in proper motion was larger than our 20\% threshold in one of the coordinates (outside the shaded area in Figure \ref{f:Gaia_binaries}).

This is consistent with the idea that unresolved components can have a significant effect on the measured astrometry of binary pairs, reinforcing the argument for loose constraints in order to ensure that such hierarchical systems are not missed. With the exception of AS 205 and HD 178911, all known binaries detectable by \textit{Gaia} would also have made a more stringent cut at $\sim$10\% in the relative difference in parallax and in one of the proper motion components. Furthermore, the 5 known binaries that are not known to have an unresolved component (blue circles in Figure \ref{f:Gaia_binaries}) also make that 10\% cut in both proper motion components. We thus conclude that most binaries should have relative discrepancies of $<$10\% in all astrometric parameters ($\pi$, $\mu_{\alpha*}$ and $\mu_\delta$), while systems agreeing to within 20\% in parallax and in one of the proper motion coordinates are likely to be hierarchical systems with an unresolved component. 

We note that such wide companions are not necessarily presently bound systems. Formerly physically associated components of a binary system may continue to travel along a nearly identical trajectory. However, we are seeking in this study companions that may have affected the formation or early evolution of inner companions and therefore also consider as bonafide any pair that previously constituted a bound system. We also point out that such a configuration would likely result in small discrepancies in the observed astrometric parameters of the individual components, an additional argument for the loose constraints considered above. In conclusion we trust that systems passing the selection criteria described above have consistent astrometric parameters and kinematics, and may be treated as binaries for the purpose of this work.


\bsp	
\label{lastpage}
\end{document}